\documentclass[notitlepage, aps,prd, reprint,preprintnumbers,onecolumn, longbibliography, showpacs]{revtex4-2}

\usepackage[%
colorlinks=true,
urlcolor=blue,
linkcolor=blue,
citecolor=blue
]{hyperref}
\usepackage{slashed, color, amsmath, amssymb, mathrsfs}
\usepackage{graphicx}
\usepackage{hyperref}
\usepackage{longtable}
\usepackage{array}
\usepackage{accents}
\usepackage{tensor}
\usepackage{longtable}
\usepackage{booktabs}
\usepackage{multirow}
\usepackage{bm}
\usepackage{subfigure}
\usepackage{float}
\graphicspath{{figures/}}

\newcommand{\pr}{\operatorname{pr}}
\allowdisplaybreaks

\begin{document}

\title[]{Bayesian method for fitting the low-energy constants in chiral perturbation theory}
\author{Hao-Xiang Pan$^{1}$}
\author{De-Kai Kong$^{1}$}
\author{Qiao-Yi Wen$^{2}$}
\author{Shao-Zhou Jiang$^{1}$}
\email{jsz@gxu.edu.cn}
\affiliation{$^{1}$ Key Laboratory for Relativistic Astrophysics, School of Physical Science and Technology, Guangxi University, Nanning 530004, People's Republic of China\\
$^{2}$ Department of Physics and Siyuan Laboratory, Jinan University, Guangzhou 510632, People's Republic of China}

%\date{\today}
\begin{abstract}
The values of the low-energy constants (LECs) are very important in the chiral perturbation theory. This paper adopts a Bayesian method with the truncation errors to globally fit eight next-to-leading order (NLO) LECs $L_i^r$ and next-to-next-leading order (NNLO) LECs $C_i^r$. With the estimation of the truncation errors, the fitting results of $L_i^r$ in the NLO and NNLO are very close. The posterior distributions of $C_i^r$ indicate the boundary-dependent relations of these $C_i^r$. Ten $C_i^r$ are weakly dependent on the boundaries and their values are reliable. The other $C_i^r$ are required more experimental data to constrain their boundaries. Some linear combinations of $C_i^r$ are also fitted with more reliable posterior distributions. If one knows some more precise values of $C_i^r$, some other $C_i^r$ can be obtained by these values. With these fitting LECs, most observables provide a good convergence, except for the $\pi K$ scattering lengths $a_0^{3/2}$ and $a_0^{1/2}$. An example is also introduced to test the improvement of the method. All the computations indicate that considering the truncation errors can improve the global fit greatly, and more prior information can obtain better fitting results. This fitting method can be extended to the other effective field theories and the perturbation theory.
\end{abstract}

\maketitle
\section{Introduction}\label{intro}\label{Sec:I}
Effective field theory (EFT) is a very important theory in dealing with interactions between particles under a low-energy scale. Chiral perturbation theory (ChPT) is a kind of EFT. It first focuses on the low-energy strong interactions between the low-energy pseudoscalar mesons and then extends to baryons and other mesons. ChPT is based on the $SU(3)_L\times SU(3)_R$ flavor symmetry in the chiral limit, in which the three lightest quarks are considered massless. The only constraints of the chiral Lagrangian are symmetries, such as charge conjugate symmetry, parity symmetry, chiral symmetry, etc. However, there are infinite independent terms satisfying these symmetries. The Weinberg power-counting scheme expands these terms by the mesonic momentum ($p$) \cite{Weinberg:1978kz}. The leading-order (LO, $\mathcal{O}(p^2)$ order) terms give the most contributions, and they are considered first. If one wants to obtain a higher precision, the terms in the next-to-leading order (NLO, $\mathcal{O}(p^4)$ order), the next-to-next-to-leading order (NNLO, $\mathcal{O}(p^6)$ order), etc. will be considered gradually. Each term contains a corresponding unknown parameter, called low-energy constant (LEC), which contains the information of the effective strong interactions. For the three-flavor ChPT, there are 2, 10+2, 90+4, and 1233+21 LECs in the LO, NLO, NNLO and next-to-next-to-next-to-leading order ($\mathcal{O}(p^8)$ order) \cite{Gasser:1983yg, Gasser:1984gg,Bijnens:1999sh, Bijnens:2018lez}, respectively. If all these LECs were known, all theoretical calculations would be obtained numerical values. However, the number of these LECs are too large, especially in the high orders. Besides, with CHPT itself, one cannot fix these LECs. The LECs are usually determined by the other approaches, such as global fit \cite{Bijnens:2011tb,Bijnens:2014lea,Yang:2020eif}, lattice QCD \cite{Can:2015exa,Can:2013tna,Bahtiyar:2016dom}, chiral quark model \cite{Yan:1992gz,Dowdall:2013rya,MILC:2010hzw,Bernard:2009ds}, resonance chiral theory \cite{Dowdall:2013rya, MILC:2010hzw, Bazavov:2009fk, Bazavov:2009bb, Bernard:2009ds}, sum rules \cite{Golterman:2014nua}, holographic QCD \cite{Colangelo:2012ipa}, dispersion relations \cite{Guo:2007ff, Guo:2007hm, Guo:2009hi}, and so on. Each method has its advantages and sphere of application. Until now, no approach can determine the exact values of these LECs. This paper only focuses on the global fit method.

There has been a lot of research based on global fits. Some LECs up to NNLO have been fitted. Ref. \cite{Bijnens:1994ie} fits $K_{\ell 4}$ form factors and $\pi \pi$ scatter lengths to get the values of $L_1^r$, $L_2^r$ and $L_3^r$. Six years later, $L_i^r(i=1, 2, 3, 5, 7, 8)$ is determined by fitting the quark mass ratio $m_s/\hat m$, the decay constant ratio $F_K/F_\pi$ and the $K_{\ell4}$ form factor \cite{Amoros:2000mc}. Another eleven years later, a new global fit appears, adding $\pi \pi$ scattering lengths ($a_0^0$ and $a_0^2$), $\pi K$ scattering lengths ($a_0^{1/2}$ and $a_0^{1/3}$) and scalar form factor threshold parameters ($\langle r^2\rangle_S^\pi$ and $c_S^\pi$). $L_{1-8}^r$ and some $C^r_i$ are obtained \cite{Bijnens:2011tb}. Ref. \cite{Bijnens:2014lea} adds some two-flavor LECs and updates the values of LECs fitted in Ref. \cite{Bijnens:2011tb}. The last two references not only fit the LECs $L_{1-8}^r$ at the NLO, but also estimate a part of the NNLO LECs $C_i^r$. However, both of them ignore the higher-order truncated contributions. Ref. \cite{Yang:2020eif} proposes a geometric sequence model to introduce the higher-order truncated contributions. Its NLO fitting values of $L_i^r$ are very close to the NNLO fitting values in Ref. \cite{Bijnens:2014lea}. This is because a physical quantity contains not only the sum of LO and NLO theoretical values, but also the sum of higher-order contributions, which sometimes cannot be ignored when compared to the NLO contributions. If the NLO fit includes the higher-order contributions, $L_i^r$ will be closer to the true values. Hence, fitting $L_i^r$ at NLO and NNLO yields closer results. This shows that the higher-order contributions indeed cannot be simply ignored in the ChPT fit. Above all references have adopted a classical statistical method to fit LECs. Theoretically, the precision of the fitting results is dependent on the amount and precision of the experimental data. In other words, more data and more precise data will lead to more precise LECs. However, there exist some problems in the classical statistics and some improvements are needed.
\begin{enumerate}
\item The geometric sequence model in Ref. \cite{Yang:2020eif} is too simple. The contribution at each order, in fact, needs not be a geometric sequence. In addition, in order to estimate the NNLO contribution, the geometric sequence itself requires the LO and NLO contributions. However, the LO contribution is sometimes zero, so the NNLO contribution cannot be estimated. In some special cases, the NNLO contribution may be larger than the NLO one, such as $\pi K$ scattering lengths $a_0^{1/2}$ and $a_0^{3/2}$ \cite{Bijnens:2011tb,Bijnens:2014lea}. Hence, Ref. \cite{Yang:2020eif} adopts a special approach to deal with this problem. In most cases, two two-flavor NLO LECs $\bar l_{2,3}$ have a bad convergence. It takes a long time to fit them, about one day with 20 cores in a CPU Intel Xeon Gold 6230. In addition, how to confirm the sign of the NNLO contribution is also a problem. These cause that the model is not consistent for all physical quantities. The model is not a universal approach.

\item The number of $C_i^r$ is much larger than the number of the input experimental data. There exists an overfitting problem in the NNLO fit. Refs. \cite{Bijnens:2011tb,Bijnens:2014lea} adopt a random walking algorithm, but the result is boundary-dependent. A Monte Carlo method is used to fit the LECs in Ref. \cite{Yang:2020eif}, but its efficiency is low. Moreover, the complicated errors of $C_i^r$ are hard to be estimated. They usually cannot be obtained as a normal distribution.

\item Although the geometric sequence model gives a reasonable result in Ref. \cite{Yang:2020eif}, this model is hard to extend to the other EFTs, because of the reasons discussed above. Furthermore, it is also hard to evaluate different models in order to select the best one, because $\chi^2/\mathrm{d.o.f.}$ (degrees of freedom) is too small and an overfitting problem exists. It is hard to select the best model from some overfitting models by $\chi^2/\mathrm{d.o.f.}$ A more universal method requires a credible quantified index. The best model can be selected by this index.

\item Refs. \cite{Bijnens:2011tb,Bijnens:2014lea, Yang:2020eif} treat the two-flavor NLO LECs $\bar l_i$ as the independent input experimental data, but some $\bar l_i$ are possibly dependent on other experimental quantities. In fact, the $\pi\pi$ scattering lengths $a_0^{0}$ and $a_0^{2}$ are dependent on $\bar l_1$, $\bar l_2$ and $\bar l_4$ \cite{Colangelo:2001df}. Hence, their covariance matrix needs to be considered.

\item The most important thing is that before a global fit, one has known something about ChPT and the fitting experimental data, but this information does not obviously embody in the fit. For example, for the NLO fitting, although the truncation errors are not known, the other references have given some approximate values of the NNLO LECs. With these NNLO LECs, even at the NLO fit, one can roughly obtain the signs of the truncation errors. If these signs are introduced into the NLO fit, the results may be more precise. Furthermore, ChPT assumes that the orders of magnitude of LECs at a given chiral order are nearly the same. If this information is considered in the global fit, the range of the unknown LECs even through the NNLO contribution can be estimated. Simply speaking, more information may lead to a more precise result.
\end{enumerate}

In addition to classical statistics, Bayesian statistics, which has been successful in artificial intelligence, can play a better role in the global fit of EFTs. Bayesian statistics can make good use of the known information to give a more reasonable result. Even when the amount of data is small, Bayesian statistics can be better than classical statistics. Ref. \cite{Schindler:2008fh} has applied Bayesian statistics to EFTs. It proposes two toy models and compares the results obtained by Bayesian and classical statistics. The advantages of Bayesian statistics in EFTs have been demonstrated. Later, Ref. \cite{Furnstahl:2014xsa} introduces Bayesian statistics into nuclear physics. A year later, a specific framework for using Bayesian statistics in EFTs appears \cite{Wesolowski:2015fqa}. Subsequently, Refs. \cite{Melendez:2017phj, Svensson:2021lzs, Ekstrom:2019twv, Wesolowski:2018lzj,Alnamlah:2020cko,Yang:2020pgi,Lovell:2020sep,Phillips:2020dmw,Bedaque:2021bja,Wesolowski:2021cni,Connell:2021qcd,Lin:2021umz,Djarv:2021hcj,Acharya:2021lrv,Odell:2021nmp,Lovell:2022pkw,Hagen:2022tqp,Papenbrock:2022vdf,Muli:2022jma,Zhai:2022ied,Fraboulet:2022zvt,Jiang:2022off,Ekstrom:2022yea,Jay:2020jkz,Catacora-Rios:2019goa,Ekstrom:2019lss,Zhang:2019odg,Luna:2019ufu,Epelbaum:2019zqc} use Bayesian statistics to calculate the magnitudes of truncation errors in the different EFTs. This paper will improve the approach in Ref. \cite{Yang:2020eif}. The new approach contains the framework of Bayesian statistics and the application of Markov Chain Monte Carlo (MCMC). Some MCMC algorithms, such as the Metropolis-Hastings algorithm \cite{Metropolis:1953am,Hastings:1970aa}, Hamiltonian Monte Carlo algorithm \cite{DUANE1987216} and No-U-turn Sampler algorithm \cite{hoffman2011nouturn}, will be used to fit the LECs with the help of the PyMC3 package \cite{Salvatier2016}. The major improvements of the new approach and the motivations of this paper are as follows.
\begin{enumerate}
\item The geometric sequence is not required in the fit. It is replaced by a Bayesian method. Generally, the new method does not require the assumptions about how ChPT converges.

\item The approach is more general. Some examples are carried out to check whether the approach works well. The parameters in the examples are completely random. Hence, this approach is not only used to fit the LECs in ChPT, but also can be applied to other EFTs and perturbation theory.

\item The cost of time for this approach is greatly reduced with the help of MCMC. A better result will be obtained within ten minutes.

\item The covariance matrix given in Ref. \cite{Colangelo:2001df} will be considered in the fit, so it is maintained.

\item The Bayesian method is applied fully in the fit. More information under some reasonable assumptions is considered if possible, such as the assumptions of the signs and the order of magnitude of the truncation errors.

\item Although the number of input values is not large enough, some clearer distribution of $C^r_i$ and some more precise values of $L^r_i$ will be obtained. In addition, the boundary dependence of $C_i^r$ can be seen more clearly.
\end{enumerate}

This paper is organized as follows: Section \ref{MCMC} gives a brief introduction to Bayesian statistics and MCMC. In Sec. \ref{Sec:II}, two Bayesian models and some evaluation criteria are introduced. One model contains truncation errors, but the other one not. Some details of the calculation are also discussed. One example is studied in Sec. \ref{ME}, in order to evaluate the above models. The input physical observables mentioned in ChPT is given in Sec. \ref{Sec:III}. In Sec. \ref{Sec:IV}, some NLO and NNLO LECs are fitted by the above models. A set of new LECs are obtained. Sec. \ref{sum} gives a summary and some discussions.

\section{Bayesian statistics and MCMC}\label{MCMC}
This section provides a brief introduction to fit data by Bayesian statistics and MCMC. More details can be found in Refs. \cite{Schindler:2008fh,Furnstahl:2014xsa}. Some content is very basic and can also be found in textbooks about probability theory and Bayesian analysis. For convenience, some parameters are given meanings in ChPT, but it has a much wider scope of applications. They can be any parameters to be fitted in a problem.

Considering a general case, some parameters need to be fitted from a set of data. $\bm{D}=(D_1,D_2,D_3, \ldots, D_m)$ denotes a set of known input data. In physics, it is usually experimental data or physical constant quantities. All $D_i$ are not assumed independent. $\bm{a}=(a_1,a_2,a_3, \ldots, a_n)$ is a parameter vector. In physics, its components are usually some parameters needed to be fitted. In this paper, $\bm{a}$ means LECs. The rest of this section will introduce an approach to fit $\bm{a}$ by Bayesian statistics and MCMC. This approach is faster than only the Bayesian statistics without MCMC.

The core of Bayesian statistics is Bayes' formula
\begin{align}
\pr(\bm{a} | \bm{D})=\frac{\pr(\bm{D} | \bm{a}) \pr(\bm{a})}{\pr(\bm{D})},\label{eqb1}
\end{align}
The meanings of Eq. \eqref{eqb1} is as follows.

\begin{enumerate}
\item $\pr(\bm{a})$ is the prior probability distribution function (PDF). It reflects the knowledge of $\bm{a}$ before $\bm{D}$ is observed. If one does not know anything about $\bm{a}$, $\pr(\bm{a})$ is usually set to a uniform distribution. Usually, experiment or/and theory can give an approximated value. At least the order of magnitude is known before fitting in most cases. Due to the introduction of $\pr(\bm{a})$, one would argue that Bayesian statistics are subjective. However, $\pr(\bm{a})$ is nothing more than some assumptions in the construction of a model. This is similar to the $\chi^2$ fit usually needing an initial value of a reasonable range.

\item $\pr(\bm{D} | \bm{a})$ is the likelihood function. It is related to $\bm{D}$ and reflects the confidence of $\bm{D}$ under the given $\bm{a}$. It can be expressed as
\begin{align}
\pr(\bm{D} | \bm{a})=
\exp{\left\{-\frac{1}{2} \left[\bm{\mu}^{\text {th }}(\bm{a})-\bar{\bm{D}}\right]^T(\bm{\Sigma}_{\bm D})^{-1}\left[\bm{\mu}^{\text{th }}(\bm{a})-\bar{\bm{D}}\right]\right\}}.\label{eqb6}
\end{align}
where $\bm{\mu}^{\mathrm{th}}(\bm{a})$ is the theoretical expected value of the data, which is dependent on $\bm{a}$. $\bar{\bm{D}}$ is the expected value of the data, i.e. the experimental central value. $\bm{\Sigma}_{\bm D}$ is the covariance matrix of $\bm{D}$. The errors and the correlation information of $\bm{D}$ are contained in $\bm{\Sigma}_{\bm D}$.

\item $\pr(\bm{a} | \bm{D})$ is the posterior PDF. It is the result of Bayesian analysis. It also reflects the full knowledge of $\bm{D}$ from a fitting model. $\pr(\bm{a} | \bm{D})$ is the PDFs of $\bm{a}$, but not only some expected values. $\pr(\bm{a} | \bm{D})$ can be viewed as an update of $\pr(\bm{a})$ after $\bm{D}$ have been observed. In addition, $\pr(\bm{a} | \bm{D})$ in one fit can be regarded as $\pr(\bm{a})$ in another fit after appending some new $\bm{D}$.

\item $\pr(\bm{D})$ is called Bayesian evidence. It is known as the marginal likelihood PDF. It means the average probability of $\bm{D}$ in the fitting model. In addition, it can also be simply treated as a normalization coefficient. Because a fit is concerned with the relative PDFs of $\bm{a}$ rather than their absolute PDFs, this normalization coefficient does not play an important role in the fit. Ignoring $\pr(\bm{D})$, Bayes' formula can be expressed in a proportional form
\begin{align}
\pr(\bm{a} | \bm{D}) \propto \pr(\bm{D} | \bm{a}) \pr(\bm{a}).\label{eqb2}
\end{align}
Hence, $\pr(\bm{D} | \bm{a}) \pr(\bm{a})$ is also called the \emph{core} of the posterior PDF.
\end{enumerate}

There are some different methods to determine $\pr(\bm{a} | \bm{D})$ without $\pr(\bm{D})$, such as MCMC. We have tried three algorithms to generate the Markov chain, i.e. Metropolis-Hasting algorithm \cite{Metropolis:1953am,Hastings:1970aa,Gregory2005}, Hamiltonian Monte Carlo algorithm \cite{DUANE1987216} and No-U-turn Sampler algorithm \cite{hoffman2011nouturn}. The last two algorithms are a bit more complicated, but they have a faster computational efficiency. The details can be found in the above references. We have checked that all these algorithms can obtain almost the same distribution. The No-U-turn Sampler algorithm is the fastest one. It costs about half the time compared to the Metropolis-Hastings algorithm.

\section{Models and Details} \label{Sec:II}
\subsection{preparation}\label{sec:preparation}
The above section gives a general approach to fit the parameter $\bm{a}$ in the known analytical relationship $\bm{\mu}^{\mathrm{th}}(\bm{a})$ by Bayesian statistics and MCMC. However, in ChPT, this approach cannot be adopted directly, because the strict theoretical relationship $\bm{\mu}^{\mathrm{th}}(\bm{a})$ is hard to be obtained. It is usually calculated order by order,
\begin{align}
\bm{\mu}^{\mathrm{th}}(\bm{a})=\bm{\mu}^{\mathrm{LO}}(\bm{a}^\mathrm{LO})+\bm{\mu}^{\mathrm{NLO}}(\bm{a}^\mathrm{LO},\bm{a}^\mathrm{NLO})+\bm{\mu}^{\mathrm{NNLO}}(\bm{a}^\mathrm{LO},\bm{a}^\mathrm{NLO},\bm{a}^\mathrm{NNLO})+\cdots,\label{Da}
\end{align}
where $\bm{\mu}^{\mathrm{LO}}$, $\bm{\mu}^{\mathrm{NLO}}$ and $\bm{\mu}^{\mathrm{NNLO}}$ are the theoretical chiral expansion of $\bm{\mu}^{\mathrm{th}}(\bm{a})$ at the LO, NLO and NNLO, respectively. $\bm{a}^\mathrm{LO}$, $\bm{a}^\mathrm{NLO}$ and $\bm{a}^\mathrm{NNLO}$ are the LO, NLO and NNLO LECs, respectively, such as $L^r_i$ and $C^r_i$. At present, the higher-order relationship $\bm{\mu}^\mathrm{HO}(\bm{a})$ (i.e. truncation error) is lacking, so this paper only considers the expansion up to the NNLO. As discussed in the introduction, $\bm{\mu}^\mathrm{HO}(\bm{a})$ may make a great impact on the results. Hence, it should be considered in the fit.

The introduction mentions that many references have discussed how to estimate the truncation errors, such as Ref. \cite{Melendez:2017phj}. However, that approach cannot be used directly in the present case. There exist some serious problems. Ref. \cite{Melendez:2017phj} knows $\bm{\mu}^{\mathrm{LO}}$, $\bm{\mu}^{\mathrm{NLO}}$ and $\bm{\mu}^{\mathrm{NNLO}}$ without errors to estimate the distribution of $\bm{\mu}^{\mathrm{th}}$. However, in the present case, $\bm{\mu}^{\mathrm{th}}$ with systematical errors and the analytical expressions of $\bm{\mu}^{\mathrm{LO}}(\bm{a}^\mathrm{LO})$, $\bm{\mu}^{\mathrm{NLO}}(\bm{a}^\mathrm{LO},\bm{a}^\mathrm{NLO})$ and $\bm{\mu}^{\mathrm{NNLO}}(\bm{a}^\mathrm{LO},\bm{a}^\mathrm{NLO},\bm{a}^\mathrm{NNLO})$ are known, but $\bm{\mu}^{\mathrm{LO}}$, $\bm{\mu}^{\mathrm{NLO}}$, $\bm{\mu}^{\mathrm{NNLO}}$, $\bm{a}^\mathrm{NLO}$, $\bm{a}^\mathrm{NNLO}$ and their distributions are needed to be fitted by $\bm{D}$ and $\bm{\Sigma_D}$. Ref. \cite{Melendez:2017phj} computes the Bayesian evidence by a multidimensional integral (Eq. (8) in Ref. \cite{Melendez:2017phj}). In several special cases, the Bayesian evidence can be integrated analytically, but it usually needs to be integrated numerically. A multi-dimensional numerical integral is usually hard to be done, and it may cost a lot of time. However, the MCMC approach avoids determining the Bayesian evidence, and the computational speed is faster. In addition, Ref. \cite{Melendez:2017phj} requires Eq. \eqref{Da} to be convergent order by order, but Refs. \cite{Bijnens:2014lea,Yang:2020eif} have already indicated $\bm{\mu}^{\mathrm{NLO}}>\bm{\mu}^{\mathrm{NNLO}}$ for some physical quantities. Hence, a new approach is needed.

Generally, in an actual fit, some of $\bm{a}^\mathrm{LO}$, $\bm{a}^\mathrm{NLO}$ and $\bm{a}^\mathrm{NNLO}$ may have dimensions, and their values may be very small or very large. For example, the NNLO LECs $C_i$ is about $10^{-3}\mathrm{GeV}^{-2}$. For convenience, they are first removed the dimensions. For example, most literature provides $C^r_i$ (defined in \cite{Bijnens:1999hw}) without dimension, but not $C_i$. Moreover, very small or very large values may lead to numerical errors. Hence, all LECs divide by an order of magnitude, in order to make them roughly 1. This can be done in an actual fit. For example, both experiment and theory can estimate $C^r_i$ is about $10^{-6}$. The order of magnitude of LECs is regarded as a prior of LECs in this paper. For convenience, all the quantities in this section are assumed to be dimensionless, and all $\bm{a}^\mathrm{LO}$, $\bm{a}^\mathrm{NLO}$ and $\bm{a}^\mathrm{NNLO}$ are assumed roughly 1. In fact, the number 1 is not very strict. As long as the number is not very large, the fit also works well. For convenience, $\bm{a}^\mathrm{LO}$ is assumed to be known, and it does not need to be fitted in this section. If one wants to fit $\bm{a}^\mathrm{LO}$, there is no difference from fitting $\bm{a}^\mathrm{NLO}$ and $\bm{a}^\mathrm{NNLO}$.

In the actual ChPT fit in this paper, the number of $D_i$ is less than the total number of $\bm{a}^\mathrm{LO}$, $\bm{a}^\mathrm{NLO}$ and $\bm{a}^\mathrm{NNLO}$. There exists an overfitting. Hence, some constraint conditions are introduced to decrease the parametric space. In order to consider the convergence of ChPT, Eq. \eqref{eqb6} need to be introduced some information about the high orders. If one has no more information about the high orders, in this paper, the parameters in Eq. \eqref{eqb6} are modified to
\begin{align}
\bm{\mu}^{\mathrm{th}}(\bm{a})&\to
\begin{pmatrix}\label{co1}
\bm{\mu}^{\mathrm{th}}(\bm{a})\\
|\bm{\mu}^\mathrm{NLO}/\bm{\mu}^\mathrm{LO}|\\
|\bm{\mu}^\mathrm{NNLO}/\bm{\mu}^\mathrm{LO}|
\end{pmatrix},\\
\bar{\bm{D}}&\to
\begin{pmatrix}\label{co2}
\bar{\bm{D}}\\
0.2\bm{I}\\
0.05\bm{I}
\end{pmatrix},\\
\bm{\Sigma}_{\bm D}&\to
\begin{pmatrix}\label{co3}
\bm{\Sigma}_{\bm D}&&\\
&0.2^2\bm{I}&\\
&&0.05^2\bm{I}
\end{pmatrix},
\end{align}
where $\bm{I}$ means an identity matrix with a suitable dimension. These changes assume that $|\mu_i^{\mathrm{NLO}}/\mu_i^{\mathrm{LO}}|$ satisfies a normal distribution $N(0.2,0.2^2)$ ($\mu=0.2,\sigma^2=0.2^2$) for any $i$ (ignore the negative part), and $|\mu_i^{\mathrm{NNLO}}/\mu_i^{\mathrm{LO}}|$ has a similar meaning. The values come from the convergence hypothesis of ChPT. According to ChPT, $|\mu^\mathrm{NLO}/\mu^\mathrm{LO}|$ is about $0.1 \sim 0.3$, and $|\mu^\mathrm{NNLO}/\mu^\mathrm{NLO}|$ is also about $0.1 \sim 0.3$. Both $0.2$ and $0.05$ are near the central values. The standard deviations are chosen the same as the expected values, in order to give a large enough possibility at a wide range, because the estimation may not be very exact. In order to make the model universal, we choose the relative difference, but not the absolute value. This is because EFT/ChPT can provide us an approximate ratio between two orders, but not their absolute values. Of course, if one knows an approximate absolute value of a special quantity at a given order, Eq. \eqref{co1} can be replaced with this absolute value. Some similar constraints about the truncation errors will be discussed in Sec. \ref{mb}. Of course, these constraints can be correspondingly modified to different values, if one has a better understanding about some physical quantities.

For convenience, only one input datum $D$ or a component form $D_i$ is discussed. If one wants to consider more than one datum $\bm D$, the discussion also works.

\subsection{Model A}\label{ma}

First, the truncation error is not considered in the fit, which is called Model A. Considering a physical quantity with an experimental value $D\pm\sigma_D$, its theoretical value is $\mu\pm\sigma$. The theoretical values to the NLO and NNLO without errors are
\begin{align}
\mu^\mathrm{(NLO)}_\mathrm A&=\mu^\mathrm{LO}_\mathrm A(\bm{a}^\mathrm{LO})+\mu^\mathrm{NLO}_\mathrm A(\bm{a}^\mathrm{LO},\bm{a}^\mathrm{NLO}),\label{eq1}\\
\mu^\mathrm{(NNLO)}_\mathrm A&=\mu^\mathrm{LO}_\mathrm A(\bm{a}^\mathrm{LO})+\mu^\mathrm{NLO}_\mathrm A(\bm{a}^\mathrm{LO},\bm{a}^\mathrm{NLO})+\mu^\mathrm{NNLO}_\mathrm A(\bm{a}^\mathrm{LO},\bm{a}^\mathrm{NLO},\bm{a}^\mathrm{NNLO}),\label{eq2}
\end{align}
respectively. The term with a superscript without a couple of parentheses means the theoretical value only at this order. For example, $\mu^\mathrm{NLO}_\mathrm A$ means the NLO theoretical value of $\mu$. Eqs. \eqref{eq1} and \eqref{eq2} are applied in the NLO and the NNLO fit, respectively.

This model assumes $\pr(a_i^\mathrm{NLO})$ is the standard normal distribution, because the magnitudes of all $\bm{a}^\mathrm{NLO}$ and $\bm{a}^\mathrm{NNLO}$ are already normalized to roughly 1. In other words, one only introduces the information about the rough magnitudes of $\bm{a}^\mathrm{NLO}$ and $\bm{a}^\mathrm{NNLO}$, but no more information is considered at present. The advantage of this assumption is that more information of $\pr(\bm{a} | \bm{D})$ can be derived from the experimental data $\bm{D}$ themselves, in order to reduce the subjectivity. In addition, Eq. \eqref{eqb6} is adopted in the fit, but not Eqs. \eqref{co1} -- \eqref{co3}. We have checked that both $\bm{a}^\mathrm{NLO}$ and $\bm{a}^\mathrm{NNLO}$ need not be very close to 1. The results change slightly, as long as their values are not very large. This is because the standard normal distribution has a not very small possibility in a wide range. The same conclusion is true for the below model.

In order to improve Model A, more information is appended. It is called Model B.

\subsection{Model B}\label{mb}
Generally, the truncation error can be simply considered as a normal distribution, and the parameters of the normal distribution are based on the known information from the knowledge of the theory. However, in some special cases, the sign of the truncation error is known, or the probability of the sign is known. This information from the sign is considered separately. Eqs. \eqref{eq1} and \eqref{eq2} are improved to
\begin{align}
\mu^\mathrm{(NLO)}_\mathrm B&=\mu^\mathrm{LO}_\mathrm B(\bm{a}^\mathrm{LO})+\mu^\mathrm{NLO}_\mathrm B(\bm{a}^\mathrm{LO},\bm{a}^\mathrm{NLO})+(2s-1) e \mu^\mathrm{LO}_\mathrm B,\label{eq3}\\
\mu^\mathrm{(NNLO)}_\mathrm B&=\mu^\mathrm{LO}_\mathrm B(\bm{a}^\mathrm{LO})+\mu^\mathrm{NLO}_\mathrm B(\bm{a}^\mathrm{LO},\bm{a}^\mathrm{NLO})+\mu^\mathrm{NNLO}_\mathrm B(\bm{a}^\mathrm{LO},\bm{a}^\mathrm{NLO},\bm{a}^\mathrm{NNLO})+(2s-1) e \mu^\mathrm{LO}_\mathrm B,\label{eq4}
\end{align}
respectively. The last terms on the right-hand side of Eqs. \eqref{eq3} and \eqref{eq4} represent the higher-order (HO) truncation error $\mu^\mathrm{HO}_\mathrm B=(2s-1) e \mu^\mathrm{LO}_\mathrm B$. For $\mu^\mathrm{(NLO)}_\mathrm B$ and $\mu^\mathrm{(NNLO)}_\mathrm B$, it means the contribution higher than the NLO and NNLO, respectively. The parameter $s$ relates to the sign of the truncation error. It is assumed to be a Bernoulli random variable with parameters 1,
\begin{align}
\pr\{s=k\}=p^k(1-p)^{1-k},\quad k=0,1.\label{dp}
\end{align}
$p$ is the probability for $s=1$. If one does not know the information of the sign, $p=0.5$. $s$ parameter can give a correct sign of the truncation error. If the estimating truncation error gives a narrow range with a wrong sign, the theoretical values will be far from the true value and the fit will be bad. The parameter $s$ is introduced to solve this problem. $s$ can change the wrong sign into a correct one. Oppositely, if the estimating truncation error gives a correct sign, or the range is too wide to cover the true truncation error, $s$ will have no impact on this case. The parameter $e$ reflects the relative magnitude of the truncation error, relative to $\mu^\mathrm{LO}_\mathrm B$. One needs not know the absolute magnitude of the truncation error. However, if the EFT is satisfied, the relative magnitude at each order can be estimated. For example, the ratio between two adjacent orders is about $p/\Lambda$, where $p$ is the momentum of the low-energy particles and $\Lambda$ is the scale of the EFT. In ChPT, $|\mu^\mathrm{NLO}_\mathrm B/\mu^\mathrm{LO}_\mathrm B|$ is about $0.1 \sim 0.3$, and $|\mu^\mathrm{NNLO}_\mathrm B/\mu^\mathrm{NLO}_\mathrm B|$ is also about $0.1 \sim 0.3$, and so on. Therefore, it can be considered that $|\mu^\mathrm{HO}_\mathrm B/\mu^\mathrm{LO}_\mathrm B|$ is about 5\% (2\%) for $\mu^\mathrm{(NLO)}_\mathrm B$ ($\mu^\mathrm{(NNLO)}_\mathrm B$). Hence, the parameter $e$ is assumed to be a Gaussian random variable
\begin{align}
\pr(e)=N(\mu_e, \sigma^2_e),
\end{align}
where $\mu_e$ is the expected magnitude of $\mu^\mathrm{HO}_\mathrm B/\mu^\mathrm{LO}_\mathrm B$, and $\sigma_e$ is its standard deviation. If one does not know more information about the truncation error, a possible and reasonable choice is $\mu_e=\sigma_e=0.05$ (0.02) for $\mu^\mathrm{(NLO)}_\mathrm B$ ($\mu^\mathrm{(NNLO)}_\mathrm B$).

The parameters $p$, $\mu_e$, $\sigma_e$, $\bm{a}^\mathrm{NLO}$ and $\bm{a}^\mathrm{NNLO}$ sometimes can be estimated through the information of the data. Hence, they can be set to another values, even though the prior PDFs of them can be also set to another form, as long as the information is accurate enough.

There are two extreme cases in Model B, which will be adopted only for model evaluation in Sec. \ref{ME}. These two cases are called Model B$_1$ and Model B$_2$, respectively.

\emph{Model B$_1$}: In this case, one knows nothing about $\mu^\mathrm{HO}_\mathrm B$, such as the sign and the rough magnitude. Only the approximate order of magnitude of $\mu^\mathrm{HO}_\mathrm B$ is known from ChPT, such as about 5\% of LO at the NLO fit. As in the discussion above, for all quantities, we set $\mu_e=\sigma_e=0.05$ (0.02) and $p=0.5$ for the NLO (NNLO) fit. At present, we do not consider more information about $\bm{a}^\mathrm{NLO}$ and $\bm{a}^\mathrm{NNLO}$. Hence, $\bm{a}^\mathrm{NLO}$ and $\bm{a}^\mathrm{NNLO}$ are set to the standard normal distribution $N(0,1)$. The convergence constraints are the same as Eqs. \eqref{co1} -- \eqref{co3}.

\emph{Model B$_2$} In this model, the magnitudes of each $\mu^\mathrm{HO}_\mathrm B$ all have a certain understanding. Hence, one can set different prior PDFs to different $\mu^\mathrm{HO}_\mathrm B$, separately. The parameters $\mu_e$, $\sigma_e$ and $p$ from different quantities can be set to different values. For example, if one knows the sign is positive, $p$ is set to 1. The priors for $\mu_e$ and $\sigma_e$ are set as
\begin{equation}
\begin{aligned}
\mu^\mathrm{NLO}_e&=|\mu^\mathrm{NLO}_\mathrm{tr}/\mu^\mathrm{LO}_\mathrm{tr}|,&\sigma_e^\mathrm{NLO}&=\max\left(0.3\mu_e,0.05\right),\\
\mu^\mathrm{NNLO}_e&=|\mu^\mathrm{NNLO}_\mathrm{tr}/\mu^\mathrm{LO}_\mathrm{tr}|,&\sigma_e^\mathrm{NNLO}&=\max\left(0.3\mu_e,0.02\right),
\end{aligned}\label{eq:emu}
\end{equation}
where the superscripts NLO (NNLO) represent the NLO (NNLO) fit, the subscript ``tr'' means true value. Because we have only adopted this model for the example in Sec. \ref{ME} to evaluate the models, all the true values are known. Similarly, the true ranges of $\bm{a}^\mathrm{NLO}$ and $\bm{a}^\mathrm{NNLO}$ are generated by some given parameters. Their true ranges are also known. Therefore, their prior ranges are given the same as their true ranges. In addition, the constraints can be set to different values for the different physical quantities.

Models B$_1$ and B$_2$ adopt two extreme priors, they are only used to fit the example in Sec. \ref{ME}. Because this example are artificial, and the true values are known, we can select none or all prior information in the fit. For the actual experimental data, the known prior information is between Model B$_1$ and Model B$_2$. For example, one may have some information about a part of $\bm{D}$, and the signs and the approximate magnitudes of $\mu^\mathrm{HO}_\mathrm B$ can be given as Model B$_2$. However, for another part of $\bm{D}$, one may have no information about their $\mu^\mathrm{HO}_\mathrm B$, because of the lack of the current theory and/or experiment. For this part of $\bm{D}$, one can only give the prior PDFs as those in Model B$_1$. Besides these two cases, one may more possibly know some information of $\mu^\mathrm{HO}_\mathrm B$. For example, $\mu^\mathrm{HO}_\mathrm B$ is more likely to be positive, or its value is possible around $1$ or $-2$. The prior PDF can be set according to this information. The fitting method of Models B, B$_1$ and B$_2$ are the same, except the prior PDFs are different. It can be expected that the general Model B is better than Model B$_1$, but worse than Model B$_2$. Therefore, in Chapter \ref{Sec:IV}, we have uniformly used Model B to represent the new model proposed in this paper.

\subsection{Calculation details}\label{cd}
This subsection discusses some special cases in the fit.

Sometimes, one needs to fit the differentiation of $\mu(t)$ numerically, which depends on another parameter $t$, such as $f'_s$ and $g'_p$ in Sec. \ref{Sec:III}. The numerical deviation $\Delta \mu=\mu(t+\Delta t)-\mu(t)$ needs to calculate the difference between the two quantities $\mu(t+\Delta t)$ and $\mu(t)$, but each quantity has an error. If one adopts Eq. \eqref{eq3} or \eqref{eq4} to determine $\mu(t+\Delta t)$ and $\mu(t)$, the estimating truncation error of $\mu'(t)$ will contain the above two errors and become large. Therefore, the truncation error of $\mu'(t)$ is estimated from $\mu^{\prime,\mathrm{LO}}$, $\mu^{\prime,\mathrm{NLO}}$ and $\mu^{\prime,\mathrm{NNLO}}$, but not the difference of Eq. \eqref{eq3} or \eqref{eq4}. In other words, $\mu'(t)$ is treated as one quantity, but not a difference. However, for physical quantities with derivative values such as $\langle r^2\rangle_S^\pi$ and $c_S^\pi$, we place the HO terms in the denominator, which absorbs the effects of higher-order errors well.

Sometimes, in the NNLO fit, the amount of $\bm{a}^\mathrm{NNLO}$ is much larger than the number of $\bm{D}$, but the total number of $\bm{a}^\mathrm{LO}$ and $\bm{a}^\mathrm{NLO}$ is less than the number of input $\bm{D}$. The NNLO fit in ChPT is in this situation. All $\bm{a}^\mathrm{LO}$, $\bm{a}^\mathrm{NLO}$ and $\bm{a}^\mathrm{NNLO}$ are fitted as follows.
\begin{enumerate}
\item All $\bm{a}^\mathrm{NNLO}$ first linearly combine into some linearly independent $\widetilde{\bm{a}}$. The number of $\widetilde{\bm{a}}$ is equal to the number of $\bm{D}$, and one $\tilde{a}_i$ only correlates to one $D_i$. This is also reasonable in ChPT, because the NNLO fit only contains the linear combinations of $C^r_i$. One can combine them to the linearly independent ones.

\item $\bm{a}^\mathrm{LO}$ and $\bm{a}^\mathrm{NLO}$ are first fitted at the NLO by Model B. The results denote to $\hat{\bm{a}}^\mathrm{LO}\pm\hat{\bm{\sigma}}^\mathrm{LO}$ and $\hat{\bm{a}}^\mathrm{NLO}\pm\hat{\bm{\sigma}}^\mathrm{NLO}$. This is called the NLO fit.

\item In the NNLO fit, $\bm{a}^\mathrm{LO}$, $\bm{a}^\mathrm{NLO}$ and $\widetilde{\bm{a}}$ are fitted simultaneously. If no more information is known, the NNLO priors of $\bm{a}^\mathrm{LO}$ and $\bm{a}^\mathrm{NLO}$ are set to some suitable normal distributions $N\left(\bm{\mu}^\mathrm{LO(NLO)},(\bm{\sigma}^\mathrm{LO(NLO)})^2\right)$, where
\begin{align}\label{mu1}
\bm{\mu}^\mathrm{LO(NLO)}=\hat{\bm{a}}^\mathrm{LO(NLO)},\quad \bm{\sigma}^\mathrm{LO(NLO)}=\max\left(\hat{\bm{a}}^\mathrm{LO(NLO)}/2,\hat{\bm{\sigma}}^\mathrm{LO/(NLO)}\right).
\end{align}
The definition of $\bm{\sigma}^\mathrm{LO(NLO)}$ chooses the maximum of the two parameters $\hat{\bm{a}}^\mathrm{LO(NLO)}/2$ and $\bm{\sigma}^\mathrm{LO(NLO)}$. This is because either of them may be very small, this definition enlarges the prior ranges of $N\left(\bm{\mu}^\mathrm{LO(NLO)},(\bm{\sigma}^\mathrm{LO(NLO)})^2\right)$, in order to improve performance. The prior PDFs of $\widetilde{\bm{a}}$ is set to the standard normal distribution, if one knows nothing about $\widetilde{\bm{a}}$. Otherwise, some more reasonable prior PDFs can be set according to the known information.

The prior PDFs of $\bm{a}^\mathrm{LO}$, $\bm{a}^\mathrm{NLO}$ and $\widetilde{\bm{a}}$ not only make good use of the information from the NLO fit, but also allow some free spaces for these parameters. Because the NLO fitting $\bm{a}^\mathrm{LO}$ and $\bm{a}^\mathrm{NLO}$ can give a reasonable order of magnitude in most cases, the NNLO fit also selects the NLO posterior PDFs to calculate the NNLO prior PDFs. In addition, the new parameter $\widetilde{\bm{a}}$ is also introduced in the NNLO fit. Hence, the NNLO fit is not a repeated fit to the data, even if some of the NLO posterior information is used. We have also tried to do the NNLO fit without the posterior PDFs from the NLO fit for the example in Sec. \ref{ME}, and set the prior of $\bm{a}^\mathrm{LO}$, $\bm{a}^\mathrm{NLO}$, and $\bm{a}^\mathrm{NNLO}$ uniformly to the standard normal distribution. However, this gives very poor results, which can deviate very far from the true values. Therefore, it is necessary to use some sensible information about LECs as a prior in the NNLO fit.

Of course, if some information about $\bm{a}^\mathrm{LO}$ and $\bm{a}^\mathrm{NLO}$ is known, one can set another sensible prior PDFs.

\item Finally, all $\bm{a}^{\mathrm{NNLO}}$ are fitted with the posterior $\widetilde{\bm{a}}$ obtained above, with some appropriate uniform distributions. The boundaries of the uniform distribution are dependent on the approximate order of magnitudes of the truncation errors. This is because the NLO research has usually been studied widely, and more information is known. However, the NNLO research is usually lacking, and the values of $\bm{a}^{\mathrm{NNLO}}$ are not quite sure. Hence, a uniform distribution can give a larger probability near the boundaries, in order to study the boundary-dependent property. After the fit, the posterior PDFs of the truncation error will be changed into better ones.
\end{enumerate}

Models B is very efficient. For the actual fit in ChPT, which will be discussed below, a personal computer with CPU Intel i3-10105 only costs about ten minutes with 4 cores. This method greatly reduces the time compared with the method in Ref. \cite{Yang:2020eif}, which costs about one day with 20-core CPU Intel Xeon Gold 6230.

All the numerical results are represented by the highest posterior density (HPD). The HPD is the minimum interval containing a certain proportion of probability density. The most common proportion is $95\%$ HPD or $98\%$ HPD, but we have chosen $68\%$ HPD. Because it is similar to $1\sigma$ interval in the classical statistics \cite{Wesolowski:2015fqa}, such as the minimum $\chi^2$ method. All the results in this paper have been compared. It indicates that the difference between $68\%$ HPD and $1\sigma$ interval is very small, most last significant digits have no difference or a difference of 1 or 2. Only very few of them have a difference of 3 or 4. No one is larger than 4. Hence, we sometimes do not distinguish them in this paper.

\subsection{Evaluation criteria}\label{EC}
In order to evaluate which model is the best, there needs an evaluation criterion. This criterion is better to be quantified. One can evaluate different models by the quantified index. Bayesian evidence is one possible criterion, but it is too simple. The widely applicable information criterion (WAIC) and leave-one-out cross-validation (LOOCV) are introduced in recent years. WAIC considers how well the data fits the model and also penalizes complex models. LOOCV splits the data into a training set and validation set and repeats many times to evaluate the model. The definitions of WAIC and LOOCV involve some related concepts and formulas, which need a long discussion. Their definitions and a more detailed explanation can be found in Refs. \cite{Gelman2013,Vehtari_2016}. Simply speaking, if Model B has larger values of WAIC and LOOCV than Model A, Model B is considered better than Model A. Of course, only a couple of these values for one model are meaningless, because one does not know how large is enough. They are only meaningful for comparing different models.

For the example in Sec. \ref{ME}, the true values of parameters $a_{\mathrm{i,tr}}$ are known. In addition to both WAIC and LOOCV, the fitting results $a_{i,\mathrm{model}}$ can be compared to the true values directly. For example, $a_{i,\mathrm{A}}$ means the expected value of $a_{i}$ is fitted by Model A. It is more intuitive to see how well the fit is. Hence, we define the following two quantities as criteria.
\begin{align}
\mathrm{Pct}_{\mathrm{model}}=&\frac{a_{i,\mathrm{model}}-a_{\mathrm{i,tr}}}{a_{i,\mathrm{tr}}}\times 100\%, \label{equ:2}\\
\mathrm{Pct\sigma}_{\mathrm{model}}=&\frac{a_{i,\mathrm{model}}-a_{i,\mathrm{tr}}}{\sigma_{i,{\mathrm{model}}}}. \label{equ:3}
\end{align}
$\mathrm{Pct}_{\mathrm{model}}$ is the relative error between the fitting value $a_{i,\mathrm{model}}$ and the true value $a_{i,\mathrm{tr}}$. It indicates how well the fitting expected value is. $\mathrm{Pct\sigma}_{\mathrm{model}}$ is the ratio of the difference between the true value and the fitting value to the fitting standard error $\sigma_{i,{\mathrm{model}}}$. It indicates how well the fitting error is. The smaller these two values are, the better the model is. These two criteria are only used for the example in Sec. \ref{ME}, because one does not know the true values in the actual fit.

In order to clarify the convergence of $\mu$, the percentages at each order are defined as Ref. \cite{Yang:2020eif}
\begin{align}
\mathrm{Pct}_{\mathrm{order}}=\frac{{\bar\mu}^{\mathrm{order}}_{\mathrm{model}}}{{\bar\mu}_{\mathrm{model}}}\times 100\%, \label{equ:1}
\end{align}
where $\mu^{\mathrm{order}}_{\mathrm{model}}$ is defined in Eqs. \eqref{eq1}--\eqref{eq4}. $\mu_{\mathrm{model}}$ means the fitting value obtained by a special model, containing all orders. The notation bar means the expected value. For example, ${\bar\mu}^{\mathrm{NLO}}_{\mathrm{A}}$ means the NLO expected contribution obtained by model A, and ${\bar\mu}_{\mathrm{A}}$ is the expected value containing all orders obtained by model A.

For the NNLO fit, the differences among WAIC, LOOCV, $\mathrm{Pct}_{\mathrm{model}}$ and $\mathrm{Pct\sigma}_{\mathrm{model}}$ among different models are small. It is more important to evaluate how well all $a^\mathrm{NNLO}_i$ are fitted, because the NNLO fitting $a^\mathrm{NLO}_i$ are usually precise enough, but $a^\mathrm{NNLO}_i$ usually have large errors. For the example in Sec. \ref{ME}, the true values of $a^\mathrm{NNLO}_i$ are known, and $a^\mathrm{NNLO}_i=\tilde{a}_i$, the fitting values can also compare to the true values directly. Usually, the contributions of $a^\mathrm{NNLO}_{i}$ do not mix with $a^\mathrm{LO}_i$ and $a^\mathrm{NLO}_i$, such as ChPT. The contributions of $a^\mathrm{NNLO}_{i}$ can be separated, called $\mu_{a^\mathrm{NNLO}_{i}}$. In order to see how well the fitting $a^\mathrm{NNLO}_i$ are, we defined
\begin{align}
\mathrm{PM}_{\mathrm{model}}=\sqrt{{\sum_{i=1}^n\left(\frac{(a^\mathrm{NNLO}_{i,\mathrm{tr}}-\bar{a}^\mathrm{NNLO}_{i,\mathrm{model}})}{a^\mathrm{NNLO}_{i,\mathrm{tr}}}{\frac{\bar{\mu}_{a^\mathrm{NNLO}_{i,\mathrm{model}}}}{\mu_{i,\mathrm{tr}}}} \right)^2}\bigg/n}. \label{equ:4}
\end{align}
The subscript ``tr'' means the true values, the subscript ``model'' means the model which are adopted, and $\mu_{i,\mathrm{tr}}$ is the true value of the $i$-th physical quantity. The notation bar means the expected value. $n$ is the number of physical quantities. In this paper $n=17$. For example, $a^\mathrm{NNLO}_{i,\mathrm{A}}$ means $a^\mathrm{NNLO}_{i}$ is fitted by Model A, $\mu_{a^\mathrm{NNLO}_{i,\mathrm{A}}}$ means only the contribution from $a^\mathrm{NNLO}_{i}$ by Model A. The first fraction on the right side of Eq. \eqref{equ:4} is the relative error of $a^\mathrm{NNLO}_i$, while the second fraction is treated as its weight. The weight represents the contribution of $\bar{\mu}_{a^\mathrm{NNLO}_{i,\mathrm{model}}}$ in $\mu_{i,\mathrm{tr}}$. The smaller the PM value is, the better the result is. A larger weight needs a more precise $\bar{a}^\mathrm{NNLO}_{i,\mathrm{model}}$ to reduce the PM value. PM value is only used in the example in Sec. \ref{ME}, because the true values of this example are known, but in the actual case, the true values are not known.

The next section will evaluate the above models by these evaluation criteria.

\section{Model evaluation}\label{ME}

In order to quantitatively demonstrate the advantage of Model B based on Bayesian statistics, this section gives an example to fit the parameters similar to LECs. The same as the actual fit of the LECs in Sec. \ref{Sec:IV}, a group of functions is generated randomly, each group containing 17 different quantities $O_i$. They are shown in Eqs. \eqref{A1} in Appendix \ref{app:A}. The power of $t$ is similar to the chiral dimension in ChPT. Taylor expanding these functions about $t$, the analytical results at each order can be obtained. The $t$, $t^2$ and $t^3$ orders correspond to LO, NLO and NNLO in ChPT, respectively. After the expansion, $t=1$. $a_i^\mathrm{LO}$, $a_i^\mathrm{NLO}$ and $a_i^\mathrm{NNLO}$ are similar to LO, NLO ($L^r_i$) and NNLO ($C^r_i$) LECs in ChPT, respectively. $b_i$ are some known constants, which are introduced to adjust the convergences of these Taylor series. All parameters $a_i^\mathrm{LO}$, $a_i^\mathrm{NLO}$, $a_i^\mathrm{NNLO}$ and $b_i$ are generated randomly and independently. For convenience, the parameters in each function are different, although they have the same name. For example, $b_1$ in $O_1$ and $O_2$ are different. The values of $b_i$ and $a_i^\mathrm{LO}$ in the example can be found in Table \ref{pA1}. In fact, the LO LECs do not appear in the actual ChPT fit in this paper. Hence, we treat them as known constants and do not fit them. This section only discusses the impact from truncation errors, but it does not mention overfitting. Hence, each $O_i$ only contains one $a_i^\mathrm{NNLO}$, i.e. $\tilde{a}_i^\mathrm{NNLO}=a_i^\mathrm{NNLO}$.

Since all the parameters $b_i$, $a_i^\mathrm{LO}$, $a_i^\mathrm{NLO}$ and $a_i^\mathrm{NNLO}$ in this example are known, all the analytical results $O_i$ can be calculated by these parameters directly. In this section, we define all the known values of these parameters as true values. The fitting values of these parameters are called theoretical values, which are fitted by the models in Sec. \ref{Sec:II}. In order to distinguish these two types of values, all the true values are marked by a subscript ``tr'', such as $a_{i,\mathrm{tr}}^\mathrm{NLO}$, and all the theoretical values are marked by the model name, such as $a_{i,\mathrm{A}}^\mathrm{NLO}$.

In order to imitate the realistic experiment, the fitting data do not adopt the true values but with some experimental errors $\sigma_i$. The imitative experimental data are generated by the distribution $N(O_{i,\mathrm{tr}},\sigma^2_i)$, $\sigma_i/O_{i,\mathrm{tr}}=0.02$  in the example \cite{Schindler:2008fh}. For convenience, these imitative experimental data are also called experimental data for short. Their values are in the third column of Table \ref{table3}  with a subscript ``exp'', respectively.

Because the above true values are known, the true values of $\mu^\mathrm{LO}$, $\mu^\mathrm{NLO}$ and $\mu^\mathrm{NNLO}$ can be also calculated analytically. The parameters of truncation errors in Model B$_2$ are set as Eq. \eqref{eq:emu} and the description above it. The values of $p$, $\mu_e$ and $\sigma_e$ are given in Table \ref{app table1}. Similarly, the true values of the LECs are also known, so their prior distribution are set to the normal distribution $N(\mu_{a_i},\sigma_{a_i}^2)$, where
\begin{align}
\mu_{a_i^\mathrm{NLO/NNLO}}=N\left(a_{i,\mathrm{tr}}^\mathrm{NLO/NNLO},\left(0.1a_{i,\mathrm{tr}}^\mathrm{NLO/NNLO}\right)^2\right),\quad
\sigma_{a_i^\mathrm{NLO/NNLO}}=0.5\mu_{a_i^\mathrm{NLO/NNLO}}.\label{eq:prlecs}
\end{align}
We have deliberately given $\mu_{a_i}$ a deviation from the true value, in order to avoid fit at the true value. The distribution parameters of $\mu_{a_i}$ and $\sigma_{a_i}$ at each order are given in Table \ref{app table2}.

\subsection{The NLO fit of the example}

The input parameters in Model B$_2$ are given in Columns 2 to 6 of Table \ref{app table1} in Appendix \ref{app:B}. After the NLO fit, we have checked that the obtained Markov chain satisfies the assumption of the detailed balance condition, and the results are reliable. All the other fits in this paper have the same conclusion.

\begin{figure}[htbp]
\centering
\includegraphics[width=1\linewidth]{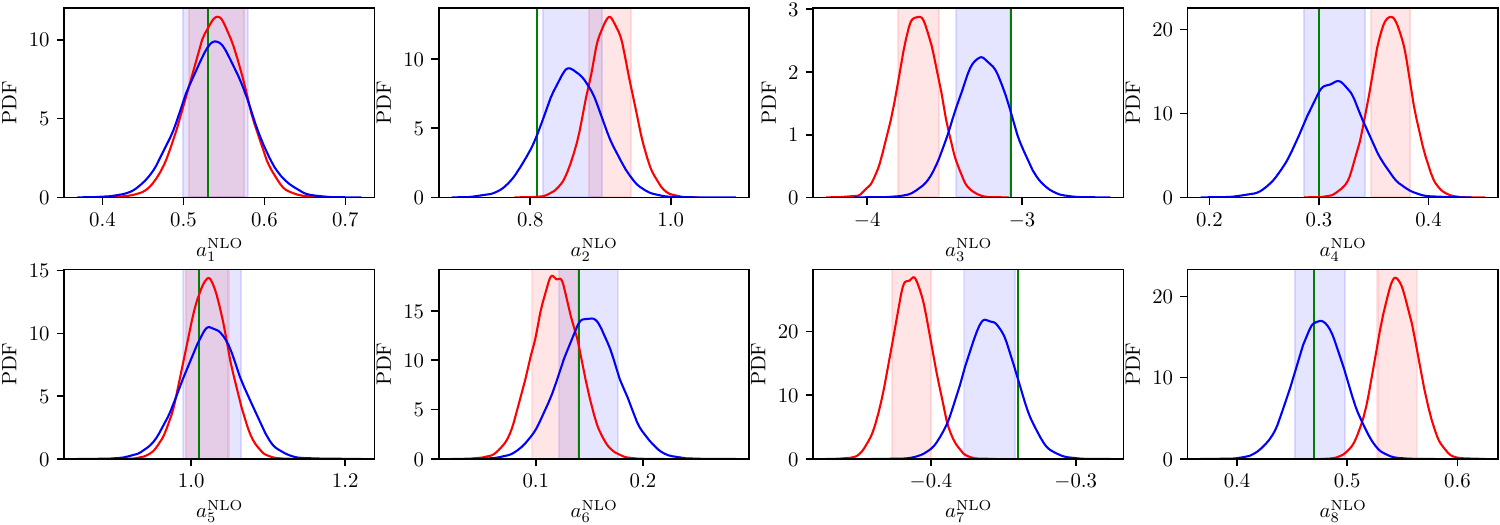}
\caption{The NLO fitting posterior PDFs of $\bm{a}_i^\mathrm{NLO}$. The red lines and the light red areas are obtained by Model A. The blue lines and the light blues area are obtained by Model B$_2$. The lines are the distribution curve of $\bm{a}_i^\mathrm{NLO}$. The light-colored areas depict the $68\%$ HPDs. The green lines denote the true value.}
\label{fig:ep4p}
\end{figure}

Figure \ref{fig:ep4p} illustrates the distributions obtained by Models A and B$_2$. The shapes of the lines are similar to normal distributions, although the details have a little difference. We have checked that the boundaries of 68\% HPD are almost the same as $1\sigma$ boundaries of a normal distribution. Hence, we sometimes do not distinguish them. It can be seen that the center values of Model B$_2$ are more closed to the true values. However, the errors of Model B$_2$ are larger than those of Model A. This is because Model B$_2$ considers the errors of the truncation errors, but Model A does not.

The numerical posterior information of $a^\mathrm{NLO}_i$ is listed in Table \ref{table1}. The WAIC and LOOCV of Model B$_2$ are the largest, but these values of Model A are the smallest. The WAIC and the LOOCV of Model B$_1$ are a bit smaller than those of Model B$_2$, but much larger than those of Model A. This means that Model B$_2$ gives the best results, but Model A is the worst. Model B$_1$ obviously improves the results of Model A, but a bit weaker than Model B$_2$. This conclusion can also be seen from $\mathrm{Pct}_{\mathrm{A, B_1, B_2}}$. However, most $|\mathrm{Pct\sigma}_{\mathrm{B_2}}|$ are still a bit larger than $|\mathrm{Pct\sigma}_{\mathrm{B_1}}|$. This is because the errors of $a^\mathrm{NLO}_{i,\mathrm{B_2}}$ are about half $a^\mathrm{NLO}_{i,\mathrm{B_1}}$. Overall, $a^\mathrm{NLO}_{i,\mathrm{B_2}}$ is closer to the true value.

\begin{table}[!htbp]
\renewcommand\arraystretch{1.3}
\caption{The NLO and the NNLO fitting results of $a^\mathrm{NLO}_i$ in the example. Row 2 is the true value of $a^\mathrm{NLO}_{i}$. Rows 3, 6 and 9 are the NLO fitting results of Model A, B$_1$ and B$_2$, respectively. Rows 12, 15 and 18 are the NNLO fitting results of Model A, B$_1$ and B$_2$, respectively. The percentage $\mathrm{Pct}_{\mathrm{A, B_1, B_2}}$ is defined in Eq. \eqref{equ:2}, and the ratio $\mathrm{Pct\sigma}_{\mathrm{A, B_1, B_2}}$ is defined in Eq. \eqref{equ:3}.}\label{table1}
\begin{ruledtabular}
\begin{tabular}{lcccccccccc}
$i$                                      & 1          & 2          & 3             & 4          & 5          & 6          & 7             & 8          & WAIC     & LOO      \\
\hline
$a^\mathrm{NLO}_{i,\mathrm{tr}}$         & 0.53       & 0.80       & $-$3.07       & 0.3        & 1.01       & 0.14       & $-$0.34       & 0.47       &          &          \\\hline
		&&&&&NLO\\
$a^\mathrm{NLO}_{i,\mathrm{A}}$     & 0.541(35)  & 0.914(30)  & $-$3.671(135) & 0.366(18)  & 1.021(28)  & 0.118(22)  & $-$0.413(14) & 0.545(18)  & $-$49.130 & $-$56.310 \\
$\mathrm{Pct}_{\mathrm{A}}$         & 2.1\%      & 14.3\%     & 19.6\%        & 22.0\%     & 1.1\%      & $-$15.7\%  & 21.5\%       & 16.0\%     &           &           \\
$\mathrm{Pct\sigma}_{\mathrm{A}}$   & 0.3        & 3.8        & $-$4.5        & 3.7        & 0.4        & $-$1.0     & $-$5.2       & 4.2        &           &           \\
$a^\mathrm{NLO}_{i,\mathrm{B_1}}$   & 0.550(121) & 0.842(79)  & $-$3.192(335) & 0.324(46)  & 0.972(68)  & 0.110(53)  & $-$0.361(34) & 0.503(43)  & 14.964    & 7.889     \\
$\mathrm{Pct}_{\mathrm{B_1}}$       & 3.8\%      & 5.2\%      & 4.0\%         & 8.0\%      & $-$3.8\%   & $-$21.4\%  & 6.2\%        & 7.0\%      &           &           \\
$\mathrm{Pct\sigma}_{\mathrm{B_1}}$ & 0.2        & 0.5        & $-$0.4        & 0.5        & $-$0.6     & $-$0.6     & $-$0.6       & 0.8        &           &           \\
$a^\mathrm{NLO}_{i,\mathrm{B_2}}$   & 0.539(41)  & 0.860(43)  & $-$3.252(175) & 0.314(28)  & 1.027(38)  & 0.149(28)  & $-$0.359(18) & 0.475(23)  & 27.307    & 23.713    \\
$\mathrm{Pct}_{\mathrm{B_2}}$       & 1.7\%      & 7.5\%      & 5.9\%         & 4.7\%      & 1.7\%      & 6.4\%      & 5.6\%        & 1.1\%      &           &           \\
$\mathrm{Pct\sigma}_{\mathrm{B_2}}$ & 0.2        & 1.4        & $-$1.0        & 0.5        & 0.4        & 0.3        & $-$1.1       & 0.2        &           &           \\ \hline
&&&&&NNLO\\
$a^\mathrm{NLO}_{i,\mathrm{A}}$     & 0.924(574) & 0.603(146) & $-$1.984(457) & 0.416(383) & 0.553(357) & 0.084(383) & $-$0.243(53) & 0.510(316) & 14.364    & 7.782     \\
$\mathrm{Pct}_{\mathrm{A}}$         & 74.34\%    & $-$24.63\% & $-$35.37\%    & 38.67\%    & $-$45.25\% & $-$40.00\% & $-$28.53\%   & 8.51\%     &           &           \\
$\mathrm{Pct\sigma}_{\mathrm{A}}$   & 0.69       & $-$1.35    & 2.38          & 0.30       & $-$1.28    & $-$0.15    & 1.83         & 0.13       &           &           \\
$a^\mathrm{NLO}_{i,\mathrm{B_1}}$   & 0.534(116) & 0.831(74)  & $-$3.042(26)  & 0.316(43)  & 0.968(62)  & 0.116(40)  & $-$0.346(27) & 0.491(42)  & 41.143    & 32.782    \\
$\mathrm{Pct}_{\mathrm{B_1}}$       & 0.75\%     & 3.87\%     & $-$0.91\%     & 5.33\%     & $-$4.16\%  & $-$17.14\% & 1.76\%       & 4.47\%     &           &           \\
$\mathrm{Pct\sigma}_{\mathrm{B_1}}$ & 0.03       & 0.42       & 0.11          & 0.37       & $-$0.68    & $-$0.60    & $-$0.22      & 0.50       &           &           \\
$a^\mathrm{NLO}_{i,\mathrm{B_2}}$   & 0.525(81)  & 0.808(35)  & $-$3.195(111) & 0.319(18)  & 0.995(37)  & 0.138(32)  & $-$0.354(12) & 0.474(24)  & 56.730    & 53.277    \\
$\mathrm{Pct}_{\mathrm{B_2}}$       & $-$0.9\%   & 1.0\%      & 4.1\%         & 6.3\%      & $-$1.5\%   & $-$1.4\%   & 4.1\%        & 0.9\%      &           &           \\
$\mathrm{Pct\sigma}_{\mathrm{B_2}}$ & $-$0.06    & 0.23       & $-$1.13       & 1.06       & $-$0.41    & $-$0.06    & $-$1.17      & 0.17       &           &

\end{tabular}
\end{ruledtabular}
\end{table}

\begin{figure}
\centering
\centering
\includegraphics[width=0.47\linewidth]{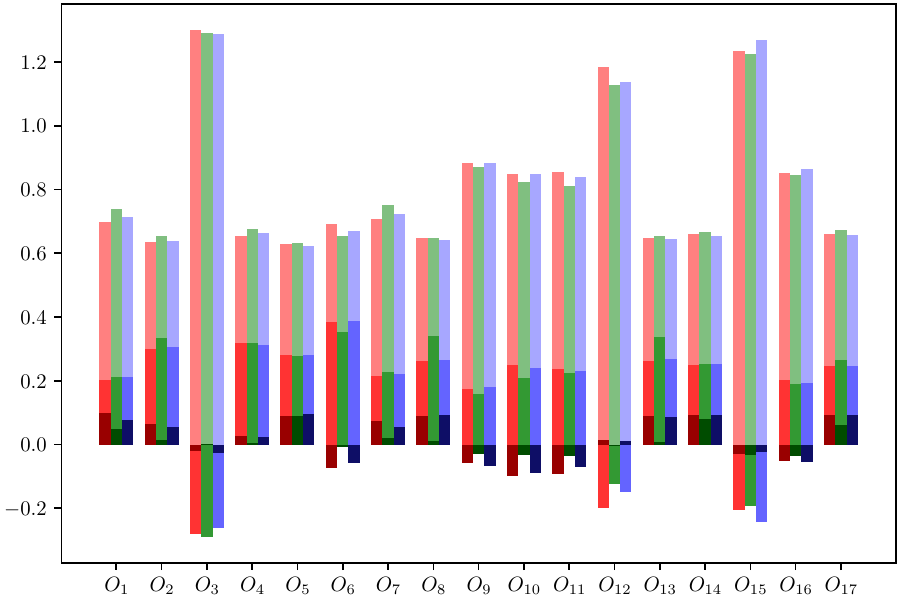}\quad
\includegraphics[width=0.47\linewidth]{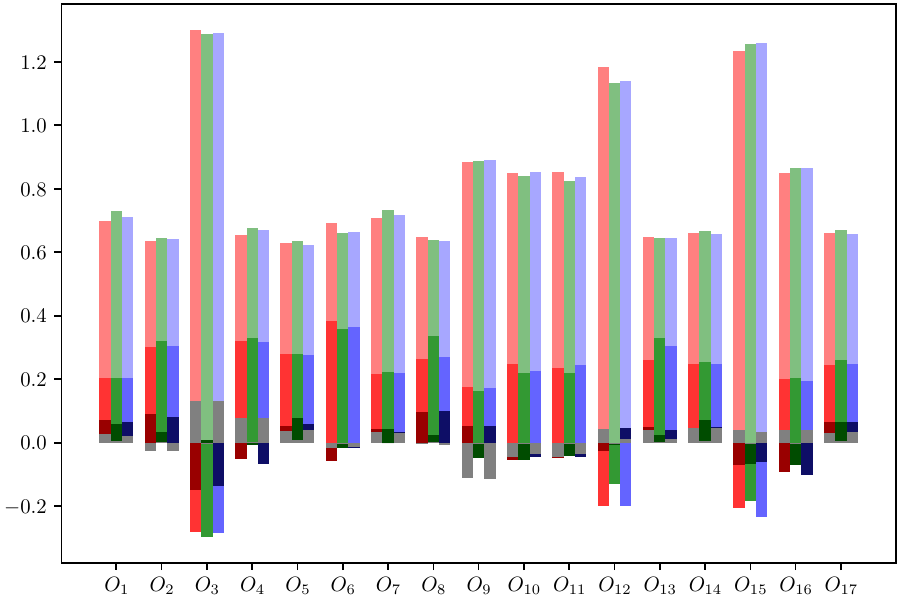}

(a)\hspace{0.45\textwidth} (b)
\caption{The proportions of $O_i$ at each order for the example. The red, green and blue strips in the figure represent the true values, the values obtained by Models B$_1$ and B$_2$, respectively. The lightest and the second lightest colors are the proportions (defined in Eq. \eqref{equ:1}) of LO and NLO, respectively. (a) The NLO fit. The darkest color is the proportion of HO. (b) The NNLO fit. The darkest color and the dark gray are the proportions of NNLO and HO, respectively. To avoid layer masking, the colors of the NNLO and the HO true values of $O_7$, $O_{11}$ and $O_{14}$ are interchanged. Similarly, the colors of $O_{11}$, $O_{12}$ and $O_{13}$ of Model B$_2$ are also interchanged.}
\label{fig:nlo-percentage}
\end{figure}

Figure \ref{fig:nlo-percentage} (a) illustrates the proportions of $O_i$ at each order. The contributions at NLO and HO from Model B$_2$ are closer to the true values than those from Model B$_1$. This is because Model B$_2$ has utilized more information compared to Model B$_1$. Despite adopting relatively less information, Model B$_1$ still satisfies convergence well in its results. However, there are noticeable differences between Models B$_1$ and B$_2$ at the HO due to some truncation errors not being accurately estimated. Nevertheless, these discrepancies have a minimal impact on the results of $a^\mathrm{NLO}_i$. Therefore, whether Model B$_1$ or Model B$_2$, their results closely approximate the true values. This indicates that even if one does not possess complete knowledge about all physical quantities' truncation errors, Model B$_1$ still yields better results compared to Model A.

Table \ref{table3} shows the comparison of the true values, the experimental values and the fitting results from Models B$_1$ and B$_2$. It can be seen that the theoretical values from both Model B$_1$ and Model B$_2$ are not obviously different from the experimental values and the true values. In particular, the theoretical values obtained by Model B$_2$ are closer to the true value than those obtained by Model B$_1$. The 1$\sigma$ errors from Model B$_1$ and Model B$_2$ are roughly equal to the experimental data, but Model B$_2$ has smaller errors. Most true values fall within 1$\sigma$ intervals of the theoretical values. A few true values are in the $1\sigma$ to $2\sigma$ intervals. No true values exceed the $2\sigma$ intervals. Table \ref{table3} also indicates that more information leads to a better result.

\begin{table}[htbp]
\caption{The comparison of the NLO fitting values for the example. The subscripts tr, exp, B$_1$ and B$_2$ in the first row represent the true values, the experimental values, the theoretical values from Model B$_1$ and Model B$_2$, respectively. The experimental values in the third column are sampled from the true values. $O_i$ is defined in Eq. \eqref{A1}.}\label{table3}
\begin{ruledtabular}
\begin{tabular}{rrrrr}
$i$ &$10^2O_{i,\mathrm {tr}}$&$10^2O_{i,\mathrm {exp}}$~~~~~&$10^2O_{i,\mathrm {B_1}}$~~~~~&$10^2O_{i,\mathrm {B_2}}$~~~~~\\\hline
1  & $-$35.800 & $-$34.637 $\pm$  0.716 & $-$33.778 $\pm$  1.786 & $-$35.023 $\pm$  1.129 \\
2  & 0.173   & 0.171   $\pm$  0.003 & 0.168   $\pm$  0.005 & 0.172   $\pm$  0.004 \\
3  & $-$0.276  & $-$0.279  $\pm$  0.006 & $-$0.279  $\pm$  0.031 & $-$0.279  $\pm$  0.012 \\
4  & 0.603   & 0.590   $\pm$  0.012 & 0.584   $\pm$  0.013 & 0.595   $\pm$  0.008 \\
5  & 27.281  & 27.753  $\pm$  0.546 & 27.124  $\pm$  0.756 & 27.600  $\pm$  0.535 \\
6  & $-$0.524  & $-$0.548  $\pm$  0.010 & $-$0.554  $\pm$  0.020 & $-$0.541  $\pm$  0.014 \\
7  & $-$1.486  & $-$1.434  $\pm$  0.030 & $-$1.403  $\pm$  0.028 & $-$1.459  $\pm$  0.023 \\
8  & $-$0.955  & $-$0.970  $\pm$  0.019 & $-$0.954  $\pm$  0.414 & $-$0.965  $\pm$  0.219 \\
9  & $-$0.227  & $-$0.226  $\pm$  0.005 & $-$0.231  $\pm$  0.008 & $-$0.227  $\pm$  0.004 \\
10 & $-$52.511 & $-$52.773 $\pm$  1.050 & $-$54.107 $\pm$  3.062 & $-$52.493 $\pm$  2.069 \\
11 & 44.936  & 46.250  $\pm$  0.899 & 47.299  $\pm$  2.319 & 45.728  $\pm$  1.351 \\
12 & $-$4.223  & $-$4.397  $\pm$  0.084 & $-$4.427  $\pm$  0.673 & $-$4.393  $\pm$  0.374 \\
13 & $-$14.674 & $-$14.769 $\pm$  0.293 & $-$14.574 $\pm$  2.295 & $-$14.782 $\pm$  1.325 \\
14 & $-$24.577 & $-$24.765 $\pm$  0.492 & $-$24.351 $\pm$  0.492 & $-$24.755 $\pm$  0.419 \\
15 & $-$15.864 & $-$15.505 $\pm$  0.317 & $-$15.974 $\pm$  1.741 & $-$15.435 $\pm$  0.893 \\
16 & 3.831   & 3.746   $\pm$  0.077 & 3.847   $\pm$  0.145 & 3.774   $\pm$  0.069 \\
17 & $-$4.193  & $-$4.208  $\pm$  0.084 & $-$4.110  $\pm$  0.146 & $-$4.203  $\pm$  0.086
\end{tabular}
\end{ruledtabular}
\end{table}

\subsection{The NNLO fit of the example}
In the NNLO fit, the priors of $a_i^\mathrm{NLO}$ and $a_i^\mathrm{NNLO}$ in Models A and B$_1$ are the same as those discussed in Sections \ref{ma} and \ref{mb}. The priors in Model B$_2$ adopt Eq. \eqref{mu1}, and the parameters are given in Columns 7 to 11 of Table \ref{app table1} in Appendix \ref{app:B}.

The numerical NNLO fitting results of $a_i^\mathrm{NLO}$ obtained by Models A, B$_1$ and B$_2$ are shown in Rows 12 to 20 of Table \ref{table1}. The NNLO fitting results of $a_i^\mathrm{NNLO}$ obtained by Models A, B$_1$ and B$_2$ are shown in Table \ref{table7}. Besides WAIC and LOO, the last row also gives the PM value defined in Eq. \eqref{equ:4}.

Table \ref{table1} shows that the best results of $a_i^\mathrm{NLO}$ are obtained by Model B$_2$. The NNLO $\mathrm{Pct}_{\mathrm{B_1}}$ ($\mathrm{Pct\sigma}_{\mathrm{B_1}}$), $\mathrm{Pct}_{\mathrm{B_2}}$ ($\mathrm{Pct\sigma}_{\mathrm{B_2}}$) and their NLO values show that most of the results are improved. There exists a significant difference between the NLO and the NNLO results. This indicates that even though the NNLO prior PDFs are calculated from the NLO posterior PDFs, the NNLO fitting $a_i^\mathrm{NLO}$ does not stay at the prior PDFs, it can change to the other ranges. In other words, the NNLO fit is not a repeated NLO fit.

Table \ref{table7} shows that there are significant differences between $\mathrm{Pct}_{\mathrm{A}}$ ($\mathrm{Pct\sigma}_{\mathrm{A}}$) and $\mathrm{Pct}_{\mathrm{B_1}}$ ($\mathrm{Pct\sigma}_{\mathrm{B_1}}$) for $a_i^\mathrm{NNLO}$. Although a few $|\mathrm{Pct_{B_1}}|$ have large values (the largest is $316.8\%$), and several $|\mathrm{Pct}\sigma_\mathrm{B_1}|$ also have large values, Model B$_1$ still has a significant improvement over Model A. This can also be noticed from their PM values, which change significantly. Similarly, Model B$_2$ also shows a more significant improvement in the results. Most $\mathrm{Pct}_{\mathrm{B_2}}$ and $\mathrm{Pct\sigma}_{\mathrm{B_2}}$ are smaller than those from Models A and B$_1$. It can be seen that for the NNLO fit, the more useful information is known, the better the fitting results are.

\begin{table}[htpb]
\caption{The NNLO fitting results of the example. Column 2 is the true value of $a^\mathrm{NNLO}_{i}$. Column 3, 6 and 9 are the results of Model A, B$_1$ and B$_2$, respectively. The percentage $\mathrm{Pct}_{\mathrm{A, B_1, B_2}}$ is defined in Eq. \eqref{equ:2}, and the ratio $\mathrm{Pct\sigma}_{\mathrm{A, B_1, B_2}}$ is defined in Eq. \eqref{equ:3}. PM is defined in Eq. \eqref{equ:4}.}\label{table7}
\begin{ruledtabular}
\begin{tabular}{lrrrrrrrrrr}
$i$         &    $a^\mathrm{NNLO}_{i,\mathrm{tr}}$    &    $a^\mathrm{NNLO}_{i,\mathrm{A}}$ &    $\mathrm{Pct}_{\mathrm{A}}$    &    $\mathrm{Pct\sigma}_{\mathrm{A}}$    & $a^\mathrm{NNLO}_{i,\mathrm{B_1}}$ &    $\mathrm{Pct}_{\mathrm{B_1}}$    & $\mathrm{Pct\sigma}_{\mathrm{B_1}}$ & $a^\mathrm{NNLO}_{i,\mathrm{B_2}}$  &    $\mathrm{Pct}_{\mathrm{B_2}}$    &    $\mathrm{Pct\sigma}_{\mathrm{B_2}}$ \\ \hline
1    & 0.02                        & 0.176  (298)& 780.0\%  & 0.5  & 0.013  (29)& $-$35.0\% & $-$0.2 & 0.017  (10)& $-$15.0\% & $-$0.3 \\
2    & 0.19                        & 0.060  (293)& $-$68.4\%  & $-$0.4 & 0.102  (46)& $-$46.3\% & $-$1.9 & 0.177  (31)& $-$6.8\%  & $-$0.4 \\
3    & $-$0.72                       & 0.351  (692)& $-$148.8\% & 1.5  & $-$0.073 (264)& $-$89.9\% & 2.5  & $-$0.703 (209)& $-$2.4\%  & 0.1  \\
4    & 0.22                        & $-$0.682 (917)& $-$410.0\% & $-$1.0 & 0.917  (735)& 316.8\% & 0.9  & 0.203  (96)& $-$7.7\%  & $-$0.2 \\
5    & $-$0.16                       & 0.018  (465)& $-$111.3\% & 0.4  & $-$0.090 (60)& $-$43.8\% & 1.2  & $-$0.137 (43)& $-$14.4\% & 0.5  \\
6    & 0.26                        & 0.035  (485)& $-$86.5\%  & $-$0.5 & 0.189  (71)& $-$27.3\% & $-$1.0 & 0.192  (58)& $-$26.2\% & $-$1.2 \\
7    & $-$0.42                       & 0.088  (645)& $-$121.0\% & 0.8  & $-$0.209 (520)& $-$50.2\% & 0.4  & $-$0.413 (165)& $-$1.7\%  & 0.0  \\
8    & $-$0.45                       & 0.016  ( 1005)& $-$103.6\% & 0.5  & $-$0.136 (188)& $-$69.8\% & 1.7  & $-$0.472 (118)& 4.9\%   & $-$0.2 \\
9    & $-$0.99                       & $-$0.822 (525)& $-$17.0\%  & 0.3  & $-$0.261 (200)& $-$73.6\% & 3.6  & $-$0.966 (208)& $-$2.4\%  & 0.1  \\
10   & $-$0.06                       & $-$0.415 (670)& 591.7\%  & $-$0.5 & $-$0.076 (59)& 26.7\%  & $-$0.3 & $-$0.083 (24)& 38.3\%  & $-$1.0 \\
11   & 0.24                        & 0.005  (993)& $-$97.9\%  & $-$0.2 & 0.163  (646)& $-$32.1\% & $-$0.1 & 0.254  (132)& 5.8\%   & 0.1  \\
12   & $-$0.18                       & $-$0.182 (605)& 1.1\%    & 0.0  & $-$0.194 (85)& 7.8\%   & $-$0.2 & $-$0.219 (51)& 21.7\%  & $-$0.8 \\
13   & 1.02                        & 0.342  (706)& $-$66.5\%  & $-$1.0 & 1.011  (71)& $-$0.9\%  & $-$0.1 & 0.997  (57)& $-$2.3\%  & $-$0.4 \\
14   & 0.29                        & $-$0.226 (181)& $-$177.9\% & $-$2.9 & 0.140  (132)& $-$51.7\% & $-$1.1 & 0.265  (90)& $-$8.6\%  & $-$0.3 \\
15   & $-$0.11                       & $-$0.297 (427)& 170.0\%  & $-$0.4 & $-$0.087 (62)& $-$20.9\% & 0.4  & $-$0.110 (35)& 0.0\%   & 0.0  \\
16   & $-$0.56                       & 0.095  (707)& $-$117.0\% & 0.9  & $-$0.870 (394)& 55.4\%  & $-$0.8 & $-$0.567 (218)& 1.2\%   & 0.0  \\
17   & 0.19                        & 0.247  (714)& 30.0\%   & 0.1  & 0.188  (112)& $-$1.1\%  & 0.0  & 0.187  (67)& $-$1.6\%  & 0.0  \\
WAIC & $-$                           & 14.364 & $-$        & $-$    & 41.143 & $-$       & $-$    & 56.730 & $-$       & $-$    \\
LOO  & $-$                           & 7.782  & $-$        & $-$    & 32.782 & $-$       & $-$    & 53.277 & $-$       & $-$    \\
PM   & $-$                           & 0.2650 & $-$        & $-$    & 0.0510 & $-$       & $-$    & 0.0177 & $-$       & $-$
\end{tabular}
\end{ruledtabular}
\end{table}

Figure \ref{fig:nlo-percentage} (b) illustrates the distributions obtained by Models A and B$_2$ at each order. Table \ref{table9} gives a comparison among the true values, the experimental values, the fitting results from both Models B$_1$ and B$_2$. Both of them indicate the same conclusion as the NLO fit. Model B$_2$ can give better predictions of the truncation errors and the theoretical values.

\begin{table}[htpb]
\caption{The comparison of the NNLO fitting values for the example. The subscripts tr, exp, B$_1$ and B$_2$ in the first row represent the true values, the experimental values, the theoretical values from Model B$_1$ and Model B$_2$, respectively. The experimental values in the third column are sampled from the true values. $O_i$ is defined in Eq. \eqref{A1}.}\label{table9}
\begin{ruledtabular}
\begin{tabular}{lrrrr}
$i$ &$10^2O_{i,\mathrm {tr}}$&$10^2O_{i,\mathrm {exp}}$~~~~~&$10^2O_{i,\mathrm {B_1}}$~~~~~&$10^2O_{i,\mathrm {B_2}}$~~~~~\\\hline
1  & $-$35.800 & $-$34.637 $\pm$  0.716 & $-$34.219 $\pm$  2.034 & $-$35.189 $\pm$  0.910 \\
2  & 0.173   & 0.171   $\pm$  0.003 & 0.170   $\pm$  0.007 & 0.172   $\pm$  0.005 \\
3  & $-$0.276  & $-$0.279  $\pm$  0.006 & $-$0.279  $\pm$  0.049 & $-$0.279  $\pm$  0.036 \\
4  & 0.603   & 0.590   $\pm$  0.012 & 0.584   $\pm$  0.032 & 0.590   $\pm$  0.015 \\
5  & 27.281  & 27.753  $\pm$  0.546 & 27.091  $\pm$  0.843 & 27.604  $\pm$  0.675 \\
6  & $-$0.524  & $-$0.548  $\pm$  0.010 & $-$0.550  $\pm$  0.028 & $-$0.547  $\pm$  0.023 \\
7  & $-$1.486  & $-$1.434  $\pm$  0.030 & $-$1.434  $\pm$  0.031 & $-$1.469  $\pm$  0.020 \\
8  & $-$0.955  & $-$0.970  $\pm$  0.019 & $-$0.968  $\pm$  0.327 & $-$0.971  $\pm$  0.144 \\
9  & $-$0.227  & $-$0.226  $\pm$  0.005 & $-$0.226  $\pm$  0.008 & $-$0.225  $\pm$  0.010 \\
10 & $-$52.511 & $-$52.773 $\pm$  1.050 & $-$53.136 $\pm$  4.369 & $-$52.212 $\pm$  2.036 \\
11 & 44.936  & 46.250  $\pm$  0.899 & 46.466  $\pm$  2.628 & 45.904  $\pm$  1.176 \\
12 & $-$4.223  & $-$4.397  $\pm$  0.084 & $-$4.406  $\pm$  0.736 & $-$4.390  $\pm$  0.402 \\
13 & $-$14.674 & $-$14.769 $\pm$  0.293 & $-$14.713 $\pm$  2.670 & $-$14.760 $\pm$  1.834 \\
14 & $-$24.577 & $-$24.765 $\pm$  0.492 & $-$24.304 $\pm$  0.614 & $-$24.692 $\pm$  0.496 \\
15 & $-$15.864 & $-$15.505 $\pm$  0.317 & $-$15.609 $\pm$  1.776 & $-$15.541 $\pm$  1.006 \\
16 & 3.831   & 3.746   $\pm$  0.077 & 3.757   $\pm$  0.154 & 3.761   $\pm$  0.098 \\
17 & $-$4.193  & $-$4.208  $\pm$  0.084 & $-$4.136  $\pm$  0.134 & $-$4.212  $\pm$  0.093
\end{tabular}
\end{ruledtabular}
\end{table}

\subsection{Discussion}

In the NLO fit, we have also removed one $O_i$ and fitted the rest. The results are almost no different from the 17-input fit. Moreover, the 16-input fit can predict the 17th quantities well. This also shows that our model has a good predictive ability.

We have fitted other examples and obtained the same conclusion. If an example converges faster than the example in this paper, but the experimental errors and the NNLO contributions are at the same order, the experimental errors will have an impact on the HO values. The NNLO fitting results are a little worse. An example of this type can be downloaded from the source file in the arXiv version of this paper.

\section{Observables and inputs}\label{Sec:III}
In order to fit the actual data in ChPT and compare the results by different methods, almost the same physical quantities are chosen as those in Refs. \cite{Yang:2020eif,Bijnens:2014lea}, besides the covariance matrix of $\pi\pi$ scattering lengths $a_0^{0}$, $a_0^{2}$ and the two-flavor LECs $\bar l_1$, $\bar l_2$ and $\bar l_4$ is considered.

In Refs. \cite{Yang:2020eif,Bijnens:2014lea}, 12 input values are used in the NLO fit, i.e., the quark mass ratio $m_s/\hat m$ \cite{Amoros:1999dp, Amoros:2000mc, Bijnens:2011tb, Bijnens2019}, the ratio of decay constants of $K$ meson and $\pi$ meson $F_K/F_\pi$ \cite{Amoros:1999dp, Bijnens:2011tb, Bijnens:2014lea, Bijnens2019}, the shape factors $F$ and $G$ at threshold and their slope $f_s$, $g_p$, $f_s^{\prime}$ and $g^{\prime}_p$ for $K_{\ell 4}$ form factors \cite{Amoros:2000mc}, $\pi\pi$ scattering lengths $a_0^{0}$ and $a_0^{2}$ \cite{Colangelo:2001df}, $\pi K$ scattering lengths $a_0^{1/2}$ and $a_0^{3/2}$ \cite{Bijnens:2003xg}, pion scalar radius $\langle r^2\rangle_S^\pi$ in the form factor $F_S^\pi(t)$. In addition, there are 5 more input values added for the NNLO fit, i.e. the pion scalar curvature $c_S^\pi$ of the pion scalar form factor \cite{Bijnens:2003xg} and four two-flavor LECs $\bar l_i\, (i=1, \ldots, 4)$ \cite{Colangelo:2001df, Gasser:2007sg}. The values of these 17 physical quantities are listed below. In this paper, both 12 and 17 inputs are considered in the NLO fit for comparison.

The values of $m_s/\hat m$ and $F_K/F_\pi$ are
\begin{align}
\frac{m_s}{\hat m}=27. 3_{-1. 3}^{+0. 7}, \quad\frac{F_K}{F_\pi}=1. 199\pm 0. 003. \label{equ:14}
\end{align}

The values of $f_s$, $g_p$, $f_s^{\prime}$ and $g^{\prime}_p$ are
\begin{align}
\nonumber
f_s=5. 712\pm 0. 032, \quad f_s^{\prime}=0. 868\pm 0. 049, \\
g_p=4. 958\pm 0. 085, \quad g^{\prime}_p=0. 508\pm 0. 122. \label{equ:15}
\end{align}

The values of $\pi\pi$ scattering lengths $a_0^{0}$, $a_0^{2}$ and the three relevant two-flavor LECs are
\begin{align}
a_0^{0}=0.220 \pm 0.005, \quad a_0^{2}=-0.0444 \pm 0.0010, \quad \bar l_1=-0.4\pm 0.6, \quad \bar l_2=4.3 \pm 0.1, \quad \bar l_4=4.4 \pm 0.2.\label{equ:16}
\end{align}
The covariance matrix of $a_0^{0}$, $a_0^{2}$ and $\bar l_1$, $\bar l_2$, $\bar l_4$ is listed in Table \ref{table a}.

\begin{table}[htbp]
\caption{The covariance matrix of $a_0^{0}$, $a_0^{2}$ and $\bar l_1$, $\bar l_2$, $\bar l_4$. This is a symmetric matrix, only the values in the upper right corner of the matrix are given \cite{Colangelo:2001df}.}\label{table a}
\begin{ruledtabular}
\begin{tabular}{lccccc}
& $\Delta a_0^0$      & $\Delta a_0^2$       & $\Delta \bar{\ell}_1$ & $\Delta \bar{\ell}_2$ & $\Delta \bar{\ell}_4$ \\ \hline
$\Delta a_0^0$  & $2.0  \times 10^{-5}$ & $3.2 \times 10^{-6}$ & $1.9 \times 10^{-4}$  & $-1.7 \times 10^{-5}$ & $4.2 \times 10^{-4}$  \\
$\Delta a_0^2$                      &                     & $9.7 \times 10^{-7}$ & $1.6 \times 10^{-4}$  & $-1.2 \times 10^{-5}$ & $-4.2 \times 10^{-6}$ \\
$\Delta \bar{\ell}_1$               &                     &						 & $3.5 \times 10^{-1}$ & $-3.3 \times 10^{-2}$ & $6.7 \times 10^{-2}$   \\
$\Delta \bar{\ell}_2$               &                     &                      &        &        $1.2 \times 10^{-2}$  & $-7.2 \times 10^{-3}$   \\
$\Delta \bar{\ell}_4$               &                     &                      &                   &      & $4.8 \times 10^{-2}$
\end{tabular}
\end{ruledtabular}
\end{table}

We have tested whether the covariance matrix is present or not, it has a slight impact on the final fitting results, because the errors of $\bar l_i$ themselves are very large. Of course, in order to make the results more statistically significant, the covariance matrix is considered in the global fit.

The experimental values of $\pi K$ scattering lengths $a_0^{1/2}m_\pi$ and $a_0^{3/2}m_\pi$ are
\begin{align}
a_0^{1/2}m_\pi=0.224\pm 0.022, \quad a_0^{3/2}m_\pi=-0.0448\pm 0.0077. \label{equ:17}
\end{align}

The experimental values of the scalar radius $\langle r^2\rangle_S^\pi$ and the pion scalar form factor $c_S^\pi$ are
\begin{align}
\langle r^2\rangle_S^\pi=0.61\pm 0.04\, \mathrm{fm}^2, \quad c_S^\pi=11\pm 1\, \mathrm{GeV}^{-4}. \label{equ:18}
\end{align}

For $\bar l_3$, the following result is adopted \cite{Gasser:2007sg}
\begin{align}
\bar l_3=3.2\pm 0.7. \label{equ:21}
\end{align}

\section{Fitting the LECs in ChPT}\label{Sec:IV}
This section adopts the Bayesian Model B mentioned in Sec. \ref{mb} to perform a global fit, in order to obtain a new set of some NLO and NNLO LECs. The truncation errors are considered in the fit. Most references in this paper indicate that all $L^r_i$ ($C^r_i$) are at the order about $10^{-3}$ ($10^{-6}$). Following the preparation in Sec. \ref{sec:preparation}, they need to be first normalized by multiplying a factor $10^{3}$ ($10^{6}$), respectively.

\subsection{The NLO fitting $L_i^r$ by Model A}\label{r1}
Although this paper does not adopt the minimum $\chi^2$ method \cite{Bijnens:2011tb,Bijnens:2014lea,Yang:2020eif} to fit $L_i^r$, it can still obtain similar results from the NLO fit by Model A. The fit does not add the covariance matrix and does not consider the truncation errors, in order to compare with the results in Ref. \cite{Bijnens:2014lea}. The fitting results with the first 12 inputs in Sec. \ref{Sec:III} are shown in Table \ref{table22}. For comparison, the results in Ref. \cite{Bijnens:2014lea} are also given. Free fit means no assumptions in the fit. Otherwise, $L^r_4$ are assumed to be some fixed values. It can be seen that these two approaches indeed give very close results. The classical statistics is very similar to the Bayesian statistic. The slight differences come from the prior of $L^r_i$. This proves that they are equivalent laterally. However, Bayesian statistic is easier to introduce extra information. The minimum $\chi^2$ method can also add some constraints in the definition of $\chi^2$ \cite{Bijnens:2011tb,Bijnens:2014lea,Yang:2020eif}, but this information is restricted. For example, the prior PDF of LECs cannot be embodied in. In addition, the modified $\chi^2$ destroys the original definition of $\chi^2$. In other words, the new $\chi^2$ may not satisfy a $\chi^2$ distribution in fact.

\begin{table}[htbp]
\caption{The NLO fit by Model A, of which some different choices of $L_4^r$. Columns 2, 4, 6 and 8 are the results from free $L_4^r$, $L_4^r\equiv0$, $L_4^r\equiv0.3$ and $L_4^r\equiv-0.3$, respectively. Columns 3, 5, 7 and 9 are the results in Ref. \cite{Bijnens:2014lea} for comparison.}\label{table22}
\begin{ruledtabular}
\begin{tabular}{lcccccccc}
LECs & Free Fit & Free Fit \cite{Bijnens:2014lea} & $\mathrm{10^3L_4^r \equiv 0}$ & $\mathrm{10^3L_4^r \equiv 0}$ \cite{Bijnens:2014lea} & $\mathrm{10^3L_4^r \equiv 0. 3}$ & $\mathrm{10^3L_4^r \equiv 0. 3}$ \cite{Bijnens:2014lea}& $\mathrm{10^3L_4^r \equiv -0. 3}$ & $\mathrm{10^3L_4^r \equiv -0. 3}$ \cite{Bijnens:2014lea}\\ \hline
$10^3L_1^r$& $1.04(09)$  & $1.11(10)$  &$0.90(09)$  &$0.98(09)$ &$0.92(09)$ &$1.00(09)$ &$0.88(09)$  &$0.95(09)$   \\
$10^3L_2^r$& $1.00(11)$  & $1.05(17)$  &$1.49(08)$  &$1.56(09)$ &$1.41(08)$ &$1.48(09)$ &$1.57(08)$  &$1.64(09)$   \\
$10^3L_3^r$& $-3.52(28)$ & $-3.82(30)$ &$-3.52(28)$ &$-3.82(30)$&$-3.52(28)$&$-3.82(30)$&$-3.52(28)$ &$-3.82(30)$  \\
$10^3L_4^r$& $1.82(25)$  & $1.87(53)$  &$\equiv0$   &$\equiv0$  &$\equiv0.3$&$\equiv0.3$&$\equiv-0.3$&$\equiv-0.3$ \\
$10^3L_5^r$& $1.24(03)$  & $1.22(06)$  &$1.25(03)$  &$1.23(06)$ &$1.24(03)$ &$1.23(06)$ &$1.25(03)$  &$1.23(06)$   \\
$10^3L_6^r$& $1.46(25)$  & $1.46(46)$  &$-0.12(05)$ &$-0.11(05)$&$0.13(06)$ &$0.14(06)$ &$-0.37(04)$ &$-0.36(05)$  \\
$10^3L_7^r$& $-0.40(14)$ & $-0.39(08)$ &$-0.19(14)$ &$-0.24(15)$&$-0.23(14)$&$-0.27(14)$&$-0.16(14)$ &$-0.21(17)$  \\
$10^3L_8^r$& $0.60(12)$  & $0.65(07)$  &$0.51(12)$  &$0.53(13)$ &$0.53(12)$ &$0.55(12)$ &$0.50(12)$  &$0.50(14)$   \\
\end{tabular}
\end{ruledtabular}
\end{table}

\subsection{The NLO fitting $L_i^r$}\label{NLO17}
In order to fit $L_i^r$, a similar approach to that of the example in Sec. \ref{ME} is adopted, but the parameters in HO are slightly different from the example. $m_s/\hat m|_1$, $m_s/\hat m|_2$, $F_K/F_\pi$, $f_s$, $g_p$, $a_0^0$, $a_0^2$, $a_0^{1/2}m_\pi$, $a_0^{3/2}m_\pi$, $\bar l_1$, $\bar l_2$, $\bar l_3$ and $\bar l_4$ are the same as the expansion in Eq. \eqref{eq3}. $f_s^{\prime}$, $g^{\prime}$, $\langle r^2\rangle_S^\pi$ and $c_S^\pi$ involve a numerical differentiation. They are estimated with the method in Sec. \ref{cd}. We have gotten some information about the higher-order experimental data and the range of the LECs, so the parameters are set in a way that is between Model B$_1$ and Model B$_2$. Therefore, from here, all data are fitted using Model B. In this subsection, besides fitting the whole 17 inputs (Model B$^{17}$), we also fit the first 12 inputs (Model B$^{12}$) in Sec. \ref{Sec:III} for comparing to Refs. \cite{Bijnens:2014lea,Yang:2020eif}.

The setting parameters can be found in Columns 2 to 7 in Table \ref{app table3} in Appendix \ref{app:B}, the parameters about $a_0^{1/2}m_\pi$ and $a_0^{3/2}m_\pi$ are given by Ref. \cite{Bijnens:2014lea}, which indicates that their convergences have been broken. The values about $f_s$ and $a_0^0$ are given by their NNLO distributions, which are statistically obtained from the ranges of $L_i^r$ and $C_i^r$ collected by Refs. \cite{Jiang:2015dba,Bijnens:2014lea,Yang:2020eif} and the references in them. The other parameters are given the same as Model B$_1$. The prior of $L_i^r$ is given in Columns 2 and 5 in Table \ref{app table4} in Appendix \ref{app:B}. They refer to the $L_i^r$ ranges given in Ref. \cite{Jiang:2015dba,Bijnens:2014lea,Yang:2020eif} and the references in them. Because the values in the different references are not very close, the prior ranges are wide enough to cover all possible ranges.

\begin{figure}[!htbp]
\centering
\includegraphics[width=0.82\linewidth]{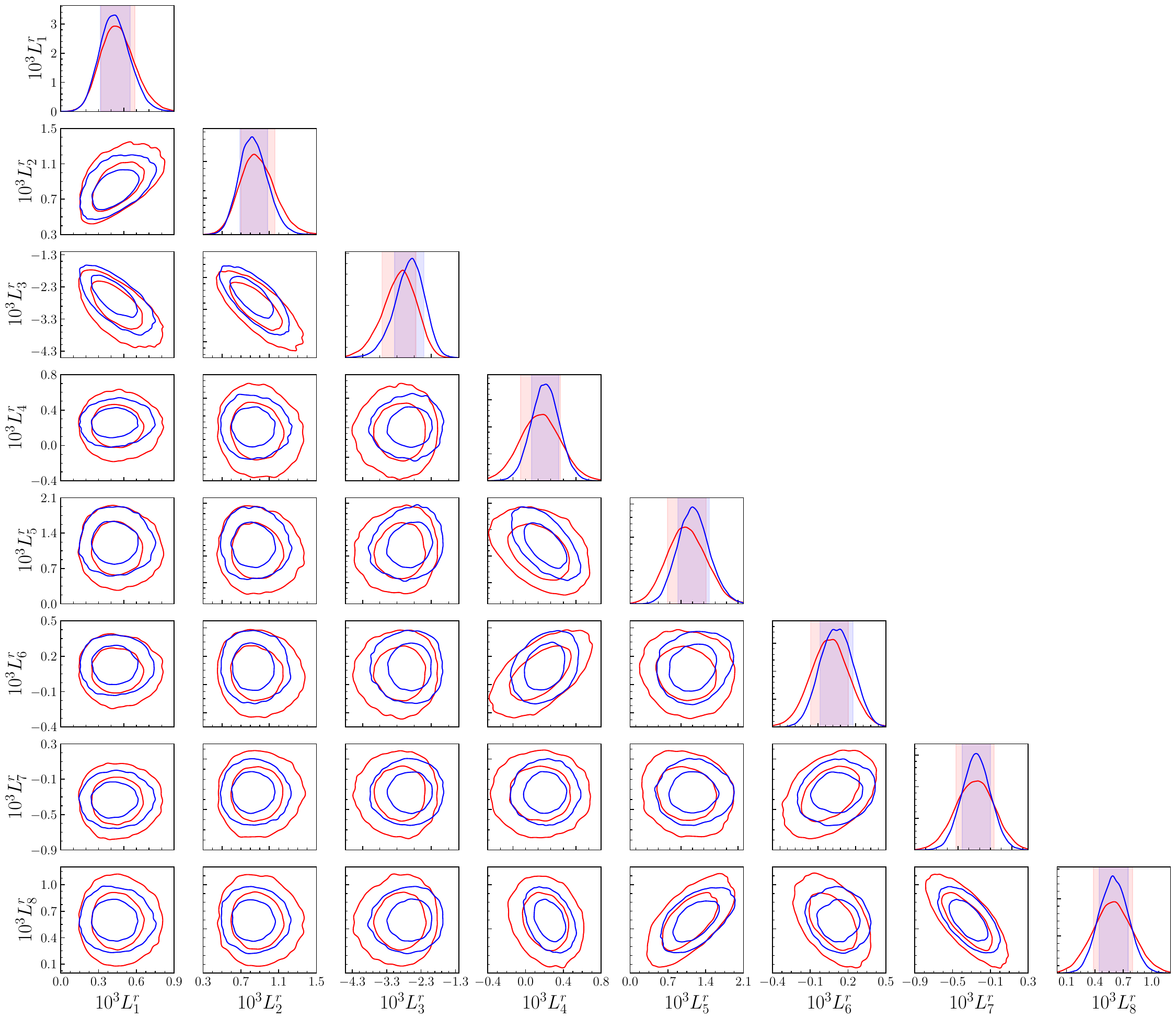}
\caption{The corner plot of the 17-input fitting $L_i^r$. The red and blue colors mean the NLO and NNLO fit, respectively. The small and large loops mean the 68\% HPD and the 95\% HPD, respectively. The light-colored areas are the 68\% HPD.}
\label{fig:nlo-corner-plots}
\end{figure}

The numerical results of both fits can be found in Table \ref{table27}. It can be seen that the results obtained by both Models B$^{12}$ and B$^{17}$ are close to the NNLO results in Ref. \cite{Bijnens:2014lea,Yang:2020eif}. Moreover, both of them also satisfy the large-$N_c$ limit, i.e. $2L_1^r-L_2^r$, $L_4^r$ and $L_6^r$ closing to zeros, although it does not give a strong prior of $L^r_4$. This shows that the contributions from truncation errors have a great impact on the NLO fit. It is also very possible that the truncation errors cannot be ignored in the NNLO fit. In addition, all theoretical errors from Model B are slightly larger than those in Ref. \cite{Bijnens:2014lea,Yang:2020eif}. This is because Ref. \cite{Yang:2020eif} does not consider the errors caused by the truncation errors. Ref. \cite{Bijnens:2014lea} even does not consider the truncation errors. Model B cannot only estimate these truncation errors, but also considers their PDFs. These PDFs lead the fitting errors to be slightly larger than those in Ref. \cite{Bijnens:2014lea,Yang:2020eif}. However, the difference is not very large, because the truncation errors are not very large. It also shows that the change between 12 and 17 inputs is not very large. The relative difference does not exceed 20\%. However, since more inputs are added, all theoretical errors became smaller. In addition, since Model B$^{12}$ and Model B$^{17}$ do not adopt the same inputs, the WAIC and LOOCV cannot be adopted as model evaluation criteria. Hence, we do not give these two values. The following discussion is based on the results of Model B$^{17}$, because the fit becomes more accurate as the input value increases. The red part in Figure \ref{fig:nlo-corner-plots} is the corner plot of $L_i^r$ with 17 inputs, from which one can see both the distributions and the potential correlations between $L_i^r$.

\begin{table}[!htbp]
\caption{The fitting results of $L_i^r$. The superscripts indicate the input number in the fit. Columns 5 to 8 are the NLO and the NNLO fitting results in Refs. \cite{Yang:2020eif,Bijnens:2014lea}, respectively.}\label{table27}
\begin{ruledtabular}
\begin{tabular}{lccccccc}
LECs  & NLO B$^{12}$  & NLO B$^{17}$  &  NNLO B$^{17}$  & NLO fit \cite{Bijnens:2014lea} & NNLO fit \cite{Bijnens:2014lea}&NLO Fit 2 \cite{Yang:2020eif}&NNLO Fit 2 \cite{Yang:2020eif}  \\ \hline
$10^3L_1^r$&$ 0.51  (15)$&$ 0.46  (14)$& $ 0.43  (12)  $&$1.00(09)$ &$ 0.53        (06)$&$ 0.44  (05)$&$ 0.43  (05)$\\
$10^3L_2^r$&$ 1.08  (22)$&$ 0.88  (18)$& $ 0.83  (15)  $&$1.48(09)$ &$ 0.81        (04)$&$ 0.84  (10)$&$ 0.74  (04)$\\
$10^3L_3^r$&$ -3.36 (61)$&$ -2.94 (49)$& $ -2.64 (44)  $&$-3.82(30)$&$ -3.07       (20)$&$ -2.84 (16)$&$ -2.74 (17)$\\
$10^3L_4^r$&$ 0.19  (18)$&$ 0.22  (16)$& $ 0.26  (11)  $&$\equiv0.3$&$ \equiv0.3 $      &$ 0.30  (33)$&$ 0.33  (08)$\\
$10^3L_5^r$&$ 1.10  (37)$&$ 1.10  (34)$& $ 1.21 (27)  $ &$1.23(06)$ &$ 1.01        (06)$&$ 0.92  (02)$&$ 0.95  (04)$\\
$10^3L_6^r$&$ 0.05  (22)$&$ 0.08  (13)$& $ 0.12  (11)  $&$0.14(06)$ &$ 0.14        (05)$&$ 0.22  (08)$&$ 0.20  (03)$\\
$10^3L_7^r$&$ -0.26 (17)$&$ -0.34 (18)$& $ -0.33 (13)  $&$-0.27(14)$&$ -0.34       (09)$&$ -0.23 (12)$&$ -0.23 (08)$\\
$10^3L_8^r$&$ 0.51  (22)$&$ 0.59  (21)$& $ 0.60  (15)  $&$0.55(12)$ &$ 0.47        (10)$&$ 0.44  (10)$&$ 0.42  (09)$\\
$\chi^2$(d.o.f.)&--&--  &--               &    --       &$           1.0(9)$&$   4.2(4)  $&$4.3(9) $    \\
\end{tabular}
\end{ruledtabular}
\end{table}

Table \ref{table28} lists the 17-input theoretical contributions at each order. $\bar{l}_i^r$ is replaced by $l_i^r$ \cite{Gasser:1983yg}, because $l_i^r$ has a better convergence, theoretically. It can be seen that most expansions at each order conform to the convergence hypothesis very well. Most LO values contribute more than 70\%, most NLO values contribute within 10\% to 23\%, and most HO values contribute less than 10\%. All these percentages are neither too large nor too small. All theoretical results agree well with the experimental data. The ratios of the adjacent two orders are about 0.2, except for $a_0^{1/2}$ and $a_0^{3/2}$, which HO contributions are larger than the NLO ones. This situation also exists in Refs. \cite{Bijnens:2014lea,Yang:2020eif}. There are two reasons. One is that the experimental values of both $a_0^{1/2}$ and $a_0^{3/2}$ are not very precise. Compared to $a_0^0$ and $a_0^2$, their errors are too large and the estimating truncation errors are not so precise. It may lead to a poor convergence. The second reason is that there indeed exist broken convergence problems in the expansions of $a_0^{1/2}$ and $a_0^{3/2}$. These two reasons are related to a more precise experiment and theoretical calculation, and we do not discuss it anymore in this paper. However, although the NLO fitting results of $a_0^{1/2}$ and $a_0^{3/2}$ in Ref. \cite{Yang:2020eif} are converged, it assumes a geometric sequence model. Ref. \cite{Bijnens:2014lea} also exists this problem. However, Model B introduces the priors and has a wider scope of application. A better prior can predict its theoretical value within a more reasonable range. In addition, the total contribution of $l_3^r$ is basically occupied by the NLO and its HO value tends towards 0. This is because the error of $l_3^r$ itself is very large, which is about 3.7 times its expected value. Therefore, the contribution of $l_3^r$ in the fit becomes very small, and the fitting expected value can be far away from the experimental expected value. Hence, adopting the experimental values of $l_3^r$ as a constraint to constrain LECs in Ref. \cite{Yang:2020eif} seems not particularly good. Model B adopts both the convergence assumption and the prior PDFs. It can handle most precise data, so most results also conform with the convergence assumption very well. Only a few results with poor convergence, because of the problem itself or the large experimental errors.

\begin{table}[htbp]
\caption{The convergences of 17 inputs. The LECs are adopted from the 17-input NLO fitting results obtained by Model B in Table \ref{table27}. The second to the fourth columns are the contributions at the LO, NLO and HO, respectively. The percentage $\mathrm{Pct}_{\mathrm{LO, NLO, HO}}$ is defined in Eq. \eqref{equ:1}. The last two columns are the theoretical estimation and the experimental inputs, respectively.}\label{table28}
\begin{ruledtabular}
\begin{tabular}{lccccc}
\multicolumn{1}{c}{Observables}&LO$|\mathrm{Pct}_\mathrm{LO}$&NLO$|\mathrm{Pct}_\mathrm{NLO}$&HO$|\mathrm{Pct}_\mathrm{HO}$&Theory&Experiment\\\hline
$m_s/\hat m|_1$            &$ 25.84  ( 96.2\%  )$&$ 0.84  ( 3.1\%   )$&$ 0.19   ( 0.7\%   )$&$ 26.9   \pm 3.1   $& $27.3_{-1.3}^{+0.7}$ \\
$m_s/\hat m|_2$            &$ 24.21  ( 88.4\%  )$&$ 3.23  ( 11.8\%  )$&$ -0.06  ( -0.2\%  )$&$ 27.4   \pm 6.3   $& $27.3_{-1.3}^{+0.7}$ \\
$F_K/F_\pi$                &$ 1.000  ( 84.1\%  )$&$ 0.184 ( 15.5\%  )$&$ 0.004  ( 0.3\%   )$&$ 1.188  \pm 0.036 $& 1.199$\pm$ 0.003     \\
$f_s$                      &$ 3.782  ( 66.2\%  )$&$ 1.267 ( 22.2\%  )$&$ 0.660  ( 11.6\%  )$&$ 5.709  \pm 0.347 $& $5.712\pm 0.032$     \\
$g_p$                      &$ 3.782  ( 77.9\%  )$&$ 0.915 ( 18.8\%  )$&$ 0.159  ( 3.3\%   )$&$ 4.856  \pm 0.191 $& $4.958\pm 0.085$     \\
$a_0^0$                    &$ 0.159  ( 72.5\%  )$&$ 0.044 ( 20.2\%  )$&$ 0.016  ( 7.4\%   )$&$ 0.2197 \pm 0.005 $& $0.2196\pm 0.0034$   \\
$10a_0^2$                  &$ -0.455 ( 104.0\% )$&$ 0.019 ( -4.4\%  )$&$ -0.002 ( 0.4\%   )$&$ -0.437 \pm 0.015 $& $-0.444\pm 0.012$    \\
$a_0^{1/2}m_\pi$           &$ 0.142  ( 63.3\%  )$&$ 0.033 ( 14.6\%  )$&$ 0.049  ( 22.1\%  )$&$ 0.224  \pm 0.014 $& $0.224\pm 0.022$     \\
$10a_0^{3/2}m_\pi$         &$ -0.709 ( 158.3\% )$&$ 0.084 ( -18.8\% )$&$ 0.177  ( -39.5\% )$&$ -0.448 \pm 0.090 $& $-0.448\pm 0.077$    \\
$10^3l_1^r$                &$ 0      ( 0.0\%   )$&$ -4.07 ( 98.6\%  )$&$ -0.06  ( 1.4\%   )$&$ -4.1   \pm 1.1   $& $-4.0\pm0.6$         \\
$10^3l_2^r$                &$ 0      ( 0.0\%   )$&$ 3.50  ( 160.2\% )$&$ -1.32  ( -60.2\% )$&$ 2.2    \pm 0.9   $& $1.9\pm0.2$          \\
$10^3l_3^r$                &$ 0      ( 0.0\%   )$&$ -0.18 ( 104.0\% )$&$ 0.01   ( -4.0\%  )$&$ -0.2   \pm 3.1   $& $0.3\pm1.1$          \\
$10^3l_4^r$                &$ 0      ( 0.0\%   )$&$ 6.07  ( 99.9\%  )$&$ 0.01   ( 0.1\%   )$&$ 6.1    \pm 2.0   $& $6.2\pm1.3$          \\
$f_s^{\prime}$             &$        $&$       $&$        $&$ 0.531  \pm 0.322 $& $ 0.868  \pm 0.049 $ \\
$g^{\prime}$               &$        $&$       $&$        $&$ 0.368  \pm 0.036 $& $ 0.508  \pm 0.122 $ \\
$\langle r^2\rangle_S^\pi$ &$        $&$       $&$        $&$ 0.60   \pm 0.13  $& $ 0.61   \pm 0.04  $ \\
$c_S^\pi$                  &$        $&$       $&$        $&$ 10     \pm 2     $& $ 11   \pm 1  $
\end{tabular}
\end{ruledtabular}
\end{table}

Refs. \cite{Bijnens:2014lea,Yang:2020eif} fit the NLO LECs only with the first 12 inputs in Sec. \ref{Sec:III}, because the remaining five physical quantities have zero value in the LO. Therefore, if the truncation errors are not considered, the NLO fit does not contain the NNLO contribution. The results would exhibit a large deviation, because the HO contributions may lead to large influences. Although Ref. \cite{Yang:2020eif} can estimate the truncation errors, it requires at least two-order values because of a geometric-sequence model. Hence, this model cannot work for these five physical quantities. At present, the Bayesian method only requires at least one-order values to estimate the truncation errors. In other words, with Model B, even physical quantities with zero LO can be used as part of data fitting in NLO. Therefore, we also perform a full fit of all 17 physical quantities at the NLO.

\subsection{The NNLO fitting $L_i^r$  and $\widetilde{C}_i$ }\label{r2}
The $C_i^r$ to be fitted at the NNLO in this paper is the same as those in Ref. \cite{Yang:2020eif}. There exist 38 $C_i^r$, while the number of observables are 17. Hence, these 38 $C_i^r$ are combined into 17 linearly independent $\widetilde{C}_i$ before the fit. The definitions of $\widetilde{C}_i$ are in Appendix A in Ref. \cite{Yang:2020eif}. In the NNLO fit, $L_i^r$ and $\widetilde{C}_i$ are fitted simultaneously using the approach mentioned in Section \ref{cd}.

The setting parameters are placed in Columns 8 to 10 in Table \ref{app table3} in Appendix \ref{app:B}. All the parameters are given as Model B$_1$, because we have known nothing about the truncation errors. The prior of $\widetilde{C}_i$ can be found in Columns 6 and 7 in Table \ref{app table4} in Appendix \ref{app:B}. They're referred to the $C_i^r$ ranges given in Table IX in Ref. \cite{Yang:2020eif}. The blue part in Figure \ref{fig:nlo-corner-plots} shows the NNLO fitting corner plot of $L_i^r$. Column 4 in Table \ref{table27} lists the NNLO fitting results of $L_i^r$. Both Figure \ref{fig:nlo-corner-plots} and Table \ref{table27} indicate that there is no significant change of the theoretical expected values between the NLO and the NNLO fit. In addition, the NNLO fitting $L_i^r$ and their correlations with smaller theoretical errors, because it is the introduction of the NNLO contributions. Table \ref{table27} also indicates that the difference between the 17 inputs at NNLO and 12 or 17 inputs at NLO are not very large, all within 20\%. This indicates that this method is stable and does not cause an obvious change of $L_i^r$ as the order increases. This is exactly one of the motivations in this paper.

Figure \ref{fig:Ctilted} in Appendix \ref{app:B} gives the posterior distributions of $\widetilde{C}_i$. The introduction of the constraints in Eqs. \eqref{co1} -- \eqref{co3} causes some $\widetilde{C}_i$ to deviate from normal distributions, but not very seriously. Table \ref{table31} shows the numerical results of $\widetilde{C}_i$. Compared with those results in Ref. \cite{Yang:2020eif}, all standard deviations are slightly larger. The reason is Eq. \eqref{eq4} considers the errors of the truncation errors and enlarges the theoretical errors.

\begin{table}[htbp]
\caption{The values and the errors of $\widetilde{C}_i$, comparing with the results in Ref.\cite{Yang:2020eif}.}\label{table31}
\begin{ruledtabular}
\begin{tabular}{cccccc}
$\widetilde{C}_i$   &   Results       & Ref. \cite{Yang:2020eif}&  $\widetilde{C}_i$     &   Results & Ref. \cite{Yang:2020eif}   \\ \hline
$\widetilde{C}_1$     & $ 0.11  (22)  $  & $0.02(12)$  & $10\widetilde{C}_{10}$   & $0.22  (32)  $   & $-0.06(13)$ \\
$\widetilde{C}_2$     & $ 0.09  (58)  $  & $0.19(34)$  & $\widetilde{C}_{11}$     & $ 0.22  (07)  $  & $0.24(02)$  \\
$10^2\widetilde{C}_3$ & $ -1.29 (77)  $  & $-0.72(42)$ & $10^3\widetilde{C}_{12}$ & $ 0.02  (05)  $  & $-0.18(01)$ \\
$10^2\widetilde{C}_4$ & $ 0.05  (08)  $  & $0.22(03)$  & $10^3\widetilde{C}_{13}$ & $ -0.13  (35)  $ & $1.02(44)$  \\
$10\widetilde{C}_5$   & $ -0.08 (04)  $  & $-0.16(02)$ & $10^4\widetilde{C}_{14}$ & $ 0.38  (27)  $  & $0.29(06)$  \\
$10^3\widetilde{C}_6$ & $ -0.76 (151)  $ & $0.26(13)$  & $10^3\widetilde{C}_{15}$ & $ -0.12 (03)  $  & $-0.11(01)$ \\
$10^2\widetilde{C}_7$ & $ -0.70 (63)  $  & $-0.42(12)$ & $10^4\widetilde{C}_{16}$ & $ -0.46 (35)  $  & $-0.56(06)$ \\
$10\widetilde{C}_8$   & $ -0.08 (22)  $  & $-0.45(09)$ & $10^4\widetilde{C}_{17}$ & $ -0.26  (22)  $ & $0.19(16)$  \\
$10^2\widetilde{C}_9$ & $ -0.46 (43)  $  & $-0.99(11)$ &                          &                  &
\end{tabular}
\end{ruledtabular}
\end{table}

Table \ref{table32} gives the theoretical contributions at each order with the NNLO fit. It can be seen that most physical quantities satisfy the chiral convergence very well, except for $a_0^{1/2}$, $10a_0^{3/2}$, $l_2^r$ and $l_3^r$. This situation also exists in the NNLO fit and in Ref. \cite{Yang:2020eif}. The reason has been discussed in Sec. \ref{NLO17}. It also leads to a large theoretical error of $l_3^r$. If a set of more precise experimental data are introduced, this problem may not exist anymore.

\begin{table}[htbp]
\caption{Same as Table \ref{table28}, except for the NNLO fit.}\label{table32}
\begin{ruledtabular}
\begin{tabular}{lcccccc}
\multicolumn{1}{c}{Observables}&\multicolumn{1}{c}{LO$|\mathrm{Pct}_\mathrm{LO}$}&\multicolumn{1}{c}{NLO$|\mathrm{Pct}_\mathrm{NLO}$}&\multicolumn{1}{c}{NNLO$|\mathrm{Pct}_\mathrm{NNLO}$}&\multicolumn{1}{c}{HO$|\mathrm{Pct}_\mathrm{HO}$}&Theory&Experiment\\\hline
$m_s/\hat m|_1$            &$ 25.84  (95.1\%)$&$ 1.45  (5.3\%  )$&$ -0.12 (-0.5\%)$&$ 0.003  (0.01\%)$&$ 27.2   \pm  4.1   $& $27.3_{-1.3}^{+0.7}$ \\
$m_s/\hat m|_2$            &$ 24.21  (88.4\%)$&$ 3.60  (13.1\% )$&$ -0.38 (-1.4\%)$&$ -0.020 (-0.07\% $&$ 27.4   \pm  10.7  $& $27.3_{-1.3}^{+0.7}$ \\
$F_K/F_\pi$                &$ 1.000  (83.2\%)$&$ 0.197 (16.4\% )$&$ 0.007 (0.6\% )$&$ -0.002 (-0.12\% $&$ 1.202  \pm  0.050 $&$ 1.199\pm 0.003$     \\
$f_s$                      &$ 3.782  (66.7\%)$&$ 1.342 (23.7\% )$&$ 0.494 (8.7\% )$&$ 0.050  (0.88\%)$&$ 5.668  \pm  0.351 $& $5.712\pm 0.032$     \\
$g_p$                      &$ 3.782  (76.9\%)$&$ 0.834 (17.0\% )$&$ 0.284 (5.8\% )$&$ 0.018  (0.37\%)$&$ 4.918  \pm  0.080 $& $4.958\pm 0.085$     \\
$a_0^0$                    &$ 0.159  (72.2\%)$&$ 0.045 (20.6\% )$&$ 0.015 (6.8\% )$&$ 0.001  (0.34\%)$&$ 0.2204  \pm  0.004 $& $0.2196\pm 0.0034$   \\
$10a_0^2$                  &$ -0.455 (104.1\% $&$ 0.020 (-4.6\% )$&$ 0.003 (-0.7\%)$&$ -0.005 (1.20\%)$&$ -0.437 \pm  0.017 $& $-0.444\pm 0.012$    \\
$a_0^{1/2}m_\pi$           &$ 0.142  (63.2\%)$&$ 0.034 (15.2\% )$&$ 0.049 (21.6\%)$&$ 0.000  (0.00\%)$&$ 0.225  \pm  0.013 $& $0.224\pm 0.022$     \\
$10a_0^{3/2}m_\pi$         &$ -0.709 (161.9\% $&$ 0.093 (-21.2\%)$&$ 0.179 (-40.8\% $&$ 0.000  (0.04\%)$&$ -0.438 \pm  0.061 $& $-0.448\pm 0.077$    \\
$10^3l_1^r$                &$ 0      (0.0\% )$&$ -3.57 (90.4\% )$&$ -0.38 (9.7\% )$&$ 0.004  (-0.11\% $&$ -4.0  \pm  0.94  $& $-4.0\pm0.6$         \\
$10^3l_2^r$                &$ 0      (0.0\% )$&$ 3.32  (174.8\%)$&$ -1.36 (-71.8\% $&$ -0.057 (-3.01\% $&$ 1.9   \pm  0.82  $& $1.9\pm0.2$          \\
$10^3l_3^r$                &$ 0      (0.0\% )$&$ -0.25 (-115.1\% $&$ 0.46  (214.6\% $&$ 0.001  (0.48\%)$&$ 0.2   \pm  3.01  $& $0.3\pm1.1$          \\
$10^3l_4^r$                &$ 0      (0.0\% )$&$ 6.85  (104.8\%)$&$ -0.27 (-4.1\%)$&$ -0.044 (-0.67\% $&$ 6.5   \pm  1.91  $& $6.2\pm1.3$          \\
$f_s^{\prime}$             &$        $&$       $&$       $&$        $&$ 0.472  \pm  0.461 $& $ 0.868  \pm 0.049 $ \\
$g^{\prime}$               &$        $&$       $&$       $&$        $&$ 0.508  \pm  0.029 $& $ 0.508  \pm 0.122 $ \\
$\langle r^2\rangle_S^\pi$ &$        $&$       $&$       $&$        $&$ 0.59   \pm  0.07  $& $ 0.61   \pm 0.04  $ \\
$c_S^\pi$                  &$        $&$       $&$       $&$        $&$ 11     \pm  1     $& $ 11   \pm 1  $

\end{tabular}
\end{ruledtabular}
\end{table}

\subsection{The NNLO fitting $C_i^r$}
This section discusses the fit about $C_i^r$. $\tilde{C}_i$, which have been determined in Table \ref{table31}, are linear combinations of $C_i^r$. Although the number of $\tilde{C}_i$ is less than the number of $C_i^r$, three $C_i^r$ can be determined by solving the linear equations \cite{Yang:2020eif}. However, some of these values will be one to two orders of magnitude times larger than those in the other references. Some constraints are required to be introduced. For the other unsolvable $C_i^r$, their distribution is solved by the Monte Carlo method \cite{Yang:2020eif}. Although the approach in Ref. \cite{Yang:2020eif} can solve this problem, its efficiency is very low, and it needs to take a lot of time. Therefore, this paper adopts the MCMC algorithms in Sec. \ref{MCMC}. We have repeated the computation many times, and some similar results are obtained. Randomness does not affect the results obtained by this method. The difference is that all the prior PDFs of parameters $C_i^r$ are all set to the different uniform distributions. The boundaries of these prior uniform distributions are the same as Eq. (38) in Ref. \cite{Yang:2020eif}. The reason to use the prior uniform distributions instead of a prior normal distribution is that we want to explore the boundary dependence of each $C_i^r$ in this overfitting problem. Normal distributions would generate fewer samples near the boundaries, and the efficiency is low.

Figure \ref{fig:c2} in Appendix \ref{app:B} illustrates the posterior distributions of $C_i^r$. It can be seen that different $C^r_i$ have different shapes. $C_i^r, (i=3, 7, 8, 10, 16, 17, 18, 20, 22, 23, 28, 30, 32, 33, 36, 63, 66, 69, 83, 88, 90)$ have a large probability near both boundaries. Their posterior PDFs are dependent on both sides. In addition, $ C_i^r, (i=2, 6, 26, 29, 34)$ only depend on one side. This can also be seen from their posterior distributions. One side has a shape similar to a half-Gaussian distribution. The constraint of these $C_i^r$ at this side is reliable, but the other side gives no constraint of these $C_i^r$. Finally, these twelve $C_i^r , (i=1, 4, 5, 11, 12, 13, 14, 15, 19, 21, 25, 31)$ give Gaussian-like posterior PDFs, so these twelve results have higher credibility. Of course, 17 data to fit 38 $C_i^r$ is far from adequate. There exists an overfitting problem. Hence, some $C_i^r$ are boundary-dependent. This property is similar to those in Ref. \cite{Yang:2020eif}.

Table \ref{table34} gives the fitting values of $C_i^r$ and compares the results in Refs. \cite{Yang:2020eif,Bijnens:2014lea,Jiang:2015dba}. The brackets “[” and “]” denote that the results are strongly dependent on the lower and the upper boundaries, respectively. The parentheses “(” and “)” denote that the results are weakly dependent on the lower and the upper boundaries, respectively. We have tried to double the boundaries, the strong-dependent boundaries deviate from the original values a lot, while weak-dependent boundaries change the original values slightly. Of course, the boundaries chosen in Ref. \cite{Yang:2020eif} are wide enough, they cover almost all results in the other references \cite{Colangelo:2012ipa,Jiang:2009uf,Bijnens:2014lea,Jiang:2015dba,Bijnens:2011tb,Kampf:2006bn,Kampf:2006bn,Jamin:2004re,Bijnens:2003uy,Cirigliano:2005xn,Unterdorfer:2008zz,Cirigliano:2006hb,Bernard:2007tk,Moussallam:2000zf}. Hence, the true values have a large probability in the intervals in Table \ref{table34}.

\begin{table}[!htbp]
\caption{The values of $C_i^r$ are in units of $10^{-6}$. The brackets ``['' and ``]'' represent strong dependence on the lower and the upper boundaries, respectively. ``('' and ``)'' represent weak dependence on the lower and the upper boundaries, respectively. The results with an asterisk mean the input boundaries on the website \cite{Bijnens2019} are very close to those in Ref. \cite{Bijnens:2014lea} (less than $10^{-10}$). The symbol ``$\equiv0$'' for the results in Ref. \cite{Jiang:2015dba} means these values are zeros in the large-$N_C$ limits.}\label{table34}
\begin{ruledtabular}
\begin{tabular}{cccccccccc}
LECs    &    Results &     Ref. \cite {Yang:2020eif}     & Ref. \cite{Bijnens:2014lea} & Ref. \cite{Jiang:2015dba} &                  LECs                  &   Results &         Ref. \cite {Yang:2020eif}   & Ref. \cite{Bijnens:2014lea} & Ref. \cite{Jiang:2015dba} \\
\hline
$C^r_1$    & 14.82$(41.49 )$    & $14[37]$      & 12$^*$     & $25.33^{+0.60}_{-1.11}$  & $C^r_{21}$ & $-$0.41$(0.82)$      & $-0.28(0.56)$ & $-$0.48    & $-0.51^{+0.09}_{-0.09}$  \\
$C^r_2$    & 3.48$(8.98  ]$     & $16(1]$       & 3.0$^*$    & $\equiv 0$               & $C^r_{22}$ & 5.88  $[ 15.71 ]$    & $14(13]$      & 9.0$^*$    & $-2.98^{+1.70}_{-2.21}$  \\
$C^r_3$    & 1.70 $[ 6.05  ]$   & $2.9[6.0]$    & 4.0$^*$    & $-0.43^{+0.09}_{-0.09}$  & $C^r_{23}$ & 0.92  $[ 3.52  ]$    & $5.6(0.9]$    & $-$1.0$^*$ & $\equiv 0$               \\
$C^r_4$    & 18.54 $(29.94)$    & $-26[16)$     & 15$^*$     & $18.11^{+0.51}_{-0.85}$  & $C^r_{25}$ & $-$21.17  $(58.67)$  & $34(33)$      & $-$11$^*$  & $-25.76^{-3.49}_{+5.02}$ \\
$C^r_5$    & $-$3.62$(19.23)$   & $-31[7)$      & $-$4.0$^*$ & $-10.88^{+0.85}_{-1.11}$ & $C^r_{26}$ & $-$4.30 $[ 42.04)$   & $31(36]$      & 10         & $23.04^{+2.98}_{-4.59}$  \\
$C^r_6$    & $-$3.43$[4.22  )$  & $-7.9[1.8)$   & $-$4.0$^*$ & $\equiv 0$               & $C^r_{28}$ & 0.45  $[ 3.95  ]$    & $-4.9[0.9)$   & $-$2.0$^*$ & $1.53^{+0.00}_{-0.09}$   \\
$C^r_7$    & 1.00 $[ 6.26 ]$    & $2.4[6.1]$    & 5.0$^*$    & $\equiv 0$               & $C^r_{29}$ & $-$23.20$[ 24.09)$   & $-49[11)$     & $-$20$^*$  & $-8.42^{-1.79}_{+2.04}$  \\
$C^r_8$    & 10.52 $[ 16.84 ]$  & $15[16]$      & 19$^*$     & $17.85^{-1.28}_{+1.36}$  & $C^r_{30}$ & 3.44  $[ 4.61  ]$    & $9.0(1.9]$    & 3.0$^*$    & $3.15^{+0.09}_{-0.17}$   \\
$C^r_{10}$ & $-3.51$$[ 13.75 ]$ & $13(6]$       & $-$0.25    & $-5.53^{+0.43}_{-0.51}$  & $C^r_{31}$ & 2.15  $(9.56)$       & $-0.71(6.70)$ & 2.0$^*$    & $-3.91^{+0.60}_{-1.11}$  \\
$C^r_{11}$ & $-$2.90$(4.17)$    & $-2.6(1.8)$   & $-$4.0$^*$ & $\equiv 0$               & $C^r_{32}$ & 1.79  $[ 3.63  ]$    & $5.6(1.9]$    & 1.7        & $1.45^{-0.17}_{+0.26}$   \\
$C^r_{12}$ & $-$6.02$(5.57)$    & $18(2)$       & $-$2.8     & $-2.89^{+0.09}_{-0.09}$  & $C^r_{33}$ & $-$0.01  $[ 3.58  ]$ & $-0.69[3.12)$ & 0.82       & $-0.43^{-0.17}_{+0.43}$  \\
$C^r_{13}$ & 1.74 $(2.06)$      & $2.2(0.9)$    & 1.5        & $\equiv 0$               & $C^r_{34}$ & 9.27  $(10.48 ]$     & $0.68(4.67)$  & 7.0$^*$    & $5.61^{-1.53}_{+2.47}$   \\
$C^r_{14}$ & $-$3.32$(3.15)$    & $-4.2(1.2)$   & $-$1.0$^*$ & $-7.40^{+1.19}_{-1.79}$  & $C^r_{36}$ & 1.27  $[ 5.12  ]$    & $4.1(4.3]$    & 2.0$^*$    & $\equiv 0$               \\
$C^r_{15}$ & $-1.30$ $(1.13)$   & $1.2(1.0)$    & $-$3.0$^*$ & $\equiv 0$               & $C^r_{63}$ & 11.09  $[ 22.82 ]$   & $-6.6[16.8)$  & --         & $21.08^{-1.79}_{+2.13}$  \\
$C^r_{16}$ & 1.10 $[ 4.20  ]$   & $-0.81(1.34)$ & 3.2        & $\equiv 0$               & $C^r_{66}$ & 3.90  $[ 26.83 ]$    & $-6.5[25.4]$  & --         & $6.80^{+0.34}_{-0.60}$   \\
$C^r_{17}$ & 0.27 $[ 2.62  ]$   & $3.6(1.6]$    & $-$1.0$^*$ & $1.45^{+0.09}_{-0.34}$   & $C^r_{69}$ & $-$1.16 $[ 19.86 ]$  & $4.6[19.0]$   & --         & $4.42^{+0.00}_{-0.09}$   \\
$C^r_{18}$ & $-$1.46$[ 5.45  ]$ & $-1.1[5.4]$   & 0.63       & $-5.10^{+0.60}_{-0.77}$  & $C^r_{83}$ & $-$1.11  $[ 20.31 ]$ & $14(16]$      & --         & $-14.79^{+1.45}_{-1.87}$ \\
$C^r_{19}$ & $-$5.42$(5.22)$    & $5.3(2.8)$    & $-$4.0$^*$ & $-2.30^{+0.77}_{-1.11}$  & $C^r_{88}$ & $-$24.09$[ 63.17 ]$  & $-38[59]$     & --         & $-14.37^{-5.78}_{+7.91}$ \\
$C^r_{20}$ & 0.53 $[ 3.45  ]$   & $-2.9[2.3)$   & 1.0        & $1.45^{-0.17}_{+0.26}$   & $C^r_{90}$ & 20.83 $[ 62.90]$     & $-35[44)$     & --         & $19.72^{-3.74}_{+4.68}$
\end{tabular}
\end{ruledtabular}
\end{table}

\section{Discussions and summary}\label{sum}
This paper proposes a more general Bayesian model (Model B) with the truncation errors. This model is based on the idea of a simple truncation-error model \cite{Yang:2020eif} and the Bayesian model framework \cite{Wesolowski:2015fqa}. Compared to Refs. \cite{Bijnens:2014lea,Yang:2020eif}, there are some advantages in Model B.
\begin{enumerate}
\item This model can transform the understanding of ChPT into the prior knowledge during the fitting process, containing the information of the LECs, the convergence of ChPT and the truncation errors. The prior information can be conveniently introduced by Eqs. \eqref{co1} -- \eqref{co3}. It does not need the other assumptions, such as the geometric-sequence assumption in Ref. \cite{Yang:2020eif}. It can also give a set of more precise NLO fitting LECs. A similar result is obtained at the NNLO fit in Ref. \cite{Bijnens:2014lea}, see Table \ref{table27}. Hence, there are good reasons to believe that the NNLO fitting LECs are also more precise, although there lacks the higher-order fitting result to be compared to.

\item With the help of the MCMC method, the distributions of the LECs can be obtained, and the computational speed is faster. The computational time of Model B is the shortest. The Bayesian method has another inherent advantage. Some clear distribution figures of LECs can be obtained, because Bayesian statistics can give more points in a given time. Therefore, one cannot only obtain the expected values and errors of the LECs, but also their distributions. Refs. \cite{Bijnens:2014lea,Yang:2020eif} cannot give the distributions of LECs, although they can give the errors.

\item Model B gives a general fitting method. It can be used to fit the other problems. The two extremes of this model (Models B$_1$ and B$_2$) have been evaluated by a toy example in Sec. \ref{ME}. It confirms that more prior information indeed gives more precise results. With the quantified evaluation criteria in Sec. \ref{EC}, one can see the improvement of the prior information more clearly. The actual ChPT fit is between the two extremes. It is better than Model A. However, Model A gives a similar result as the minimum $\chi^2$ method in Refs. \cite{Bijnens:2014lea}.

\item For the NNLO LECs $C^r_i$, more smooth PDFs are given, comparing to Ref. \cite{Yang:2020eif} (Ref. \cite{Bijnens:2014lea} does not give PDFs). With these PDFs, one can see how the $C^r_i$ depend on the boundaries more intuitively.

\item There also exist some slight improvements in this paper. The covariance matrix given in Ref. \cite{Colangelo:2001df} is considered. The results are insensitive to the initial conditions, compared to Ref. \cite{Bijnens:2014lea}.
\end{enumerate}

In order to test the effectiveness of the model, one example is randomly generated, in order to imitate the actual ChPT. Some parameters $a_i$ and some quantities $O_i$ are introduced, which imitate the LECs and the experimental data, respectively. The exact values of $a_i$ and $O_i$ are known, and they are treated as the true values. Model A, which does not consider the truncation errors, is also introduced, in order to compare two ideal cases of Model B. One case knows nothing about the truncation errors, except the orders of magnitude. The other one knows the distributions of the truncation errors. The fitting results indicate that the prior information of the truncation errors can improve the fit greatly, even though this information is not so precise. Hence, Model B is adopted to fit the actual ChPT data.

In the actual ChPT fit, it indicates that the Bayesian method without the truncation errors are similar to the classical statistics. In other words, the classical statistics can be treated as a special case of Bayesian statistics. However, Bayesian statistics can be applied more widely. With the help of Model B, some $L_i^r$ and $C_i^r$ (defined in \cite{Bijnens:1999hw}) are fitted at the NLO and the NNLO. The fitting $L_i^r$ are almost unchanged between the NLO and the NNLO fit. The change between 12 and 17 input data are also small, but all the theoretical errors decrease for the 17 inputs, because of the more precise estimation of the truncation errors. Model B also solves a problem in the free fit, which leads to $L^r_4$ and $L^r_6$ being very large, but they are zeros in the large-$N_C$ limit. Because the number of $C_i^r$ to be fitted is larger than the number of the experimental data, some independent $\widetilde{C}_i^r$ are fitted first, which are the linear combinations of the $C_i^r$ to be fitted. From the posterior PDFs of $C_i^r$, the reliable intervals of twelve $C_i^r$ are obtained, and five $C_i^r$ are only constrained with the upper or the lower boundary of the intervals, and the other 21 $C_i^r$ are strongly dependent on both boundaries. It needs more experimental data to confirm these uncertain $C_i^r$. Because all the $\widetilde{C}_i^r$ does not exist overfitting, they are more precise than $C_i^r$. If one knows some more values of these $C_i^r$, some other $C_i^r$ can be restrained by these $\widetilde{C}_i^r$. For the physical quantities to be fitted, most theoretical contributions are well convergent, except $a_0^{1/2}$ and $a_0^{3/2}$. It possibly comes from the large experimental errors, or some of these quantities are indeed not convergent. This needs more precise experimental data and theoretical calculations in the future. It can be seen that Model B can estimate the truncation errors very well.

Some input parameters are very rough, such as Eqs. \eqref{co2} and \eqref{co3}. A more precise estimation beyond the simple convergence assumption will be studied in the future work. In addition, if more analytical and experimental results are introduced, the results should be more precise. However, the NNLO theoretical calculation is complicated. It needs to be studied in the future. In addition, this approach can also be used to fit the other LECs, such as pion-nucleon, meson-baryon chiral LECs. However, both their experimental data and theoretical results are less than the mesonic LECs at present.

In conclusion, truncation errors usually cannot be ignored in the global fit, and some prior information can improve the fit greatly, even though this information is sometimes not very exact. Model B provides a feasible implementation scheme. A new set of more reliable $L_i^r$ and $C_i^r$ are fitted by Model B. This model cannot only fit LECs in ChPT, but also fit other parameters in the other EFTs and the perturbation theory.

\section{Acknowledgments}
Pan thanks Qin-He Yang for providing the original program. This project was supported by the Guangxi Science Foundation under Grants No.2022GXNSFAA035489.

\appendix
\section{One testing example}\label{app:A}
Eq. \eqref{A1} gives the functions of the example Sec. \ref{Sec:II}. For convenience, the parameters with the same name in the different functions are different. The values of $b_i$ and $a_i^\mathrm{LO}$ can be found in Table \ref{pA1}. The values of $a_i^\mathrm{NLO}$ and $a_i^\mathrm{NNLO}$ are given in the second row in Tables \ref{table1} and the second column \ref{table7}, respectively, which are marked by a subscript ``tr''.
\begin{align}
O_1={}&b_1\exp(a^\mathrm{NNLO}_{1}t^3 + a^\mathrm{NLO}_4b_2t^2 + a^\mathrm{NLO}_{7}b_3t^2 - a^\mathrm{LO}_1t) - b_1, \label{A1}\\
O_2={}&b_1\sin(b_2\exp(b_5)) - b_1\sin(b_2\exp(-a^\mathrm{NNLO}_{2}b_3t^3 - a^\mathrm{NLO}_{8}b_4t^2 + b_5\exp(-a^\mathrm{NLO}_{1}b_6t^2) - a^\mathrm{LO}_2t)),\notag\\
O_3={}&b_1\ln(-a^\mathrm{NNLO}_{3}b_3t^3 - a^\mathrm{NLO}_{1}b_2t^2 - a^\mathrm{NLO}_{6}b_4t^2 - a^\mathrm{LO}_3t + 1),\notag\\
O_4={}&b_1\exp(1 - b_2) - b_1\exp(-a^\mathrm{NNLO}_{4}b_7t^3 + a^\mathrm{NNLO}_{4}t^3 - a^\mathrm{NLO}_{1}b_6t^2 + a^\mathrm{NLO}_{3}b_5t^2 - a^\mathrm{NLO}_{4}b_4t^2 - b_2\cos(a^\mathrm{NLO}_{3}b_3t^2) \notag\\
&+ a^\mathrm{LO}_4t + 1),\notag\\
O_5={}&-b_1\ln(a^\mathrm{NNLO}_{5}b_4t^3 - a^\mathrm{NNLO}_{5}b_6t^3 - a^\mathrm{NLO}_{6}b_2b_3t^2 - a^\mathrm{NLO}_{8}b_5t^2 - a^\mathrm{LO}_5t + 1),\notag\\
O_6={}&b_1\ln(b_2\ln(-a^\mathrm{NNLO}_{6}b_4t^3 + a^\mathrm{NLO}_{2}b_3t^2 + a^\mathrm{NLO}_{6}b_5t^2 + a^\mathrm{LO}_6t + 1) + 1),\notag\\
O_7={}&-b_1\exp(b_2) + b_1\exp(a^\mathrm{NLO}_{1}b_6t^2 + b_2\exp(a^\mathrm{NNLO}_{7}t^3 + a^\mathrm{NLO}_{5}b_3t^2 + a^\mathrm{NLO}_{5}t^2 + a^\mathrm{NLO}_{8}t^2) + b_4\sin(a^\mathrm{NLO}_{4}b_5t^2) + a^\mathrm{LO}_7t),\notag\\
O_8={}&b_1\exp(-b_2\exp(-a^\mathrm{NNLO}_{8}b_3t^3 + a^\mathrm{NLO}_{3}b_4t^2 + a^\mathrm{NLO}_{7}t^2 - b_5a^\mathrm{LO}_8t)) - b_1\exp(-b_2), \notag\\
O_9={}&b_1\ln(b_2 + 1) - b_1\ln(a^\mathrm{NNLO}_{9}b_4t^3 + a^\mathrm{NNLO}_{9}b_5t^3 + a^\mathrm{NNLO}_{9}b_8t^3 + b_2\exp(a^\mathrm{NLO}_{5}b_3t^2) - b_6\sin(a^\mathrm{NNLO}_{9}b_7t^3) + a^\mathrm{LO}_9t + 1),\notag\\
O_{10}={}&-b_1\sin(b_2 + b_4\sin(b_5)) + b_1\sin(-a^\mathrm{NNLO}_{10}t^3 - a^\mathrm{NLO}_{4}t^2 + b_2\exp(-a^\mathrm{NLO}_{5}b_3t^2) + b_4\sin(b_5\exp(a^\mathrm{NLO}_{2}b_6t^2)) + a^\mathrm{LO}_{10}t),\notag\\
O_{11}={}&b_1\ln(-b_2\sin(a^\mathrm{NNLO}_{11}b_6t^3 + a^\mathrm{NLO}_{2}b_3t^2 - a^\mathrm{NLO}_{3}b_4t^2 + a^\mathrm{NLO}_{7}b_5t^2 - a^\mathrm{LO}_{11}t) + 1),\notag\\
O_{12}={}&b_1\ln(b_2\sin(-a^\mathrm{NNLO}_{12}b_6t^3 + a^\mathrm{NNLO}_{12}t^3 + a^\mathrm{NLO}_{2}b_3t^2 + a^\mathrm{NLO}_{4}t^2 + a^\mathrm{NLO}_{5}t^2 + a^\mathrm{NLO}_{6}b_4b_5t^2 + a^\mathrm{LO}_{12}t) + 1),\notag\\
O_{13}={}&b_1\exp(-a^\mathrm{NLO}_{8}b_8t^2 - b_2\sin(a^\mathrm{NLO}_{4}b_3t^2) - b_4\exp(a^\mathrm{NNLO}_{13}b_6t^3 - a^\mathrm{NLO}_{5}b_5t^2 + a^\mathrm{NLO}_{6}b_7t^2 + a^\mathrm{LO}_{13}t) + a^\mathrm{LO}_{13}t) - b_1\exp(-b_4),\notag\\
O_{14}={}&-b_1\ln(-a^\mathrm{NNLO}_{14}b_4t^3 - b_2\ln(a^\mathrm{NNLO}_{14}b_3t^3 + 1) - b_5\sin(a^\mathrm{NLO}_{2}b_6t^2) + a^\mathrm{LO}_{14}t + 1), \notag\\
O_{15}={}&-b_1\exp(b_3) + b_1\exp(-a^\mathrm{NNLO}_{15}b_2t^3 - a^\mathrm{NLO}_{7}b_5t^2 - a^\mathrm{NLO}_{7}b_7t^2 - a^\mathrm{NLO}_{8}b_6t^2 + b_3\cos(a^\mathrm{NLO}_{7}b_4t^2) - a^\mathrm{LO}_{15}t),\notag\\
O_{16}={}&-b_1\sin(\ln(b_6 + 1) + 1) - b_1\sin(a^\mathrm{NNLO}_{16}b_5t^3 + a^\mathrm{NLO}_{3}b_2t^2 + a^\mathrm{NLO}_{5}b_4t^2 + a^\mathrm{LO}_{16}t - \ln(b_6\exp(a^\mathrm{NLO}_{1}b_7t^2) + 1) \notag\\
&- \exp(-a^\mathrm{NLO}_{3}b_3t^2)),\notag\\
O_{17}={}&b_1\ln(-b_2\exp(b_3) - b_5\sin(b_6) + 1) - b_1\ln(-b_2\exp(b_3\exp(-a^\mathrm{NLO}_{8}b_4t^2)) - b_5\sin(b_6\exp(-C_{17}b_7t^3)) + a^\mathrm{LO}_{17}t + 1).\notag
\end{align}

\begin{table}[!htbp]
\caption{The values of parameters $b_i$ and $a_i^\mathrm{LO}$ in Eq. \eqref{A1}. Because the values are exact, more significant digits are given.}\label{pA1}
\begin{ruledtabular}
\begin{tabular}{lrrrrrrrrr}
 & $10^2b_{1}$ & $10^2b_{2}$ & $10^2b_{3}$ & $10^2b_{4}$ & $10^2b_{5}$ & $10^2b_{6}$ & $10^2b_{7}$ & $10^2b_{8}$ & $10^2a_i^\mathrm{LO}$           \\ \hline
O$_{1}$  & -50.00000 & -50.00000 & -50.00000  &           &           &           &           &           & -50.00000  \\
O$_{2}$  & 10.00000  & 10.00000  & 10.00000   & 10.00000  & 10.00000  & 10.00000  &           &           & 10.00000   \\
O$_{3}$  & -0.27574  & -81.82470 & -20.69689  & -95.19898 &           &           &           &           & -130.30649 \\
O$_{4}$  & -0.15389  & 55.22236  & -22.53571  & 10.47685  & 14.87959  & 10.84243  & 85.28317  &           & 164.01310  \\
O$_{5}$  & 32.80394  & 52.16045  & 52.16045   & 42.66749  & 12.43968  & 77.33250  &           &           & 52.35691   \\
O$_{6}$  & -10.25934 & -10.95043 & -11.71888  & -28.67356 & -26.84030 &           &           &           & -32.24206  \\
O$_{7}$  & -2.39804  & 2.37745   & -9.62388   & 9.39379   & 9.39265   & 0.57071   &           &           & 42.90526   \\
O$_{8}$  & -24.83947 & 69.61634  & -2.52600   & -10.10231 & 7.19485   &           &           &           & 99.63693   \\
O$_{9}$  & 0.51431   & 99.49478  & 30.47334   & 32.08646  & 32.08646  & 113.26784 & 96.25052  & 32.08646  & 77.78096   \\
O$_{10}$ & -69.04271 & -69.23904 & 38.65428   & 20.36290  & 10.78947  & 10.06250  &           &           & 82.44726   \\
O$_{11}$ & -62.98961 & -66.28358 & 42.37512   & -8.89931  & 13.20056  & 6.67356   &           &           & 91.91503   \\
O$_{12}$ & 50.00000  & 10.00000  & -135.00000 & 10.00000  & 10.00000  &           &           &           & -100.00000 \\
O$_{13}$ & 74.39589  & 103.26600 & 103.32687  & 156.81033 & 95.54550  & 83.85802  & 100.61745 & 105.09613 & 107.97725  \\
O$_{14}$ & -33.59097 & -54.18416 & -54.20951  & -41.78655 & -28.19762 & -28.53647 &           &           & -48.27921  \\
O$_{15}$ & 28.77264  & 26.35989  & 40.37381   & 49.81383  & 40.01344  & 63.80493  & 40.01344  &           & 45.46315   \\
O$_{16}$ & -9.19703  & 3.17944   & 1.51912    & 5.66408   & -5.66251  & -8.10996  & 10.10760  &           & 58.11734   \\
O$_{17}$ & -29.07789 & -62.10099 & -44.02866  & -48.43771 & -31.06314 & -34.52566 & -40.99873 &           & -12.31774
\end{tabular}
\end{ruledtabular}
\end{table}

\section{Some tables and figures for the fits}\label{app:B}
~
\begin{table}[!htbp]
\caption{The fitting parameters and the priors in Model B$_2$. The subscripts NLO and NNLO represent the NLO and NNLO fit, respectively. The definitions of these parameters are in Eqs. \eqref{eq3} -- \eqref{dp} and the text below them.}\label{app table1}
\begin{ruledtabular}
\begin{tabular}{lcccrccccrc}
$i$&$\mu_{e,\mathrm{NLO}}$&$\sigma_{e,\mathrm{NLO}}$&$p_\mathrm{NLO}$ &$\mu_{a_i^\mathrm{NLO}}$&$\sigma_{a_i^\mathrm{NLO}}$
&$\mu_{e,\mathrm{NNLO}}$&$\sigma_{e,\mathrm{NNLO}}$&$p_\mathrm{NNLO}$ &$\mu_{a_i^\mathrm{NNLO}}$&$\sigma_{a_i^\mathrm{NNLO}}$
   \\\hline
1  & 0.140 & 0.050 & 1 & 0.616  & 0.308 & 0.040 & 0.020 & 1 & 0.023    & 0.012 \\
2  & 0.100 & 0.050 & 1 & 0.751  & 0.376 & 0.040 & 0.020 & 0 & 0.178    & 0.089 \\
3  & 0.020 & 0.050 & 0 &$-$3.232& 1.616 & 0.100 & 0.030 & 1 & $-$0.758 & 0.379 \\
4  & 0.040 & 0.050 & 1 & 0.268  & 0.134 & 0.120 & 0.036 & 1 & 0.196    & 0.098 \\
5  & 0.140 & 0.050 & 1 & 1.097  & 0.549 & 0.060 & 0.020 & 1 & $-$0.146 & 0.073 \\
6  & 0.110 & 0.050 & 0 & 0.108  & 0.054 & 0.020 & 0.020 & 0 & 0.200    & 0.100 \\
7  & 0.110 & 0.050 & 1 &$-$0.281& 0.140 & 0.060 & 0.020 & 1 & $-$0.347 & 0.173 \\
8  & 0.140 & 0.050 & 1 & 0.434  & 0.217 & 0.010 & 0.020 & 0 & $-$0.484 & 0.242 \\
9  & 0.070 & 0.050 & 0 &        &       & 0.130 & 0.039 & 0 & $-$0.958 & 0.479 \\
10 & 0.120 & 0.050 & 0 &        &       & 0.050 & 0.020 & 0 & $-$0.061 & 0.031 \\
11 & 0.110 & 0.050 & 0 &        &       & 0.060 & 0.020 & 0 & 0.275    & 0.138 \\
12 & 0.010 & 0.050 & 1 &        &       & 0.040 & 0.020 & 1 & $-$0.217 & 0.109 \\
13 & 0.140 & 0.050 & 1 &        &       & 0.060 & 0.020 & 1 & 0.987    & 0.494 \\
14 & 0.140 & 0.050 & 1 &        &       & 0.070 & 0.021 & 1 & 0.279    & 0.139 \\
15 & 0.020 & 0.050 & 0 &        &       & 0.030 & 0.020 & 1 & $-$0.098 & 0.049 \\
16 & 0.060 & 0.050 & 0 &        &       & 0.050 & 0.020 & 1 & $-$0.622 & 0.311 \\
17 & 0.140 & 0.050 & 1 &        &       & 0.050 & 0.020 & 1 & 0.187    & 0.093
\end{tabular}
\end{ruledtabular}
\end{table}

\begin{table}[!htbp]
	\caption{The prior of the LECs in Model B$_2$. The subscripts NLO and NNLO represent the NLO and NNLO fit, respectively.}\label{app table2}
	\begin{ruledtabular}
		\begin{tabular}{lrrrr}
			$i$&$\mu_{a_i^\mathrm{NLO}}$&$\sigma_{a_i^\mathrm{NLO}}$
			&$\mu_{a_i^\mathrm{NNLO}}$&$\sigma_{a_i^\mathrm{NNLO}}$		\\\hline
			1  & 0.616  & 0.308 & 0.023    & 0.012  \\
			2  & 0.751  & 0.376 & 0.178    & 0.089  \\
			3  &$-$3.232& 1.616 & $-$0.758 & 0.379  \\
			4  & 0.268  & 0.134 & 0.196    & 0.098  \\
			5  & 1.097  & 0.549 & $-$0.146 & 0.073  \\
			6  & 0.108  & 0.054 & 0.200    & 0.100  \\
			7  &$-$0.281& 0.140 & $-$0.347 & 0.173  \\
			8  & 0.434  & 0.217 & $-$0.484 & 0.242  \\
			9  &        &       & $-$0.958 & 0.479  \\
			10 &        &       & $-$0.061 & 0.031  \\
			11 &        &       & 0.275    & 0.138  \\
			12 &        &       & $-$0.217 & 0.109  \\
			13 &        &       & 0.987    & 0.494  \\
			14 &        &       & 0.279    & 0.139  \\
			15 &        &       & $-$0.098 & 0.049  \\
			16 &        &       & $-$0.622 & 0.311  \\
			17 &        &       & 0.187    & 0.093
		\end{tabular}
	\end{ruledtabular}
\end{table}

\begin{table}[!htbp]
\caption{The parameters for fitting $L_i^r$ and $\widetilde{C}_i$. The superscripts 12 and 17 represent the fit with 12 and 17 inputs, respectively. The subscripts NLO and NNLO represent the NLO and NNLO fitting, respectively. The definitions of these parameters are in Eqs. \eqref{eq3} -- \eqref{dp} and the text below them.} \label{app table3}
\begin{ruledtabular}
\begin{tabular}{lccccccccc}
Quantity&$\mu_{e,\mathrm{NLO}}^{12}$&$\sigma_{e,\mathrm{NLO}}^{12}$&$p_\mathrm{NLO}^{12}$
        &$\mu_{e,\mathrm{NLO}}^{17}$&$\sigma_{e,\mathrm{NLO}}^{17}$&$p_\mathrm{NLO}^{17}$
        &$\mu_{e,\mathrm{NNLO}}^{17}$&$\sigma_{e,\mathrm{NNLO}}^{17}$&$p_\mathrm{NNLO}^{17}$\\
\hline
$m_s/\hat m|_1$            & 0.050 & 0.050 & 0.5 & 0.050 & 0.050 & 0.5 & 0.020 & 0.020 & 0.5 \\
$m_s/\hat m|_2$            & 0.050 & 0.050 & 0.5 & 0.050 & 0.050 & 0.5 & 0.020 & 0.020 & 0.5 \\
$F_K/F_\pi$                & 0.050 & 0.050 & 0.5 & 0.050 & 0.050 & 0.5 & 0.020 & 0.020 & 0.5 \\
$f_s$                      & 0.150 & 0.050 & 1   & 0.150 & 0.050 & 1   & 0.020 & 0.020 & 0.5 \\
$g_p$                      & 0.050 & 0.050 & 0.5 & 0.050 & 0.050 & 0.5 & 0.020 & 0.020 & 0.5 \\
$f_s^{\prime}$             & 0.050 & 0.050 & 0.5 & 0.050 & 0.050 & 0.5 & 0.020 & 0.020 & 0.5 \\
$g^{\prime}$               & 0.050 & 0.050 & 0.5 & 0.050 & 0.050 & 0.5 & 0.020 & 0.020 & 0.5 \\
$a_0^0$                    & 0.100 & 0.050 & 1   & 0.100 & 0.050 & 1   & 0.020 & 0.020 & 0.5 \\
$10a_0^2$                  & 0.050 & 0.050 & 0.5 & 0.050 & 0.050 & 0.5 & 0.020 & 0.020 & 0.5 \\
$a_0^{1/2}m_\pi$           & 0.350 & 0.105 & 1   & 0.350 & 0.105 & 1   & 0.020 & 0.020 & 0.5 \\
$10a_0^{3/2}m_\pi$         & 0.250 & 0.075 & 0   & 0.250 & 0.075 & 0   & 0.020 & 0.020 & 0.5 \\
$\langle r^2\rangle_S^\pi$ & 0.200 & 0.060 & 0.5 & 0.200 & 0.060 & 0.5 & 0.020 & 0.020 & 0.5 \\
$c_S^\pi$                  &       &       &     & 0.200 & 0.060 & 0.5 & 0.020 & 0.020 & 0.5 \\
$\bar l_1$                 &       &       &     & 0.200 & 0.060 & 0.5 & 0.050 & 0.050 & 0.5 \\
$\bar l_2$                 &       &       &     & 0.200 & 0.060 & 0.5 & 0.050 & 0.050 & 0.5 \\
$\bar l_3$                 &       &       &     & 0.200 & 0.060 & 0.5 & 0.050 & 0.050 & 0.5 \\
$\bar l_4$                 &       &       &     & 0.200 & 0.060 & 0.5 & 0.050 & 0.050 & 0.5
\end{tabular}
\end{ruledtabular}
\end{table}

\begin{table}[!htbp]
	\caption{The parameters of the priors of $L_i^r$ and $\widetilde{C}_i$. Their definition is above Eq. \eqref{eq:prlecs}. The superscripts 12 and 17 represent the fit with 12 and 17 inputs, respectively. The subscripts NLO and NNLO represent the NLO and the NNLO fits, respectively.} \label{app table4}
	\begin{ruledtabular}
		\begin{tabular}{lrrrrrr}
			i&$\mu_{L_i^r,\mathrm{NLO}}^{12}$&$\sigma_{L_i^r,\mathrm{NLO}}^{12}$
			&$\mu_{L_i^r,\mathrm{NLO}}^{17}$&$\sigma_{L_i^r,\mathrm{NLO}}^{17}$
			&$\mu_{\widetilde{C}_i,\mathrm{NNLO}}^{17}$&$\sigma_{\widetilde{C}_i,\mathrm{NNLO}}^{17}$\\
			\hline
			1  & 0.500  & 0.300 & 0.500  & 0.300 & 0.077  & 0.989 \\
			2  & 1.000  & 0.500 & 1.000  & 0.500 & 0.190  & 3.102 \\
			3  & $-$3.000 & 1.000 & $-$3.000 & 1.000 & $-$1.073 & 0.007 \\
			4  & 0.200  & 0.200 & 0.200  & 0.200 & 0.068  & 0.095 \\
			5  & 1.000  & 0.500 & 1.000  & 0.500 & $-$0.040 & 0.071 \\
			6  & 0.000  & 0.300 & 0.000  & 0.300 & 0.806  & 1.772 \\
			7  & $-$0.300 & 0.300 & $-$0.300 & 0.300 & $-$0.561 & 1.581 \\
			8  & 0.500  & 0.500 & 0.500  & 0.500 & $-$0.043 & 0.250 \\
			9  &        &       &        &       & $-$0.173 & 0.497 \\
			10 &        &       &        &       & $-$0.405 & 1.819 \\
			11 &        &       &        &       & 0.066  & 0.227 \\
			12 &        &       &        &       & 0.013  & 0.053 \\
			13 &        &       &        &       & $-$0.244 & 0.388 \\
			14 &        &       &        &       & 0.007  & 0.693 \\
			15 &        &       &        &       & $-$0.022 & 0.072 \\
			16 &        &       &        &       & 0.214  & 1.122 \\
			17 &        &       &        &       & $-$0.207 & 0.222
		\end{tabular}
	\end{ruledtabular}
\end{table}

\begin{figure}[htbp]
	\centering
	\includegraphics[width=1\linewidth]{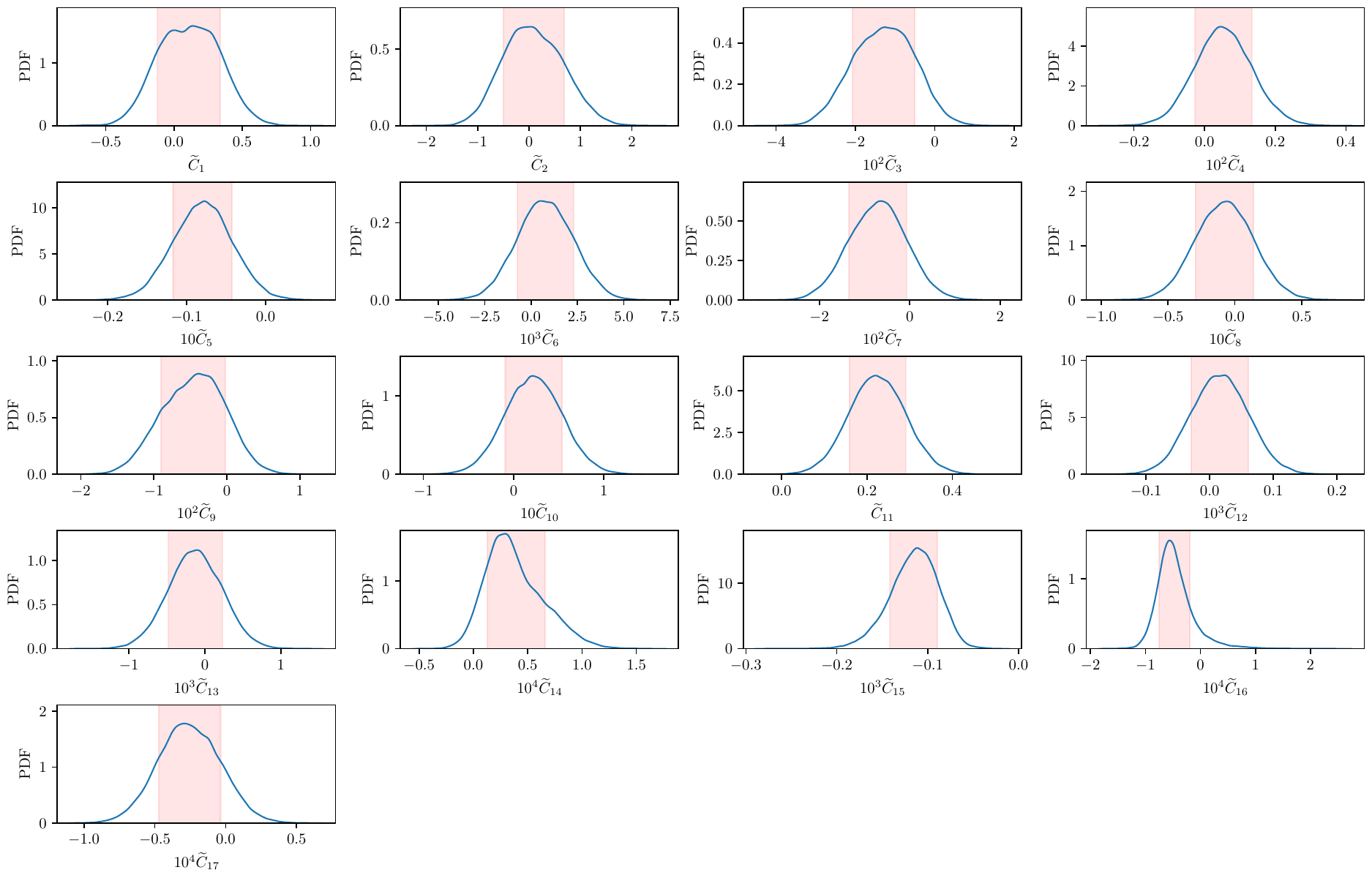}
	\caption{The posterior distributions of the NNLO fitting $\tilde{C}_i$. The vertical coordinate is the posterior PDF and the horizontal coordinate is the value of $\tilde{C}_i$. The pink shaded area depicts the 68\% HPD. The blue line is the distribution curve of $L^r_i$.}
	\label{fig:Ctilted}
\end{figure}
\begin{figure}[htbp]
	\centering
	\includegraphics[width=1\linewidth]{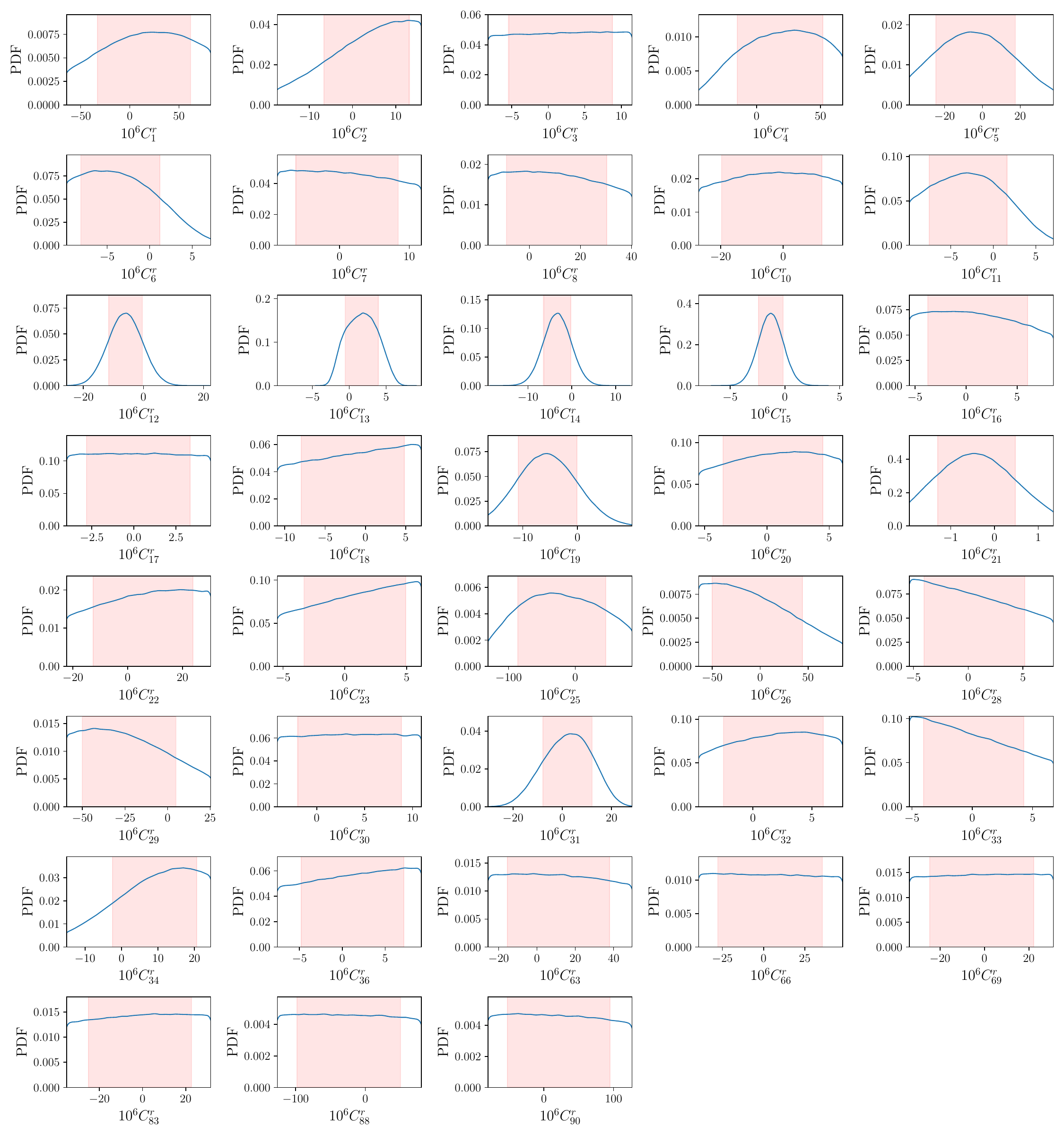}
	\caption{The posterior distributions of $C_i^r$. The horizontal axis represents the value of $C_i^r$, the upper and the lower boundaries are given in Eq. (38) in Ref. \cite{Yang:2020eif}. The vertical coordinate is the posterior PDF. The pink shaded area depicts the 68\% HPD. The blue line is the distribution curve of $C^r_i$.}
	\label{fig:c2}
\end{figure}

\clearpage
\bibliography{ref}

\begin{thebibliography}{80}%
\makeatletter
\providecommand \@ifxundefined [1]{%
 \@ifx{#1\undefined}
}%
\providecommand \@ifnum [1]{%
 \ifnum #1\expandafter \@firstoftwo
 \else \expandafter \@secondoftwo
 \fi
}%
\providecommand \@ifx [1]{%
 \ifx #1\expandafter \@firstoftwo
 \else \expandafter \@secondoftwo
 \fi
}%
\providecommand \natexlab [1]{#1}%
\providecommand \enquote  [1]{``#1''}%
\providecommand \bibnamefont  [1]{#1}%
\providecommand \bibfnamefont [1]{#1}%
\providecommand \citenamefont [1]{#1}%
\providecommand \href@noop [0]{\@secondoftwo}%
\providecommand \href [0]{\begingroup \@sanitize@url \@href}%
\providecommand \@href[1]{\@@startlink{#1}\@@href}%
\providecommand \@@href[1]{\endgroup#1\@@endlink}%
\providecommand \@sanitize@url [0]{\catcode `\\12\catcode `\$12\catcode
  `\&12\catcode `\#12\catcode `\^12\catcode `\_12\catcode `\%12\relax}%
\providecommand \@@startlink[1]{}%
\providecommand \@@endlink[0]{}%
\providecommand \url  [0]{\begingroup\@sanitize@url \@url }%
\providecommand \@url [1]{\endgroup\@href {#1}{\urlprefix }}%
\providecommand \urlprefix  [0]{URL }%
\providecommand \Eprint [0]{\href }%
\providecommand \doibase [0]{https://doi.org/}%
\providecommand \selectlanguage [0]{\@gobble}%
\providecommand \bibinfo  [0]{\@secondoftwo}%
\providecommand \bibfield  [0]{\@secondoftwo}%
\providecommand \translation [1]{[#1]}%
\providecommand \BibitemOpen [0]{}%
\providecommand \bibitemStop [0]{}%
\providecommand \bibitemNoStop [0]{.\EOS\space}%
\providecommand \EOS [0]{\spacefactor3000\relax}%
\providecommand \BibitemShut  [1]{\csname bibitem#1\endcsname}%
\let\auto@bib@innerbib\@empty
%</preamble>
\bibitem [{\citenamefont {Weinberg}(1979)}]{Weinberg:1978kz}%
  \BibitemOpen
  \bibfield  {author} {\bibinfo {author} {\bibfnamefont {S.}~\bibnamefont
  {Weinberg}},\ }\bibfield  {title} {\bibinfo {title} {{Phenomenological
  Lagrangians}},\ }\href {https://doi.org/10.1016/0378-4371(79)90223-1}
  {\bibfield  {journal} {\bibinfo  {journal} {Physica A}\ }\textbf {\bibinfo
  {volume} {96}},\ \bibinfo {pages} {327} (\bibinfo {year} {1979})}\BibitemShut
  {NoStop}%
\bibitem [{\citenamefont {Gasser}\ and\ \citenamefont
  {Leutwyler}(1984)}]{Gasser:1983yg}%
  \BibitemOpen
  \bibfield  {author} {\bibinfo {author} {\bibfnamefont {J.}~\bibnamefont
  {Gasser}}\ and\ \bibinfo {author} {\bibfnamefont {H.}~\bibnamefont
  {Leutwyler}},\ }\bibfield  {title} {\bibinfo {title} {{Chiral perturbation
  theory to one loop}},\ }\href {https://doi.org/10.1016/0003-4916(84)90242-2}
  {\bibfield  {journal} {\bibinfo  {journal} {Annals Phys. (N.Y)}\ }\textbf
  {\bibinfo {volume} {158}},\ \bibinfo {pages} {142} (\bibinfo {year}
  {1984})}\BibitemShut {NoStop}%
%%CITATION = APNYA,158,142;%%
\bibitem [{\citenamefont {Gasser}\ and\ \citenamefont
  {Leutwyler}(1985)}]{Gasser:1984gg}%
  \BibitemOpen
  \bibfield  {author} {\bibinfo {author} {\bibfnamefont {J.}~\bibnamefont
  {Gasser}}\ and\ \bibinfo {author} {\bibfnamefont {H.}~\bibnamefont
  {Leutwyler}},\ }\bibfield  {title} {\bibinfo {title} {{Chiral perturbation
  theory: Expansions in the mass of the strange quark}},\ }\href
  {https://doi.org/10.1016/0550-3213(85)90492-4} {\bibfield  {journal}
  {\bibinfo  {journal} {Nucl. Phys.}\ }\textbf {\bibinfo {volume} {B250}},\
  \bibinfo {pages} {465} (\bibinfo {year} {1985})}\BibitemShut {NoStop}%
%%CITATION = NUPHA,B250,465;%%
\bibitem [{\citenamefont {Bijnens}\ \emph {et~al.}(1999)\citenamefont
  {Bijnens}, \citenamefont {Colangelo},\ and\ \citenamefont
  {Ecker}}]{Bijnens:1999sh}%
  \BibitemOpen
  \bibfield  {author} {\bibinfo {author} {\bibfnamefont {J.}~\bibnamefont
  {Bijnens}}, \bibinfo {author} {\bibfnamefont {G.}~\bibnamefont {Colangelo}},\
  and\ \bibinfo {author} {\bibfnamefont {G.}~\bibnamefont {Ecker}},\ }\bibfield
   {title} {\bibinfo {title} {{The mesonic chiral Lagrangian of order $p^6$}},\
  }\href {https://doi.org/10.1088/1126-6708/1999/02/020} {\bibfield  {journal}
  {\bibinfo  {journal} {J.High Energy Phys.}\ }\textbf {\bibinfo {volume}
  {02}},\ \bibinfo {pages} {020} (\bibinfo {year} {1999})},\ \Eprint
  {https://arxiv.org/abs/hep-ph/9902437} {arXiv:hep-ph/9902437 [hep-ph]}
  \BibitemShut {NoStop}%
%%CITATION = HEP-PH/9902437;%%
\bibitem [{\citenamefont {Bijnens}\ \emph {et~al.}(2019)\citenamefont
  {Bijnens}, \citenamefont {Hermansson-Truedsson},\ and\ \citenamefont
  {Wang}}]{Bijnens:2018lez}%
  \BibitemOpen
  \bibfield  {author} {\bibinfo {author} {\bibfnamefont {J.}~\bibnamefont
  {Bijnens}}, \bibinfo {author} {\bibfnamefont {N.}~\bibnamefont
  {Hermansson-Truedsson}},\ and\ \bibinfo {author} {\bibfnamefont
  {S.}~\bibnamefont {Wang}},\ }\bibfield  {title} {\bibinfo {title} {{The order
  $p^8$ mesonic chiral Lagrangian}},\ }\href
  {https://doi.org/10.1007/JHEP01(2019)102} {\bibfield  {journal} {\bibinfo
  {journal} {J.High Energy Phys.}\ }\textbf {\bibinfo {volume} {01}},\ \bibinfo
  {pages} {102} (\bibinfo {year} {2019})},\ \Eprint
  {https://arxiv.org/abs/1810.06834} {arXiv:1810.06834 [hep-ph]} \BibitemShut
  {NoStop}%
%%CITATION = ARXIV:1810.06834;%%
\bibitem [{\citenamefont {Bijnens}\ and\ \citenamefont
  {Jemos}(2012)}]{Bijnens:2011tb}%
  \BibitemOpen
  \bibfield  {author} {\bibinfo {author} {\bibfnamefont {J.}~\bibnamefont
  {Bijnens}}\ and\ \bibinfo {author} {\bibfnamefont {I.}~\bibnamefont
  {Jemos}},\ }\bibfield  {title} {\bibinfo {title} {{A new global fit of the
  $L^r_i$ at next-to-next-to-leading order in chiral perturbation theory}},\
  }\href {https://doi.org/10.1016/j.nuclphysb.2011.09.013} {\bibfield
  {journal} {\bibinfo  {journal} {Nucl. Phys.}\ }\textbf {\bibinfo {volume}
  {B854}},\ \bibinfo {pages} {631} (\bibinfo {year} {2012})},\ \Eprint
  {https://arxiv.org/abs/1103.5945} {arXiv:1103.5945 [hep-ph]} \BibitemShut
  {NoStop}%
%%CITATION = ARXIV:1103.5945;%%
\bibitem [{\citenamefont {Bijnens}\ and\ \citenamefont
  {Ecker}(2014)}]{Bijnens:2014lea}%
  \BibitemOpen
  \bibfield  {author} {\bibinfo {author} {\bibfnamefont {J.}~\bibnamefont
  {Bijnens}}\ and\ \bibinfo {author} {\bibfnamefont {G.}~\bibnamefont
  {Ecker}},\ }\bibfield  {title} {\bibinfo {title} {{Mesonic low-energy
  constants}},\ }\href {https://doi.org/10.1146/annurev-nucl-102313-025528}
  {\bibfield  {journal} {\bibinfo  {journal} {Ann. Rev. Nucl. Part. Sci.}\
  }\textbf {\bibinfo {volume} {64}},\ \bibinfo {pages} {149} (\bibinfo {year}
  {2014})},\ \Eprint {https://arxiv.org/abs/1405.6488} {arXiv:1405.6488
  [hep-ph]} \BibitemShut {NoStop}%
%%CITATION = ARXIV:1405.6488;%%
\bibitem [{\citenamefont {Yang}\ \emph {et~al.}(2020)\citenamefont {Yang},
  \citenamefont {Guo}, \citenamefont {Ge}, \citenamefont {Huang}, \citenamefont
  {Liu},\ and\ \citenamefont {Jiang}}]{Yang:2020eif}%
  \BibitemOpen
  \bibfield  {author} {\bibinfo {author} {\bibfnamefont {Q.-H.}\ \bibnamefont
  {Yang}}, \bibinfo {author} {\bibfnamefont {W.}~\bibnamefont {Guo}}, \bibinfo
  {author} {\bibfnamefont {F.-J.}\ \bibnamefont {Ge}}, \bibinfo {author}
  {\bibfnamefont {B.}~\bibnamefont {Huang}}, \bibinfo {author} {\bibfnamefont
  {H.}~\bibnamefont {Liu}},\ and\ \bibinfo {author} {\bibfnamefont {S.-Z.}\
  \bibnamefont {Jiang}},\ }\bibfield  {title} {\bibinfo {title} {{New method
  for fitting the low-energy constants in chiral perturbation theory}},\ }\href
  {https://doi.org/10.1103/PhysRevD.102.094009} {\bibfield  {journal} {\bibinfo
   {journal} {Phys. Rev. D}\ }\textbf {\bibinfo {volume} {102}},\ \bibinfo
  {pages} {094009} (\bibinfo {year} {2020})},\ \Eprint
  {https://arxiv.org/abs/2004.06085} {arXiv:2004.06085 [hep-ph]} \BibitemShut
  {NoStop}%
\bibitem [{\citenamefont {Can}\ \emph {et~al.}(2015)\citenamefont {Can},
  \citenamefont {Erkol}, \citenamefont {Oka},\ and\ \citenamefont
  {Takahashi}}]{Can:2015exa}%
  \BibitemOpen
  \bibfield  {author} {\bibinfo {author} {\bibfnamefont {K.~U.}\ \bibnamefont
  {Can}}, \bibinfo {author} {\bibfnamefont {G.}~\bibnamefont {Erkol}}, \bibinfo
  {author} {\bibfnamefont {M.}~\bibnamefont {Oka}},\ and\ \bibinfo {author}
  {\bibfnamefont {T.~T.}\ \bibnamefont {Takahashi}},\ }\bibfield  {title}
  {\bibinfo {title} {{Look inside charmed-strange baryons from lattice QCD}},\
  }\href {https://doi.org/10.1103/PhysRevD.92.114515} {\bibfield  {journal}
  {\bibinfo  {journal} {Phys. Rev. D}\ }\textbf {\bibinfo {volume} {92}},\
  \bibinfo {pages} {114515} (\bibinfo {year} {2015})},\ \Eprint
  {https://arxiv.org/abs/1508.03048} {arXiv:1508.03048 [hep-lat]} \BibitemShut
  {NoStop}%
\bibitem [{\citenamefont {Can}\ \emph {et~al.}(2014)\citenamefont {Can},
  \citenamefont {Erkol}, \citenamefont {Isildak}, \citenamefont {Oka},\ and\
  \citenamefont {Takahashi}}]{Can:2013tna}%
  \BibitemOpen
  \bibfield  {author} {\bibinfo {author} {\bibfnamefont {K.~U.}\ \bibnamefont
  {Can}}, \bibinfo {author} {\bibfnamefont {G.}~\bibnamefont {Erkol}}, \bibinfo
  {author} {\bibfnamefont {B.}~\bibnamefont {Isildak}}, \bibinfo {author}
  {\bibfnamefont {M.}~\bibnamefont {Oka}},\ and\ \bibinfo {author}
  {\bibfnamefont {T.~T.}\ \bibnamefont {Takahashi}},\ }\bibfield  {title}
  {\bibinfo {title} {{Electromagnetic structure of charmed baryons in Lattice
  QCD}},\ }\href {https://doi.org/10.1007/JHEP05(2014)125} {\bibfield
  {journal} {\bibinfo  {journal} {J.High Energy Phys.}\ }\textbf {\bibinfo
  {volume} {05}},\ \bibinfo {pages} {125} (\bibinfo {year} {2014})},\ \Eprint
  {https://arxiv.org/abs/1310.5915} {arXiv:1310.5915 [hep-lat]} \BibitemShut
  {NoStop}%
\bibitem [{\citenamefont {Bahtiyar}\ \emph {et~al.}(2017)\citenamefont
  {Bahtiyar}, \citenamefont {Can}, \citenamefont {Erkol}, \citenamefont {Oka},\
  and\ \citenamefont {Takahashi}}]{Bahtiyar:2016dom}%
  \BibitemOpen
  \bibfield  {author} {\bibinfo {author} {\bibfnamefont {H.}~\bibnamefont
  {Bahtiyar}}, \bibinfo {author} {\bibfnamefont {K.~U.}\ \bibnamefont {Can}},
  \bibinfo {author} {\bibfnamefont {G.}~\bibnamefont {Erkol}}, \bibinfo
  {author} {\bibfnamefont {M.}~\bibnamefont {Oka}},\ and\ \bibinfo {author}
  {\bibfnamefont {T.~T.}\ \bibnamefont {Takahashi}},\ }\bibfield  {title}
  {\bibinfo {title} {{$\Xi_c \gamma \rightarrow\Xi^\prime_c$ transition in
  lattice QCD}},\ }\href {https://doi.org/10.1016/j.physletb.2017.06.022}
  {\bibfield  {journal} {\bibinfo  {journal} {Phys. Lett. B}\ }\textbf
  {\bibinfo {volume} {772}},\ \bibinfo {pages} {121} (\bibinfo {year}
  {2017})},\ \Eprint {https://arxiv.org/abs/1612.05722} {arXiv:1612.05722
  [hep-lat]} \BibitemShut {NoStop}%
\bibitem [{\citenamefont {Yan}\ \emph {et~al.}(1992)\citenamefont {Yan},
  \citenamefont {Cheng}, \citenamefont {Cheung}, \citenamefont {Lin},
  \citenamefont {Lin},\ and\ \citenamefont {Yu}}]{Yan:1992gz}%
  \BibitemOpen
  \bibfield  {author} {\bibinfo {author} {\bibfnamefont {T.-M.}\ \bibnamefont
  {Yan}}, \bibinfo {author} {\bibfnamefont {H.-Y.}\ \bibnamefont {Cheng}},
  \bibinfo {author} {\bibfnamefont {C.-Y.}\ \bibnamefont {Cheung}}, \bibinfo
  {author} {\bibfnamefont {G.-L.}\ \bibnamefont {Lin}}, \bibinfo {author}
  {\bibfnamefont {Y.~C.}\ \bibnamefont {Lin}},\ and\ \bibinfo {author}
  {\bibfnamefont {H.-L.}\ \bibnamefont {Yu}},\ }\bibfield  {title} {\bibinfo
  {title} {{Heavy quark symmetry and chiral dynamics}},\ }\href
  {https://doi.org/10.1103/PhysRevD.46.1148} {\bibfield  {journal} {\bibinfo
  {journal} {Phys. Rev. D}\ }\textbf {\bibinfo {volume} {46}},\ \bibinfo
  {pages} {1148} (\bibinfo {year} {1992})},\ \bibinfo {note} {[Erratum:
  Phys.Rev.D 55, 5851 (1997)]}\BibitemShut {NoStop}%
\bibitem [{\citenamefont {Dowdall}\ \emph {et~al.}(2013)\citenamefont
  {Dowdall}, \citenamefont {Davies}, \citenamefont {Lepage},\ and\
  \citenamefont {McNeile}}]{Dowdall:2013rya}%
  \BibitemOpen
  \bibfield  {author} {\bibinfo {author} {\bibfnamefont {R.~J.}\ \bibnamefont
  {Dowdall}}, \bibinfo {author} {\bibfnamefont {C.~T.~H.}\ \bibnamefont
  {Davies}}, \bibinfo {author} {\bibfnamefont {G.~P.}\ \bibnamefont {Lepage}},\
  and\ \bibinfo {author} {\bibfnamefont {C.}~\bibnamefont {McNeile}},\
  }\bibfield  {title} {\bibinfo {title} {{$V_{us}$ from $\pi$ and $K$ decay
  constants in full lattice QCD with physical $u$, $d$, $s$ and $c$ quarks}},\
  }\href {https://doi.org/10.1103/PhysRevD.88.074504} {\bibfield  {journal}
  {\bibinfo  {journal} {Phys. Rev. D}\ }\textbf {\bibinfo {volume} {88}},\
  \bibinfo {pages} {074504} (\bibinfo {year} {2013})},\ \Eprint
  {https://arxiv.org/abs/1303.1670} {arXiv:1303.1670 [hep-lat]} \BibitemShut
  {NoStop}%
%%CITATION = ARXIV:1303.1670;%%
\bibitem [{\citenamefont {Bazavov}\ \emph
  {et~al.}(2010{\natexlab{a}})\citenamefont {Bazavov} \emph
  {et~al.}}]{MILC:2010hzw}%
  \BibitemOpen
  \bibfield  {author} {\bibinfo {author} {\bibfnamefont {A.}~\bibnamefont
  {Bazavov}} \emph {et~al.} (\bibinfo {collaboration} {MILC}),\ }\bibfield
  {title} {\bibinfo {title} {{Results for light pseudoscalar mesons}},\ }\href
  {https://doi.org/10.22323/1.105.0074} {\bibfield  {journal} {\bibinfo
  {journal} {PoS}\ }\textbf {\bibinfo {volume} {LATTICE2010}},\ \bibinfo
  {pages} {074} (\bibinfo {year} {2010}{\natexlab{a}})},\ \Eprint
  {https://arxiv.org/abs/1012.0868} {arXiv:1012.0868 [hep-lat]} \BibitemShut
  {NoStop}%
\bibitem [{\citenamefont {Bernard}\ and\ \citenamefont
  {Passemar}(2010)}]{Bernard:2009ds}%
  \BibitemOpen
  \bibfield  {author} {\bibinfo {author} {\bibfnamefont {V.}~\bibnamefont
  {Bernard}}\ and\ \bibinfo {author} {\bibfnamefont {E.}~\bibnamefont
  {Passemar}},\ }\bibfield  {title} {\bibinfo {title} {{Chiral extrapolation of
  the strangeness changing $K \pi$ form factor}},\ }\href
  {https://doi.org/10.1007/JHEP04(2010)001} {\bibfield  {journal} {\bibinfo
  {journal} {J.High Energy Phys.}\ }\textbf {\bibinfo {volume} {04}},\ \bibinfo
  {pages} {001} (\bibinfo {year} {2010})},\ \Eprint
  {https://arxiv.org/abs/0912.3792} {arXiv:0912.3792 [hep-ph]} \BibitemShut
  {NoStop}%
%%CITATION = ARXIV:0912.3792;%%
\bibitem [{\citenamefont {Bazavov}\ \emph {et~al.}(2009)\citenamefont {Bazavov}
  \emph {et~al.}}]{Bazavov:2009fk}%
  \BibitemOpen
  \bibfield  {author} {\bibinfo {author} {\bibfnamefont {A.}~\bibnamefont
  {Bazavov}} \emph {et~al.} (\bibinfo {collaboration} {MILC}),\ }\bibfield
  {title} {\bibinfo {title} {{MILC results for light pseudoscalars}},\
  }\bibfield  {booktitle} {\emph {\bibinfo {booktitle} {{Proceedings, 6th
  International Workshop on Chiral dymamics: Bern, Switzerland, July 6-10,
  2009}}},\ }\href {https://doi.org/10.22323/1.086.0007} {\bibfield  {journal}
  {\bibinfo  {journal} {PoS}\ }\textbf {\bibinfo {volume} {CD09}},\ \bibinfo
  {pages} {007} (\bibinfo {year} {2009})},\ \Eprint
  {https://arxiv.org/abs/0910.2966} {arXiv:0910.2966 [hep-ph]} \BibitemShut
  {NoStop}%
%%CITATION = ARXIV:0910.2966;%%
\bibitem [{\citenamefont {Bazavov}\ \emph
  {et~al.}(2010{\natexlab{b}})\citenamefont {Bazavov} \emph
  {et~al.}}]{Bazavov:2009bb}%
  \BibitemOpen
  \bibfield  {author} {\bibinfo {author} {\bibfnamefont {A.}~\bibnamefont
  {Bazavov}} \emph {et~al.} (\bibinfo {collaboration} {MILC}),\ }\bibfield
  {title} {\bibinfo {title} {{Nonperturbative QCD simulations with 2+1 flavors
  of improved staggered quarks}},\ }\href
  {https://doi.org/10.1103/RevModPhys.82.1349} {\bibfield  {journal} {\bibinfo
  {journal} {Rev. Mod. Phys.}\ }\textbf {\bibinfo {volume} {82}},\ \bibinfo
  {pages} {1349} (\bibinfo {year} {2010}{\natexlab{b}})},\ \Eprint
  {https://arxiv.org/abs/0903.3598} {arXiv:0903.3598 [hep-lat]} \BibitemShut
  {NoStop}%
%%CITATION = ARXIV:0903.3598;%%
\bibitem [{\citenamefont {Golterman}\ \emph {et~al.}(2014)\citenamefont
  {Golterman}, \citenamefont {Maltman},\ and\ \citenamefont
  {Peris}}]{Golterman:2014nua}%
  \BibitemOpen
  \bibfield  {author} {\bibinfo {author} {\bibfnamefont {M.}~\bibnamefont
  {Golterman}}, \bibinfo {author} {\bibfnamefont {K.}~\bibnamefont {Maltman}},\
  and\ \bibinfo {author} {\bibfnamefont {S.}~\bibnamefont {Peris}},\ }\bibfield
   {title} {\bibinfo {title} {{NNLO low-energy constants from flavor-breaking
  chiral sum rules based on hadronic $\tau$-decay data}},\ }\href
  {https://doi.org/10.1103/PhysRevD.89.054036} {\bibfield  {journal} {\bibinfo
  {journal} {Phys. Rev. D}\ }\textbf {\bibinfo {volume} {89}},\ \bibinfo
  {pages} {054036} (\bibinfo {year} {2014})},\ \Eprint
  {https://arxiv.org/abs/1402.1043} {arXiv:1402.1043 [hep-ph]} \BibitemShut
  {NoStop}%
%%CITATION = ARXIV:1402.1043;%%
\bibitem [{\citenamefont {Colangelo}\ \emph {et~al.}(2012)\citenamefont
  {Colangelo}, \citenamefont {Sanz-Cillero},\ and\ \citenamefont
  {Zuo}}]{Colangelo:2012ipa}%
  \BibitemOpen
  \bibfield  {author} {\bibinfo {author} {\bibfnamefont {P.}~\bibnamefont
  {Colangelo}}, \bibinfo {author} {\bibfnamefont {J.~J.}\ \bibnamefont
  {Sanz-Cillero}},\ and\ \bibinfo {author} {\bibfnamefont {F.}~\bibnamefont
  {Zuo}},\ }\bibfield  {title} {\bibinfo {title} {{Holography, chiral
  Lagrangian and form factor relations}},\ }\href
  {https://doi.org/10.1007/JHEP11(2012)012} {\bibfield  {journal} {\bibinfo
  {journal} {J.High Energy Phys.}\ }\textbf {\bibinfo {volume} {11}},\ \bibinfo
  {pages} {012} (\bibinfo {year} {2012})},\ \Eprint
  {https://arxiv.org/abs/1207.5744} {arXiv:1207.5744 [hep-ph]} \BibitemShut
  {NoStop}%
%%CITATION = ARXIV:1207.5744;%%
\bibitem [{\citenamefont {Guo}\ \emph {et~al.}(2007)\citenamefont {Guo},
  \citenamefont {Sanz~Cillero},\ and\ \citenamefont {Zheng}}]{Guo:2007ff}%
  \BibitemOpen
  \bibfield  {author} {\bibinfo {author} {\bibfnamefont {Z.-H.}\ \bibnamefont
  {Guo}}, \bibinfo {author} {\bibfnamefont {J.~J.}\ \bibnamefont
  {Sanz~Cillero}},\ and\ \bibinfo {author} {\bibfnamefont {H.-Q.}\ \bibnamefont
  {Zheng}},\ }\bibfield  {title} {\bibinfo {title} {{Partial waves and large
  $N_C$ resonance sum rules}},\ }\href
  {https://doi.org/10.1088/1126-6708/2007/06/030} {\bibfield  {journal}
  {\bibinfo  {journal} {Journal of High Energy Physics}\ }\textbf {\bibinfo
  {volume} {06}},\ \bibinfo {pages} {030} (\bibinfo {year} {2007})},\ \Eprint
  {https://arxiv.org/abs/hep-ph/0701232} {arXiv:hep-ph/0701232} \BibitemShut
  {NoStop}%
\bibitem [{\citenamefont {Guo}\ \emph {et~al.}(2008)\citenamefont {Guo},
  \citenamefont {Sanz-Cillero},\ and\ \citenamefont {Zheng}}]{Guo:2007hm}%
  \BibitemOpen
  \bibfield  {author} {\bibinfo {author} {\bibfnamefont {Z.~H.}\ \bibnamefont
  {Guo}}, \bibinfo {author} {\bibfnamefont {J.~J.}\ \bibnamefont
  {Sanz-Cillero}},\ and\ \bibinfo {author} {\bibfnamefont {H.~Q.}\ \bibnamefont
  {Zheng}},\ }\bibfield  {title} {\bibinfo {title} {{$O(p^6)$ extension of the
  large-$N_C$ partial wave dispersion relations}},\ }\href
  {https://doi.org/10.1016/j.physletb.2008.01.073} {\bibfield  {journal}
  {\bibinfo  {journal} {Phys. Lett.}\ }\textbf {\bibinfo {volume} {B661}},\
  \bibinfo {pages} {342} (\bibinfo {year} {2008})},\ \Eprint
  {https://arxiv.org/abs/0710.2163} {arXiv:0710.2163 [hep-ph]} \BibitemShut
  {NoStop}%
%%CITATION = ARXIV:0710.2163;%%
\bibitem [{\citenamefont {Guo}\ and\ \citenamefont
  {Sanz-Cillero}(2009)}]{Guo:2009hi}%
  \BibitemOpen
  \bibfield  {author} {\bibinfo {author} {\bibfnamefont {Z.-H.}\ \bibnamefont
  {Guo}}\ and\ \bibinfo {author} {\bibfnamefont {J.~J.}\ \bibnamefont
  {Sanz-Cillero}},\ }\bibfield  {title} {\bibinfo {title} {{$\pi\pi$-scattering
  lengths at $O(p^6)$ revisited}},\ }\href
  {https://doi.org/10.1103/PhysRevD.79.096006} {\bibfield  {journal} {\bibinfo
  {journal} {Phys. Rev. D}\ }\textbf {\bibinfo {volume} {79}},\ \bibinfo
  {pages} {096006} (\bibinfo {year} {2009})},\ \Eprint
  {https://arxiv.org/abs/0903.0782} {arXiv:0903.0782 [hep-ph]} \BibitemShut
  {NoStop}%
%%CITATION = ARXIV:0903.0782;%%
\bibitem [{\citenamefont {Bijnens}\ \emph {et~al.}(1994)\citenamefont
  {Bijnens}, \citenamefont {Colangelo},\ and\ \citenamefont
  {Gasser}}]{Bijnens:1994ie}%
  \BibitemOpen
  \bibfield  {author} {\bibinfo {author} {\bibfnamefont {J.}~\bibnamefont
  {Bijnens}}, \bibinfo {author} {\bibfnamefont {G.}~\bibnamefont {Colangelo}},\
  and\ \bibinfo {author} {\bibfnamefont {J.}~\bibnamefont {Gasser}},\
  }\bibfield  {title} {\bibinfo {title} {{$K_{l4}$ decays beyond one loop}},\
  }\href {https://doi.org/10.1016/0550-3213(94)90634-3} {\bibfield  {journal}
  {\bibinfo  {journal} {Nucl. Phys.}\ }\textbf {\bibinfo {volume} {B427}},\
  \bibinfo {pages} {427} (\bibinfo {year} {1994})},\ \Eprint
  {https://arxiv.org/abs/hep-ph/9403390} {arXiv:hep-ph/9403390 [hep-ph]}
  \BibitemShut {NoStop}%
%%CITATION = HEP-PH/9403390;%%
\bibitem [{\citenamefont {Amor\'{o}s}\ \emph
  {et~al.}(2000{\natexlab{a}})\citenamefont {Amor\'{o}s}, \citenamefont
  {Bijnens},\ and\ \citenamefont {Talavera}}]{Amoros:2000mc}%
  \BibitemOpen
  \bibfield  {author} {\bibinfo {author} {\bibfnamefont {G.}~\bibnamefont
  {Amor\'{o}s}}, \bibinfo {author} {\bibfnamefont {J.}~\bibnamefont
  {Bijnens}},\ and\ \bibinfo {author} {\bibfnamefont {P.}~\bibnamefont
  {Talavera}},\ }\bibfield  {title} {\bibinfo {title} {{$K_{\ell 4}$
  form-factors and $\pi-\pi$ scattering}},\ }\href
  {https://doi.org/10.1016/S0550-3213(00)00366-7,
  10.1016/S0550-3213(01)00025-6} {\bibfield  {journal} {\bibinfo  {journal}
  {Nucl. Phys.}\ }\textbf {\bibinfo {volume} {B585}},\ \bibinfo {pages} {293}
  (\bibinfo {year} {2000}{\natexlab{a}})},\ \bibinfo {note} {[Erratum: Nucl.
  Phys.B598,665(2001)]},\ \Eprint {https://arxiv.org/abs/hep-ph/0003258}
  {arXiv:hep-ph/0003258 [hep-ph]} \BibitemShut {NoStop}%
%%CITATION = HEP-PH/0003258;%%
\bibitem [{\citenamefont {Colangelo}\ \emph {et~al.}(2001)\citenamefont
  {Colangelo}, \citenamefont {Gasser},\ and\ \citenamefont
  {Leutwyler}}]{Colangelo:2001df}%
  \BibitemOpen
  \bibfield  {author} {\bibinfo {author} {\bibfnamefont {G.}~\bibnamefont
  {Colangelo}}, \bibinfo {author} {\bibfnamefont {J.}~\bibnamefont {Gasser}},\
  and\ \bibinfo {author} {\bibfnamefont {H.}~\bibnamefont {Leutwyler}},\
  }\bibfield  {title} {\bibinfo {title} {{$\pi \pi$ scattering}},\ }\href
  {https://doi.org/10.1016/S0550-3213(01)00147-X} {\bibfield  {journal}
  {\bibinfo  {journal} {Nucl. Phys.}\ }\textbf {\bibinfo {volume} {B603}},\
  \bibinfo {pages} {125} (\bibinfo {year} {2001})},\ \Eprint
  {https://arxiv.org/abs/hep-ph/0103088} {arXiv:hep-ph/0103088 [hep-ph]}
  \BibitemShut {NoStop}%
%%CITATION = HEP-PH/0103088;%%
\bibitem [{\citenamefont {Schindler}\ and\ \citenamefont
  {Phillips}(2009)}]{Schindler:2008fh}%
  \BibitemOpen
  \bibfield  {author} {\bibinfo {author} {\bibfnamefont {M.~R.}\ \bibnamefont
  {Schindler}}\ and\ \bibinfo {author} {\bibfnamefont {D.~R.}\ \bibnamefont
  {Phillips}},\ }\bibfield  {title} {\bibinfo {title} {{Bayesian Methods for
  Parameter Estimation in Effective Field Theories}},\ }\href
  {https://doi.org/10.1016/j.aop.2008.09.003} {\bibfield  {journal} {\bibinfo
  {journal} {Annals Phys.}\ }\textbf {\bibinfo {volume} {324}},\ \bibinfo
  {pages} {682} (\bibinfo {year} {2009})},\ \bibinfo {note} {[Erratum: Annals
  Phys. 324, 2051--2055 (2009)]},\ \Eprint {https://arxiv.org/abs/0808.3643}
  {arXiv:0808.3643 [hep-ph]} \BibitemShut {NoStop}%
\bibitem [{\citenamefont {Furnstahl}\ \emph {et~al.}(2015)\citenamefont
  {Furnstahl}, \citenamefont {Phillips},\ and\ \citenamefont
  {Wesolowski}}]{Furnstahl:2014xsa}%
  \BibitemOpen
  \bibfield  {author} {\bibinfo {author} {\bibfnamefont {R.~J.}\ \bibnamefont
  {Furnstahl}}, \bibinfo {author} {\bibfnamefont {D.~R.}\ \bibnamefont
  {Phillips}},\ and\ \bibinfo {author} {\bibfnamefont {S.}~\bibnamefont
  {Wesolowski}},\ }\bibfield  {title} {\bibinfo {title} {{A recipe for EFT
  uncertainty quantification in nuclear physics}},\ }\href
  {https://doi.org/10.1088/0954-3899/42/3/034028} {\bibfield  {journal}
  {\bibinfo  {journal} {J. Phys. G}\ }\textbf {\bibinfo {volume} {42}},\
  \bibinfo {pages} {034028} (\bibinfo {year} {2015})},\ \Eprint
  {https://arxiv.org/abs/1407.0657} {arXiv:1407.0657 [nucl-th]} \BibitemShut
  {NoStop}%
\bibitem [{\citenamefont {Wesolowski}\ \emph {et~al.}(2016)\citenamefont
  {Wesolowski}, \citenamefont {Klco}, \citenamefont {Furnstahl}, \citenamefont
  {Phillips},\ and\ \citenamefont {Thapaliya}}]{Wesolowski:2015fqa}%
  \BibitemOpen
  \bibfield  {author} {\bibinfo {author} {\bibfnamefont {S.}~\bibnamefont
  {Wesolowski}}, \bibinfo {author} {\bibfnamefont {N.}~\bibnamefont {Klco}},
  \bibinfo {author} {\bibfnamefont {R.~J.}\ \bibnamefont {Furnstahl}}, \bibinfo
  {author} {\bibfnamefont {D.~R.}\ \bibnamefont {Phillips}},\ and\ \bibinfo
  {author} {\bibfnamefont {A.}~\bibnamefont {Thapaliya}},\ }\bibfield  {title}
  {\bibinfo {title} {{Bayesian parameter estimation for effective field
  theories}},\ }\href {https://doi.org/10.1088/0954-3899/43/7/074001}
  {\bibfield  {journal} {\bibinfo  {journal} {J. Phys. G}\ }\textbf {\bibinfo
  {volume} {43}},\ \bibinfo {pages} {074001} (\bibinfo {year} {2016})},\
  \Eprint {https://arxiv.org/abs/1511.03618} {arXiv:1511.03618 [nucl-th]}
  \BibitemShut {NoStop}%
\bibitem [{\citenamefont {Melendez}\ \emph {et~al.}(2017)\citenamefont
  {Melendez}, \citenamefont {Wesolowski},\ and\ \citenamefont
  {Furnstahl}}]{Melendez:2017phj}%
  \BibitemOpen
  \bibfield  {author} {\bibinfo {author} {\bibfnamefont {J.~A.}\ \bibnamefont
  {Melendez}}, \bibinfo {author} {\bibfnamefont {S.}~\bibnamefont
  {Wesolowski}},\ and\ \bibinfo {author} {\bibfnamefont {R.~J.}\ \bibnamefont
  {Furnstahl}},\ }\bibfield  {title} {\bibinfo {title} {{Bayesian truncation
  errors in chiral effective field theory: Nucleon-nucleon observables}},\
  }\href {https://doi.org/10.1103/PhysRevC.96.024003} {\bibfield  {journal}
  {\bibinfo  {journal} {Phys. Rev. C}\ }\textbf {\bibinfo {volume} {96}},\
  \bibinfo {pages} {024003} (\bibinfo {year} {2017})},\ \Eprint
  {https://arxiv.org/abs/1704.03308} {arXiv:1704.03308 [nucl-th]} \BibitemShut
  {NoStop}%
%%CITATION = ARXIV:1704.03308;%%
\bibitem [{\citenamefont {Svensson}\ \emph {et~al.}(2022)\citenamefont
  {Svensson}, \citenamefont {Ekstr\"om},\ and\ \citenamefont
  {Forss\'en}}]{Svensson:2021lzs}%
  \BibitemOpen
  \bibfield  {author} {\bibinfo {author} {\bibfnamefont {I.}~\bibnamefont
  {Svensson}}, \bibinfo {author} {\bibfnamefont {A.}~\bibnamefont
  {Ekstr\"om}},\ and\ \bibinfo {author} {\bibfnamefont {C.}~\bibnamefont
  {Forss\'en}},\ }\bibfield  {title} {\bibinfo {title} {{Bayesian parameter
  estimation in chiral effective field theory using the Hamiltonian Monte Carlo
  method}},\ }\href {https://doi.org/10.1103/PhysRevC.105.014004} {\bibfield
  {journal} {\bibinfo  {journal} {Phys. Rev. C}\ }\textbf {\bibinfo {volume}
  {105}},\ \bibinfo {pages} {014004} (\bibinfo {year} {2022})},\ \Eprint
  {https://arxiv.org/abs/2110.04011} {arXiv:2110.04011 [nucl-th]} \BibitemShut
  {NoStop}%
\bibitem [{\citenamefont {Ekstr\"om}\ \emph {et~al.}(2019)\citenamefont
  {Ekstr\"om}, \citenamefont {Forss\'en}, \citenamefont {Dimitrakakis},
  \citenamefont {Dubhashi}, \citenamefont {Johansson}, \citenamefont
  {Muhammad}, \citenamefont {Salomonsson},\ and\ \citenamefont
  {Schliep}}]{Ekstrom:2019twv}%
  \BibitemOpen
  \bibfield  {author} {\bibinfo {author} {\bibfnamefont {A.}~\bibnamefont
  {Ekstr\"om}}, \bibinfo {author} {\bibfnamefont {C.}~\bibnamefont
  {Forss\'en}}, \bibinfo {author} {\bibfnamefont {C.}~\bibnamefont
  {Dimitrakakis}}, \bibinfo {author} {\bibfnamefont {D.}~\bibnamefont
  {Dubhashi}}, \bibinfo {author} {\bibfnamefont {H.~T.}\ \bibnamefont
  {Johansson}}, \bibinfo {author} {\bibfnamefont {A.~S.}\ \bibnamefont
  {Muhammad}}, \bibinfo {author} {\bibfnamefont {H.}~\bibnamefont
  {Salomonsson}},\ and\ \bibinfo {author} {\bibfnamefont {A.}~\bibnamefont
  {Schliep}},\ }\bibfield  {title} {\bibinfo {title} {{Bayesian optimization in
  ab initio nuclear physics}},\ }\href
  {https://doi.org/10.1088/1361-6471/ab2b14} {\bibfield  {journal} {\bibinfo
  {journal} {J. Phys. G}\ }\textbf {\bibinfo {volume} {46}},\ \bibinfo {pages}
  {095101} (\bibinfo {year} {2019})},\ \Eprint
  {https://arxiv.org/abs/1902.00941} {arXiv:1902.00941 [nucl-th]} \BibitemShut
  {NoStop}%
\bibitem [{\citenamefont {Wesolowski}\ \emph {et~al.}(2019)\citenamefont
  {Wesolowski}, \citenamefont {Furnstahl}, \citenamefont {Melendez},\ and\
  \citenamefont {Phillips}}]{Wesolowski:2018lzj}%
  \BibitemOpen
  \bibfield  {author} {\bibinfo {author} {\bibfnamefont {S.}~\bibnamefont
  {Wesolowski}}, \bibinfo {author} {\bibfnamefont {R.~J.}\ \bibnamefont
  {Furnstahl}}, \bibinfo {author} {\bibfnamefont {J.~A.}\ \bibnamefont
  {Melendez}},\ and\ \bibinfo {author} {\bibfnamefont {D.~R.}\ \bibnamefont
  {Phillips}},\ }\bibfield  {title} {\bibinfo {title} {{Exploring Bayesian
  parameter estimation for chiral effective field theory using
  nucleon\textendash{}nucleon phase shifts}},\ }\href
  {https://doi.org/10.1088/1361-6471/aaf5fc} {\bibfield  {journal} {\bibinfo
  {journal} {J. Phys. G}\ }\textbf {\bibinfo {volume} {46}},\ \bibinfo {pages}
  {045102} (\bibinfo {year} {2019})},\ \Eprint
  {https://arxiv.org/abs/1808.08211} {arXiv:1808.08211 [nucl-th]} \BibitemShut
  {NoStop}%
\bibitem [{\citenamefont {Alnamlah}\ \emph {et~al.}(2021)\citenamefont
  {Alnamlah}, \citenamefont {P\'erez},\ and\ \citenamefont
  {Phillips}}]{Alnamlah:2020cko}%
  \BibitemOpen
  \bibfield  {author} {\bibinfo {author} {\bibfnamefont {I.~K.}\ \bibnamefont
  {Alnamlah}}, \bibinfo {author} {\bibfnamefont {E.~A.~C.}\ \bibnamefont
  {P\'erez}},\ and\ \bibinfo {author} {\bibfnamefont {D.~R.}\ \bibnamefont
  {Phillips}},\ }\bibfield  {title} {\bibinfo {title} {{Effective field theory
  approach to rotational bands in odd-mass nuclei}},\ }\href
  {https://doi.org/10.1103/PhysRevC.104.064311} {\bibfield  {journal} {\bibinfo
   {journal} {Phys. Rev. C}\ }\textbf {\bibinfo {volume} {104}},\ \bibinfo
  {pages} {064311} (\bibinfo {year} {2021})},\ \Eprint
  {https://arxiv.org/abs/2011.01083} {arXiv:2011.01083 [nucl-th]} \BibitemShut
  {NoStop}%
\bibitem [{\citenamefont {Yang}\ \emph {et~al.}(2021)\citenamefont {Yang},
  \citenamefont {Ekstr\"om}, \citenamefont {Forss\'en},\ and\ \citenamefont
  {Hagen}}]{Yang:2020pgi}%
  \BibitemOpen
  \bibfield  {author} {\bibinfo {author} {\bibfnamefont {C.~J.}\ \bibnamefont
  {Yang}}, \bibinfo {author} {\bibfnamefont {A.}~\bibnamefont {Ekstr\"om}},
  \bibinfo {author} {\bibfnamefont {C.}~\bibnamefont {Forss\'en}},\ and\
  \bibinfo {author} {\bibfnamefont {G.}~\bibnamefont {Hagen}},\ }\bibfield
  {title} {\bibinfo {title} {{Power counting in chiral effective field theory
  and nuclear binding}},\ }\href {https://doi.org/10.1103/PhysRevC.103.054304}
  {\bibfield  {journal} {\bibinfo  {journal} {Phys. Rev. C}\ }\textbf {\bibinfo
  {volume} {103}},\ \bibinfo {pages} {054304} (\bibinfo {year} {2021})},\
  \Eprint {https://arxiv.org/abs/2011.11584} {arXiv:2011.11584 [nucl-th]}
  \BibitemShut {NoStop}%
\bibitem [{\citenamefont {Lovell}\ \emph {et~al.}(2020)\citenamefont {Lovell},
  \citenamefont {Nunes}, \citenamefont {Catacora-Rios},\ and\ \citenamefont
  {King}}]{Lovell:2020sep}%
  \BibitemOpen
  \bibfield  {author} {\bibinfo {author} {\bibfnamefont {A.~E.}\ \bibnamefont
  {Lovell}}, \bibinfo {author} {\bibfnamefont {F.~M.}\ \bibnamefont {Nunes}},
  \bibinfo {author} {\bibfnamefont {M.}~\bibnamefont {Catacora-Rios}},\ and\
  \bibinfo {author} {\bibfnamefont {G.~B.}\ \bibnamefont {King}},\ }\bibfield
  {title} {\bibinfo {title} {{Recent advances in the quantification of
  uncertainties in reaction theory}},\ }\href
  {https://doi.org/10.1088/1361-6471/abba72} {\bibfield  {journal} {\bibinfo
  {journal} {J. Phys. G}\ }\textbf {\bibinfo {volume} {48}},\ \bibinfo {pages}
  {014001} (\bibinfo {year} {2020})},\ \Eprint
  {https://arxiv.org/abs/2012.09012} {arXiv:2012.09012 [nucl-th]} \BibitemShut
  {NoStop}%
\bibitem [{\citenamefont {Phillips}\ \emph {et~al.}(2021)\citenamefont
  {Phillips} \emph {et~al.}}]{Phillips:2020dmw}%
  \BibitemOpen
  \bibfield  {author} {\bibinfo {author} {\bibfnamefont {D.~R.}\ \bibnamefont
  {Phillips}} \emph {et~al.},\ }\bibfield  {title} {\bibinfo {title} {{Get on
  the BAND Wagon: A Bayesian Framework for Quantifying Model Uncertainties in
  Nuclear Dynamics}},\ }\href {https://doi.org/10.1088/1361-6471/abf1df}
  {\bibfield  {journal} {\bibinfo  {journal} {J. Phys. G}\ }\textbf {\bibinfo
  {volume} {48}},\ \bibinfo {pages} {072001} (\bibinfo {year} {2021})},\
  \Eprint {https://arxiv.org/abs/2012.07704} {arXiv:2012.07704 [nucl-th]}
  \BibitemShut {NoStop}%
\bibitem [{\citenamefont {Bedaque}\ \emph {et~al.}(2021)\citenamefont {Bedaque}
  \emph {et~al.}}]{Bedaque:2021bja}%
  \BibitemOpen
  \bibfield  {author} {\bibinfo {author} {\bibfnamefont {P.}~\bibnamefont
  {Bedaque}} \emph {et~al.},\ }\bibfield  {title} {\bibinfo {title} {{A.I. for
  nuclear physics}},\ }\href {https://doi.org/10.1140/epja/s10050-020-00290-x}
  {\bibfield  {journal} {\bibinfo  {journal} {Eur. Phys. J. A}\ }\textbf
  {\bibinfo {volume} {57}},\ \bibinfo {pages} {100} (\bibinfo {year}
  {2021})}\BibitemShut {NoStop}%
\bibitem [{\citenamefont {Wesolowski}\ \emph {et~al.}(2021)\citenamefont
  {Wesolowski}, \citenamefont {Svensson}, \citenamefont {Ekstr\"om},
  \citenamefont {Forss\'en}, \citenamefont {Furnstahl}, \citenamefont
  {Melendez},\ and\ \citenamefont {Phillips}}]{Wesolowski:2021cni}%
  \BibitemOpen
  \bibfield  {author} {\bibinfo {author} {\bibfnamefont {S.}~\bibnamefont
  {Wesolowski}}, \bibinfo {author} {\bibfnamefont {I.}~\bibnamefont
  {Svensson}}, \bibinfo {author} {\bibfnamefont {A.}~\bibnamefont {Ekstr\"om}},
  \bibinfo {author} {\bibfnamefont {C.}~\bibnamefont {Forss\'en}}, \bibinfo
  {author} {\bibfnamefont {R.~J.}\ \bibnamefont {Furnstahl}}, \bibinfo {author}
  {\bibfnamefont {J.~A.}\ \bibnamefont {Melendez}},\ and\ \bibinfo {author}
  {\bibfnamefont {D.~R.}\ \bibnamefont {Phillips}},\ }\bibfield  {title}
  {\bibinfo {title} {{Rigorous constraints on three-nucleon forces in chiral
  effective field theory from fast and accurate calculations of few-body
  observables}},\ }\href {https://doi.org/10.1103/PhysRevC.104.064001}
  {\bibfield  {journal} {\bibinfo  {journal} {Phys. Rev. C}\ }\textbf {\bibinfo
  {volume} {104}},\ \bibinfo {pages} {064001} (\bibinfo {year} {2021})},\
  \Eprint {https://arxiv.org/abs/2104.04441} {arXiv:2104.04441 [nucl-th]}
  \BibitemShut {NoStop}%
\bibitem [{\citenamefont {Connell}\ \emph {et~al.}(2021)\citenamefont
  {Connell}, \citenamefont {Billig},\ and\ \citenamefont
  {Phillips}}]{Connell:2021qcd}%
  \BibitemOpen
  \bibfield  {author} {\bibinfo {author} {\bibfnamefont {M.~A.}\ \bibnamefont
  {Connell}}, \bibinfo {author} {\bibfnamefont {I.}~\bibnamefont {Billig}},\
  and\ \bibinfo {author} {\bibfnamefont {D.~R.}\ \bibnamefont {Phillips}},\
  }\bibfield  {title} {\bibinfo {title} {{Does Bayesian model averaging improve
  polynomial extrapolations? Two toy problems as tests}},\ }\href
  {https://doi.org/10.1088/1361-6471/ac215a} {\bibfield  {journal} {\bibinfo
  {journal} {J. Phys. G}\ }\textbf {\bibinfo {volume} {48}},\ \bibinfo {pages}
  {104001} (\bibinfo {year} {2021})},\ \Eprint
  {https://arxiv.org/abs/2106.05906} {arXiv:2106.05906 [stat.ME]} \BibitemShut
  {NoStop}%
\bibitem [{\citenamefont {Lin}\ \emph {et~al.}(2021)\citenamefont {Lin},
  \citenamefont {Hammer},\ and\ \citenamefont {Mei\ss{}ner}}]{Lin:2021umz}%
  \BibitemOpen
  \bibfield  {author} {\bibinfo {author} {\bibfnamefont {Y.-H.}\ \bibnamefont
  {Lin}}, \bibinfo {author} {\bibfnamefont {H.-W.}\ \bibnamefont {Hammer}},\
  and\ \bibinfo {author} {\bibfnamefont {U.-G.}\ \bibnamefont {Mei\ss{}ner}},\
  }\bibfield  {title} {\bibinfo {title} {{Dispersion-theoretical analysis of
  the electromagnetic form factors of the nucleon: Past, present and future}},\
  }\href {https://doi.org/10.1140/epja/s10050-021-00562-0} {\bibfield
  {journal} {\bibinfo  {journal} {Eur. Phys. J. A}\ }\textbf {\bibinfo {volume}
  {57}},\ \bibinfo {pages} {255} (\bibinfo {year} {2021})},\ \Eprint
  {https://arxiv.org/abs/2106.06357} {arXiv:2106.06357 [hep-ph]} \BibitemShut
  {NoStop}%
\bibitem [{\citenamefont {Dj\"arv}\ \emph {et~al.}(2022)\citenamefont
  {Dj\"arv}, \citenamefont {Ekstr\"om}, \citenamefont {Forss\'en},\ and\
  \citenamefont {Johansson}}]{Djarv:2021hcj}%
  \BibitemOpen
  \bibfield  {author} {\bibinfo {author} {\bibfnamefont {T.}~\bibnamefont
  {Dj\"arv}}, \bibinfo {author} {\bibfnamefont {A.}~\bibnamefont {Ekstr\"om}},
  \bibinfo {author} {\bibfnamefont {C.}~\bibnamefont {Forss\'en}},\ and\
  \bibinfo {author} {\bibfnamefont {H.~T.}\ \bibnamefont {Johansson}},\
  }\bibfield  {title} {\bibinfo {title} {{Bayesian predictions for A=6 nuclei
  using eigenvector continuation emulators}},\ }\href
  {https://doi.org/10.1103/PhysRevC.105.014005} {\bibfield  {journal} {\bibinfo
   {journal} {Phys. Rev. C}\ }\textbf {\bibinfo {volume} {105}},\ \bibinfo
  {pages} {014005} (\bibinfo {year} {2022})},\ \Eprint
  {https://arxiv.org/abs/2108.13313} {arXiv:2108.13313 [nucl-th]} \BibitemShut
  {NoStop}%
\bibitem [{\citenamefont {Acharya}\ and\ \citenamefont
  {Bacca}(2022)}]{Acharya:2021lrv}%
  \BibitemOpen
  \bibfield  {author} {\bibinfo {author} {\bibfnamefont {B.}~\bibnamefont
  {Acharya}}\ and\ \bibinfo {author} {\bibfnamefont {S.}~\bibnamefont
  {Bacca}},\ }\bibfield  {title} {\bibinfo {title} {{Gaussian process error
  modeling for chiral effective-field-theory calculations of
  np\ensuremath{\leftrightarrow}d\ensuremath{\gamma} at low energies}},\ }\href
  {https://doi.org/10.1016/j.physletb.2022.137011} {\bibfield  {journal}
  {\bibinfo  {journal} {Phys. Lett. B}\ }\textbf {\bibinfo {volume} {827}},\
  \bibinfo {pages} {137011} (\bibinfo {year} {2022})},\ \Eprint
  {https://arxiv.org/abs/2109.13972} {arXiv:2109.13972 [nucl-th]} \BibitemShut
  {NoStop}%
\bibitem [{\citenamefont {Odell}\ \emph {et~al.}(2022)\citenamefont {Odell},
  \citenamefont {Brune}, \citenamefont {Phillips}, \citenamefont {deBoer},\
  and\ \citenamefont {Paneru}}]{Odell:2021nmp}%
  \BibitemOpen
  \bibfield  {author} {\bibinfo {author} {\bibfnamefont {D.}~\bibnamefont
  {Odell}}, \bibinfo {author} {\bibfnamefont {C.~R.}\ \bibnamefont {Brune}},
  \bibinfo {author} {\bibfnamefont {D.~R.}\ \bibnamefont {Phillips}}, \bibinfo
  {author} {\bibfnamefont {R.~J.}\ \bibnamefont {deBoer}},\ and\ \bibinfo
  {author} {\bibfnamefont {S.~N.}\ \bibnamefont {Paneru}},\ }\bibfield  {title}
  {\bibinfo {title} {{Performing Bayesian Analyses With AZURE2 Using BRICK: An
  Application to the 7Be System}},\ }\href
  {https://doi.org/10.3389/fphy.2022.888476} {\bibfield  {journal} {\bibinfo
  {journal} {Front. in Phys.}\ }\textbf {\bibinfo {volume} {10}},\ \bibinfo
  {pages} {888476} (\bibinfo {year} {2022})},\ \Eprint
  {https://arxiv.org/abs/2112.12838} {arXiv:2112.12838 [nucl-th]} \BibitemShut
  {NoStop}%
\bibitem [{\citenamefont {Lovell}\ \emph {et~al.}(2022)\citenamefont {Lovell},
  \citenamefont {Mohan}, \citenamefont {Sprouse},\ and\ \citenamefont
  {Mumpower}}]{Lovell:2022pkw}%
  \BibitemOpen
  \bibfield  {author} {\bibinfo {author} {\bibfnamefont {A.~E.}\ \bibnamefont
  {Lovell}}, \bibinfo {author} {\bibfnamefont {A.~T.}\ \bibnamefont {Mohan}},
  \bibinfo {author} {\bibfnamefont {T.~M.}\ \bibnamefont {Sprouse}},\ and\
  \bibinfo {author} {\bibfnamefont {M.~R.}\ \bibnamefont {Mumpower}},\
  }\bibfield  {title} {\bibinfo {title} {{Nuclear masses learned from a
  probabilistic neural network}},\ }\href
  {https://doi.org/10.1103/PhysRevC.106.014305} {\bibfield  {journal} {\bibinfo
   {journal} {Phys. Rev. C}\ }\textbf {\bibinfo {volume} {106}},\ \bibinfo
  {pages} {014305} (\bibinfo {year} {2022})},\ \Eprint
  {https://arxiv.org/abs/2201.00676} {arXiv:2201.00676 [nucl-th]} \BibitemShut
  {NoStop}%
\bibitem [{\citenamefont {Hagen}\ \emph {et~al.}(2022)\citenamefont {Hagen},
  \citenamefont {Novario}, \citenamefont {Sun}, \citenamefont {Papenbrock},
  \citenamefont {Jansen}, \citenamefont {Lietz}, \citenamefont {Duguet},\ and\
  \citenamefont {Tichai}}]{Hagen:2022tqp}%
  \BibitemOpen
  \bibfield  {author} {\bibinfo {author} {\bibfnamefont {G.}~\bibnamefont
  {Hagen}}, \bibinfo {author} {\bibfnamefont {S.~J.}\ \bibnamefont {Novario}},
  \bibinfo {author} {\bibfnamefont {Z.~H.}\ \bibnamefont {Sun}}, \bibinfo
  {author} {\bibfnamefont {T.}~\bibnamefont {Papenbrock}}, \bibinfo {author}
  {\bibfnamefont {G.~R.}\ \bibnamefont {Jansen}}, \bibinfo {author}
  {\bibfnamefont {J.~G.}\ \bibnamefont {Lietz}}, \bibinfo {author}
  {\bibfnamefont {T.}~\bibnamefont {Duguet}},\ and\ \bibinfo {author}
  {\bibfnamefont {A.}~\bibnamefont {Tichai}},\ }\bibfield  {title} {\bibinfo
  {title} {{Angular-momentum projection in coupled-cluster theory: Structure of
  Mg34}},\ }\href {https://doi.org/10.1103/PhysRevC.105.064311} {\bibfield
  {journal} {\bibinfo  {journal} {Phys. Rev. C}\ }\textbf {\bibinfo {volume}
  {105}},\ \bibinfo {pages} {064311} (\bibinfo {year} {2022})},\ \Eprint
  {https://arxiv.org/abs/2201.07298} {arXiv:2201.07298 [nucl-th]} \BibitemShut
  {NoStop}%
\bibitem [{\citenamefont {Papenbrock}(2022)}]{Papenbrock:2022vdf}%
  \BibitemOpen
  \bibfield  {author} {\bibinfo {author} {\bibfnamefont {T.}~\bibnamefont
  {Papenbrock}},\ }\bibfield  {title} {\bibinfo {title} {{Effective field
  theory of pairing rotations}},\ }\href
  {https://doi.org/10.1103/PhysRevC.105.044322} {\bibfield  {journal} {\bibinfo
   {journal} {Phys. Rev. C}\ }\textbf {\bibinfo {volume} {105}},\ \bibinfo
  {pages} {044322} (\bibinfo {year} {2022})},\ \Eprint
  {https://arxiv.org/abs/2202.13146} {arXiv:2202.13146 [nucl-th]} \BibitemShut
  {NoStop}%
\bibitem [{\citenamefont {Muli}\ \emph {et~al.}(2022)\citenamefont {Muli},
  \citenamefont {Acharya}, \citenamefont {Hernandez},\ and\ \citenamefont
  {Bacca}}]{Muli:2022jma}%
  \BibitemOpen
  \bibfield  {author} {\bibinfo {author} {\bibfnamefont {S.~S.~L.}\
  \bibnamefont {Muli}}, \bibinfo {author} {\bibfnamefont {B.}~\bibnamefont
  {Acharya}}, \bibinfo {author} {\bibfnamefont {O.~J.}\ \bibnamefont
  {Hernandez}},\ and\ \bibinfo {author} {\bibfnamefont {S.}~\bibnamefont
  {Bacca}},\ }\bibfield  {title} {\bibinfo {title} {{Bayesian analysis of
  nuclear polarizability corrections to the Lamb shift of muonic H-atoms and
  He-ions}},\ }\href {https://doi.org/10.1088/1361-6471/ac81e0} {\bibfield
  {journal} {\bibinfo  {journal} {J. Phys. G}\ }\textbf {\bibinfo {volume}
  {49}},\ \bibinfo {pages} {105101} (\bibinfo {year} {2022})},\ \Eprint
  {https://arxiv.org/abs/2203.10792} {arXiv:2203.10792 [nucl-th]} \BibitemShut
  {NoStop}%
\bibitem [{\citenamefont {Zhai}\ \emph {et~al.}(2022)\citenamefont {Zhai},
  \citenamefont {Liu}, \citenamefont {Lu},\ and\ \citenamefont
  {Geng}}]{Zhai:2022ied}%
  \BibitemOpen
  \bibfield  {author} {\bibinfo {author} {\bibfnamefont {Q.-Y.}\ \bibnamefont
  {Zhai}}, \bibinfo {author} {\bibfnamefont {M.-Z.}\ \bibnamefont {Liu}},
  \bibinfo {author} {\bibfnamefont {J.-X.}\ \bibnamefont {Lu}},\ and\ \bibinfo
  {author} {\bibfnamefont {L.-S.}\ \bibnamefont {Geng}},\ }\bibfield  {title}
  {\bibinfo {title} {{Zcs(3985) in next-to-leading-order chiral effective field
  theory: The first truncation uncertainty analysis}},\ }\href
  {https://doi.org/10.1103/PhysRevD.106.034026} {\bibfield  {journal} {\bibinfo
   {journal} {Phys. Rev. D}\ }\textbf {\bibinfo {volume} {106}},\ \bibinfo
  {pages} {034026} (\bibinfo {year} {2022})},\ \Eprint
  {https://arxiv.org/abs/2205.03878} {arXiv:2205.03878 [hep-ph]} \BibitemShut
  {NoStop}%
\bibitem [{\citenamefont {Fraboulet}\ and\ \citenamefont
  {Ebran}(2023)}]{Fraboulet:2022zvt}%
  \BibitemOpen
  \bibfield  {author} {\bibinfo {author} {\bibfnamefont {K.}~\bibnamefont
  {Fraboulet}}\ and\ \bibinfo {author} {\bibfnamefont {J.-P.}\ \bibnamefont
  {Ebran}},\ }\bibfield  {title} {\bibinfo {title} {{Addressing energy density
  functionals in the language of path-integrals I: comparative study of
  diagrammatic techniques applied to the (0~+~0)-D O(N)-symmetric $\varphi
  ^{4}$-theory}},\ }\href {https://doi.org/10.1140/epja/s10050-023-00933-9}
  {\bibfield  {journal} {\bibinfo  {journal} {Eur. Phys. J. A}\ }\textbf
  {\bibinfo {volume} {59}},\ \bibinfo {pages} {91} (\bibinfo {year} {2023})},\
  \Eprint {https://arxiv.org/abs/2208.13044} {arXiv:2208.13044 [nucl-th]}
  \BibitemShut {NoStop}%
\bibitem [{\citenamefont {Jiang}\ and\ \citenamefont
  {Forss\'en}(2022)}]{Jiang:2022off}%
  \BibitemOpen
  \bibfield  {author} {\bibinfo {author} {\bibfnamefont {W.}~\bibnamefont
  {Jiang}}\ and\ \bibinfo {author} {\bibfnamefont {C.}~\bibnamefont
  {Forss\'en}},\ }\bibfield  {title} {\bibinfo {title} {{Bayesian probability
  updates using sampling/importance resampling: Applications in nuclear
  theory}},\ }\href {https://doi.org/10.3389/fphy.2022.1058809} {\bibfield
  {journal} {\bibinfo  {journal} {Front. in Phys.}\ }\textbf {\bibinfo {volume}
  {10}},\ \bibinfo {pages} {1058809} (\bibinfo {year} {2022})},\ \Eprint
  {https://arxiv.org/abs/2210.02507} {arXiv:2210.02507 [nucl-th]} \BibitemShut
  {NoStop}%
\bibitem [{\citenamefont {Ekstr\"om}\ \emph {et~al.}(2023)\citenamefont
  {Ekstr\"om}, \citenamefont {Forss\'en}, \citenamefont {Hagen}, \citenamefont
  {Jansen}, \citenamefont {Jiang},\ and\ \citenamefont
  {Papenbrock}}]{Ekstrom:2022yea}%
  \BibitemOpen
  \bibfield  {author} {\bibinfo {author} {\bibfnamefont {A.}~\bibnamefont
  {Ekstr\"om}}, \bibinfo {author} {\bibfnamefont {C.}~\bibnamefont
  {Forss\'en}}, \bibinfo {author} {\bibfnamefont {G.}~\bibnamefont {Hagen}},
  \bibinfo {author} {\bibfnamefont {G.~R.}\ \bibnamefont {Jansen}}, \bibinfo
  {author} {\bibfnamefont {W.}~\bibnamefont {Jiang}},\ and\ \bibinfo {author}
  {\bibfnamefont {T.}~\bibnamefont {Papenbrock}},\ }\bibfield  {title}
  {\bibinfo {title} {{What is ab initio in nuclear theory?}},\ }\href
  {https://doi.org/10.3389/fphy.2023.1129094} {\bibfield  {journal} {\bibinfo
  {journal} {Front. Phys.}\ }\textbf {\bibinfo {volume} {11}},\ \bibinfo
  {pages} {1129094} (\bibinfo {year} {2023})},\ \Eprint
  {https://arxiv.org/abs/2212.11064} {arXiv:2212.11064 [nucl-th]} \BibitemShut
  {NoStop}%
\bibitem [{\citenamefont {Jay}\ and\ \citenamefont {Neil}(2021)}]{Jay:2020jkz}%
  \BibitemOpen
  \bibfield  {author} {\bibinfo {author} {\bibfnamefont {W.~I.}\ \bibnamefont
  {Jay}}\ and\ \bibinfo {author} {\bibfnamefont {E.~T.}\ \bibnamefont {Neil}},\
  }\bibfield  {title} {\bibinfo {title} {{Bayesian model averaging for analysis
  of lattice field theory results}},\ }\href
  {https://doi.org/10.1103/PhysRevD.103.114502} {\bibfield  {journal} {\bibinfo
   {journal} {Phys. Rev. D}\ }\textbf {\bibinfo {volume} {103}},\ \bibinfo
  {pages} {114502} (\bibinfo {year} {2021})},\ \Eprint
  {https://arxiv.org/abs/2008.01069} {arXiv:2008.01069 [stat.ME]} \BibitemShut
  {NoStop}%
\bibitem [{\citenamefont {Catacora-Rios}\ \emph {et~al.}(2019)\citenamefont
  {Catacora-Rios}, \citenamefont {King}, \citenamefont {Lovell},\ and\
  \citenamefont {Nunes}}]{Catacora-Rios:2019goa}%
  \BibitemOpen
  \bibfield  {author} {\bibinfo {author} {\bibfnamefont {M.}~\bibnamefont
  {Catacora-Rios}}, \bibinfo {author} {\bibfnamefont {G.~B.}\ \bibnamefont
  {King}}, \bibinfo {author} {\bibfnamefont {A.~E.}\ \bibnamefont {Lovell}},\
  and\ \bibinfo {author} {\bibfnamefont {F.~M.}\ \bibnamefont {Nunes}},\
  }\bibfield  {title} {\bibinfo {title} {{Exploring experimental conditions to
  reduce uncertainties in the optical potential}},\ }\href
  {https://doi.org/10.1103/PhysRevC.100.064615} {\bibfield  {journal} {\bibinfo
   {journal} {Phys. Rev. C}\ }\textbf {\bibinfo {volume} {100}},\ \bibinfo
  {pages} {064615} (\bibinfo {year} {2019})},\ \Eprint
  {https://arxiv.org/abs/2112.00873} {arXiv:2112.00873 [nucl-th]} \BibitemShut
  {NoStop}%
\bibitem [{\citenamefont {Ekstr\"om}\ and\ \citenamefont
  {Hagen}(2019)}]{Ekstrom:2019lss}%
  \BibitemOpen
  \bibfield  {author} {\bibinfo {author} {\bibfnamefont {A.}~\bibnamefont
  {Ekstr\"om}}\ and\ \bibinfo {author} {\bibfnamefont {G.}~\bibnamefont
  {Hagen}},\ }\bibfield  {title} {\bibinfo {title} {{Global sensitivity
  analysis of bulk properties of an atomic nucleus}},\ }\href
  {https://doi.org/10.1103/PhysRevLett.123.252501} {\bibfield  {journal}
  {\bibinfo  {journal} {Phys. Rev. Lett.}\ }\textbf {\bibinfo {volume} {123}},\
  \bibinfo {pages} {252501} (\bibinfo {year} {2019})},\ \Eprint
  {https://arxiv.org/abs/1910.02922} {arXiv:1910.02922 [nucl-th]} \BibitemShut
  {NoStop}%
\bibitem [{\citenamefont {Zhang}\ \emph {et~al.}(2020)\citenamefont {Zhang},
  \citenamefont {Nollett},\ and\ \citenamefont {Phillips}}]{Zhang:2019odg}%
  \BibitemOpen
  \bibfield  {author} {\bibinfo {author} {\bibfnamefont {X.}~\bibnamefont
  {Zhang}}, \bibinfo {author} {\bibfnamefont {K.~M.}\ \bibnamefont {Nollett}},\
  and\ \bibinfo {author} {\bibfnamefont {D.~R.}\ \bibnamefont {Phillips}},\
  }\bibfield  {title} {\bibinfo {title} {{$S$-factor and scattering-parameter
  extractions from ${}^{3}\mathrm{He} +{}^{4}\mathrm{He} \rightarrow
  {}^{7}\mathrm{Be} + \gamma$}},\ }\href
  {https://doi.org/10.1088/1361-6471/ab6a71} {\bibfield  {journal} {\bibinfo
  {journal} {J. Phys. G}\ }\textbf {\bibinfo {volume} {47}},\ \bibinfo {pages}
  {054002} (\bibinfo {year} {2020})},\ \Eprint
  {https://arxiv.org/abs/1909.07287} {arXiv:1909.07287 [nucl-th]} \BibitemShut
  {NoStop}%
\bibitem [{\citenamefont {Luna}\ and\ \citenamefont
  {Papenbrock}(2019)}]{Luna:2019ufu}%
  \BibitemOpen
  \bibfield  {author} {\bibinfo {author} {\bibfnamefont {B.~K.}\ \bibnamefont
  {Luna}}\ and\ \bibinfo {author} {\bibfnamefont {T.}~\bibnamefont
  {Papenbrock}},\ }\bibfield  {title} {\bibinfo {title} {{Low-energy bound
  states, resonances, and scattering of light ions}},\ }\href
  {https://doi.org/10.1103/PhysRevC.100.054307} {\bibfield  {journal} {\bibinfo
   {journal} {Phys. Rev. C}\ }\textbf {\bibinfo {volume} {100}},\ \bibinfo
  {pages} {054307} (\bibinfo {year} {2019})},\ \Eprint
  {https://arxiv.org/abs/1907.11345} {arXiv:1907.11345 [nucl-th]} \BibitemShut
  {NoStop}%
\bibitem [{\citenamefont {Epelbaum}\ \emph {et~al.}(2020)\citenamefont
  {Epelbaum} \emph {et~al.}}]{Epelbaum:2019zqc}%
  \BibitemOpen
  \bibfield  {author} {\bibinfo {author} {\bibfnamefont {E.}~\bibnamefont
  {Epelbaum}} \emph {et~al.},\ }\bibfield  {title} {\bibinfo {title} {{Towards
  high-order calculations of three-nucleon scattering in chiral effective field
  theory}},\ }\href {https://doi.org/10.1140/epja/s10050-020-00102-2}
  {\bibfield  {journal} {\bibinfo  {journal} {Eur. Phys. J. A}\ }\textbf
  {\bibinfo {volume} {56}},\ \bibinfo {pages} {92} (\bibinfo {year} {2020})},\
  \Eprint {https://arxiv.org/abs/1907.03608} {arXiv:1907.03608 [nucl-th]}
  \BibitemShut {NoStop}%
\bibitem [{\citenamefont {Metropolis}\ \emph {et~al.}(1953)\citenamefont
  {Metropolis}, \citenamefont {Rosenbluth}, \citenamefont {Rosenbluth},
  \citenamefont {Teller},\ and\ \citenamefont {Teller}}]{Metropolis:1953am}%
  \BibitemOpen
  \bibfield  {author} {\bibinfo {author} {\bibfnamefont {N.}~\bibnamefont
  {Metropolis}}, \bibinfo {author} {\bibfnamefont {A.~W.}\ \bibnamefont
  {Rosenbluth}}, \bibinfo {author} {\bibfnamefont {M.~N.}\ \bibnamefont
  {Rosenbluth}}, \bibinfo {author} {\bibfnamefont {A.~H.}\ \bibnamefont
  {Teller}},\ and\ \bibinfo {author} {\bibfnamefont {E.}~\bibnamefont
  {Teller}},\ }\bibfield  {title} {\bibinfo {title} {{Equation of state
  calculations by fast computing machines}},\ }\href
  {https://doi.org/10.1063/1.1699114} {\bibfield  {journal} {\bibinfo
  {journal} {J. Chem. Phys.}\ }\textbf {\bibinfo {volume} {21}},\ \bibinfo
  {pages} {1087} (\bibinfo {year} {1953})}\BibitemShut {NoStop}%
\bibitem [{\citenamefont {Hastings}(1970)}]{Hastings:1970aa}%
  \BibitemOpen
  \bibfield  {author} {\bibinfo {author} {\bibfnamefont {W.~K.}\ \bibnamefont
  {Hastings}},\ }\bibfield  {title} {\bibinfo {title} {{Monte Carlo Sampling
  Methods Using Markov Chains and Their Applications}},\ }\href
  {https://doi.org/10.1093/biomet/57.1.97} {\bibfield  {journal} {\bibinfo
  {journal} {Biometrika}\ }\textbf {\bibinfo {volume} {57}},\ \bibinfo {pages}
  {97} (\bibinfo {year} {1970})}\BibitemShut {NoStop}%
\bibitem [{\citenamefont {Duane}\ \emph {et~al.}(1987)\citenamefont {Duane},
  \citenamefont {Kennedy}, \citenamefont {Pendleton},\ and\ \citenamefont
  {Roweth}}]{DUANE1987216}%
  \BibitemOpen
  \bibfield  {author} {\bibinfo {author} {\bibfnamefont {S.}~\bibnamefont
  {Duane}}, \bibinfo {author} {\bibfnamefont {A.}~\bibnamefont {Kennedy}},
  \bibinfo {author} {\bibfnamefont {B.~J.}\ \bibnamefont {Pendleton}},\ and\
  \bibinfo {author} {\bibfnamefont {D.}~\bibnamefont {Roweth}},\ }\bibfield
  {title} {\bibinfo {title} {Hybrid monte carlo},\ }\href
  {https://doi.org/https://doi.org/10.1016/0370-2693(87)91197-X} {\bibfield
  {journal} {\bibinfo  {journal} {Physics Letters B}\ }\textbf {\bibinfo
  {volume} {195}},\ \bibinfo {pages} {216} (\bibinfo {year}
  {1987})}\BibitemShut {NoStop}%
\bibitem [{\citenamefont {Homan}\ and\ \citenamefont
  {Gelman}(2014)}]{hoffman2011nouturn}%
  \BibitemOpen
  \bibfield  {author} {\bibinfo {author} {\bibfnamefont {M.~D.}\ \bibnamefont
  {Homan}}\ and\ \bibinfo {author} {\bibfnamefont {A.}~\bibnamefont {Gelman}},\
  }\bibfield  {title} {\bibinfo {title} {The no-u-turn sampler: Adaptively
  setting path lengths in hamiltonian monte carlo},\ }\href@noop {} {\bibfield
  {journal} {\bibinfo  {journal} {J. Mach. Learn. Res.}\ }\textbf {\bibinfo
  {volume} {15}},\ \bibinfo {pages} {1593–1623} (\bibinfo {year}
  {2014})}\BibitemShut {NoStop}%
\bibitem [{\citenamefont {Salvatier}\ \emph {et~al.}(2016)\citenamefont
  {Salvatier}, \citenamefont {Wiecki},\ and\ \citenamefont
  {Fonnesbeck}}]{Salvatier2016}%
  \BibitemOpen
  \bibfield  {author} {\bibinfo {author} {\bibfnamefont {J.}~\bibnamefont
  {Salvatier}}, \bibinfo {author} {\bibfnamefont {T.~V.}\ \bibnamefont
  {Wiecki}},\ and\ \bibinfo {author} {\bibfnamefont {C.}~\bibnamefont
  {Fonnesbeck}},\ }\bibfield  {title} {\bibinfo {title} {Probabilistic
  programming in python using {PyMC}3},\ }\href
  {https://doi.org/10.7717/peerj-cs.55} {\bibfield  {journal} {\bibinfo
  {journal} {{PeerJ} Computer Science}\ }\textbf {\bibinfo {volume} {2}},\
  \bibinfo {pages} {e55} (\bibinfo {year} {2016})}\BibitemShut {NoStop}%
\bibitem [{\citenamefont {Gregory}(2005)}]{Gregory2005}%
  \BibitemOpen
  \bibfield  {author} {\bibinfo {author} {\bibfnamefont {P.}~\bibnamefont
  {Gregory}},\ }\href@noop {} {\emph {\bibinfo {title} {Bayesian Logical Data
  Analysis for the Physical Sciences}}}\ (\bibinfo  {publisher} {Cambridge
  University Press},\ \bibinfo {address} {Cambridge},\ \bibinfo {year}
  {2005})\BibitemShut {NoStop}%
\bibitem [{\citenamefont {Bijnens}\ \emph {et~al.}(2000)\citenamefont
  {Bijnens}, \citenamefont {Colangelo},\ and\ \citenamefont
  {Ecker}}]{Bijnens:1999hw}%
  \BibitemOpen
  \bibfield  {author} {\bibinfo {author} {\bibfnamefont {J.}~\bibnamefont
  {Bijnens}}, \bibinfo {author} {\bibfnamefont {G.}~\bibnamefont {Colangelo}},\
  and\ \bibinfo {author} {\bibfnamefont {G.}~\bibnamefont {Ecker}},\ }\bibfield
   {title} {\bibinfo {title} {{Renormalization of chiral perturbation theory to
  order $p^6$}},\ }\href {https://doi.org/10.1006/aphy.1999.5982} {\bibfield
  {journal} {\bibinfo  {journal} {Annals Phys.}\ }\textbf {\bibinfo {volume}
  {280}},\ \bibinfo {pages} {100} (\bibinfo {year} {2000})},\ \Eprint
  {https://arxiv.org/abs/hep-ph/9907333} {arXiv:hep-ph/9907333 [hep-ph]}
  \BibitemShut {NoStop}%
%%CITATION = HEP-PH/9907333;%%
\bibitem [{\citenamefont {Gelman}\ \emph {et~al.}(2013)\citenamefont {Gelman},
  \citenamefont {Carlin}, \citenamefont {Stern}, \citenamefont {Dunson},
  \citenamefont {Vehtari},\ and\ \citenamefont {Rubin}}]{Gelman2013}%
  \BibitemOpen
  \bibfield  {author} {\bibinfo {author} {\bibfnamefont {A.}~\bibnamefont
  {Gelman}}, \bibinfo {author} {\bibfnamefont {J.~B.}\ \bibnamefont {Carlin}},
  \bibinfo {author} {\bibfnamefont {H.~S.}\ \bibnamefont {Stern}}, \bibinfo
  {author} {\bibfnamefont {D.~B.}\ \bibnamefont {Dunson}}, \bibinfo {author}
  {\bibfnamefont {A.}~\bibnamefont {Vehtari}},\ and\ \bibinfo {author}
  {\bibfnamefont {D.~B.}\ \bibnamefont {Rubin}},\ }\href@noop {} {\emph
  {\bibinfo {title} {Bayesian Data Analysis}}},\ \bibinfo {edition} {3rd}\ ed.\
  (\bibinfo  {publisher} {CPC Press},\ \bibinfo {address} {Boca Raton},\
  \bibinfo {year} {2013})\BibitemShut {NoStop}%
\bibitem [{\citenamefont {Vehtari}\ \emph {et~al.}(2016)\citenamefont
  {Vehtari}, \citenamefont {Gelman},\ and\ \citenamefont
  {Gabry}}]{Vehtari_2016}%
  \BibitemOpen
  \bibfield  {author} {\bibinfo {author} {\bibfnamefont {A.}~\bibnamefont
  {Vehtari}}, \bibinfo {author} {\bibfnamefont {A.}~\bibnamefont {Gelman}},\
  and\ \bibinfo {author} {\bibfnamefont {J.}~\bibnamefont {Gabry}},\ }\bibfield
   {title} {\bibinfo {title} {Practical bayesian model evaluation using
  leave-one-out cross-validation and {WAIC}},\ }\href
  {https://doi.org/10.1007/s11222-016-9696-4} {\bibfield  {journal} {\bibinfo
  {journal} {Statistics and Computing}\ }\textbf {\bibinfo {volume} {27}},\
  \bibinfo {pages} {1413} (\bibinfo {year} {2016})}\BibitemShut {NoStop}%
\bibitem [{\citenamefont {Amor\'{o}s}\ \emph
  {et~al.}(2000{\natexlab{b}})\citenamefont {Amor\'{o}s}, \citenamefont
  {Bijnens},\ and\ \citenamefont {Talavera}}]{Amoros:1999dp}%
  \BibitemOpen
  \bibfield  {author} {\bibinfo {author} {\bibfnamefont {G.}~\bibnamefont
  {Amor\'{o}s}}, \bibinfo {author} {\bibfnamefont {J.}~\bibnamefont
  {Bijnens}},\ and\ \bibinfo {author} {\bibfnamefont {P.}~\bibnamefont
  {Talavera}},\ }\bibfield  {title} {\bibinfo {title} {{Two-point functions at
  two loops in three flavor chiral perturbation theory}},\ }\href
  {https://doi.org/10.1016/S0550-3213(99)00674-4} {\bibfield  {journal}
  {\bibinfo  {journal} {Nucl. Phys.}\ }\textbf {\bibinfo {volume} {B568}},\
  \bibinfo {pages} {319} (\bibinfo {year} {2000}{\natexlab{b}})},\ \Eprint
  {https://arxiv.org/abs/hep-ph/9907264} {arXiv:hep-ph/9907264 [hep-ph]}
  \BibitemShut {NoStop}%
%%CITATION = HEP-PH/9907264;%%
\bibitem [{\citenamefont {Bijnens}(2019)}]{Bijnens2019}%
  \BibitemOpen
  \bibfield  {author} {\bibinfo {author} {\bibfnamefont {J.}~\bibnamefont
  {Bijnens}},\ }\href@noop {} {\bibinfo {title} {{Chiral perturbation
  theory}}},\ \bibinfo {howpublished}
  {\url{http://home.thep.lu.se/~bijnens/chpt/}} (\bibinfo {year}
  {2019})\BibitemShut {NoStop}%
\bibitem [{\citenamefont {Bijnens}\ and\ \citenamefont
  {Dhonte}(2003)}]{Bijnens:2003xg}%
  \BibitemOpen
  \bibfield  {author} {\bibinfo {author} {\bibfnamefont {J.}~\bibnamefont
  {Bijnens}}\ and\ \bibinfo {author} {\bibfnamefont {P.}~\bibnamefont
  {Dhonte}},\ }\bibfield  {title} {\bibinfo {title} {{Scalar form-factors in
  $SU(3)$ chiral perturbation theory}},\ }\href
  {https://doi.org/10.1088/1126-6708/2003/10/061} {\bibfield  {journal}
  {\bibinfo  {journal} {J.High Energy Phys.}\ }\textbf {\bibinfo {volume}
  {10}},\ \bibinfo {pages} {061} (\bibinfo {year} {2003})},\ \Eprint
  {https://arxiv.org/abs/hep-ph/0307044} {arXiv:hep-ph/0307044 [hep-ph]}
  \BibitemShut {NoStop}%
%%CITATION = HEP-PH/0307044;%%
\bibitem [{\citenamefont {Gasser}\ \emph {et~al.}(2007)\citenamefont {Gasser},
  \citenamefont {Haefeli}, \citenamefont {Ivanov},\ and\ \citenamefont
  {Schmid}}]{Gasser:2007sg}%
  \BibitemOpen
  \bibfield  {author} {\bibinfo {author} {\bibfnamefont {J.}~\bibnamefont
  {Gasser}}, \bibinfo {author} {\bibfnamefont {C.}~\bibnamefont {Haefeli}},
  \bibinfo {author} {\bibfnamefont {M.~A.}\ \bibnamefont {Ivanov}},\ and\
  \bibinfo {author} {\bibfnamefont {M.}~\bibnamefont {Schmid}},\ }\bibfield
  {title} {\bibinfo {title} {{Integrating out strange quarks in ChPT}},\ }\href
  {https://doi.org/10.1016/j.physletb.2007.06.058} {\bibfield  {journal}
  {\bibinfo  {journal} {Phys. Lett.}\ }\textbf {\bibinfo {volume} {B652}},\
  \bibinfo {pages} {21} (\bibinfo {year} {2007})},\ \Eprint
  {https://arxiv.org/abs/0706.0955} {arXiv:0706.0955 [hep-ph]} \BibitemShut
  {NoStop}%
%%CITATION = ARXIV:0706.0955;%%
\bibitem [{\citenamefont {Jiang}\ \emph {et~al.}(2015)\citenamefont {Jiang},
  \citenamefont {Wei}, \citenamefont {Chen},\ and\ \citenamefont
  {Wang}}]{Jiang:2015dba}%
  \BibitemOpen
  \bibfield  {author} {\bibinfo {author} {\bibfnamefont {S.-Z.}\ \bibnamefont
  {Jiang}}, \bibinfo {author} {\bibfnamefont {Z.-L.}\ \bibnamefont {Wei}},
  \bibinfo {author} {\bibfnamefont {Q.-S.}\ \bibnamefont {Chen}},\ and\
  \bibinfo {author} {\bibfnamefont {Q.}~\bibnamefont {Wang}},\ }\bibfield
  {title} {\bibinfo {title} {{Computation of the $O(p^6)$ order low-energy
  constants: An update}},\ }\href {https://doi.org/10.1103/PhysRevD.92.025014}
  {\bibfield  {journal} {\bibinfo  {journal} {Phys. Rev. D}\ }\textbf {\bibinfo
  {volume} {92}},\ \bibinfo {pages} {025014} (\bibinfo {year} {2015})},\
  \Eprint {https://arxiv.org/abs/1502.05087} {arXiv:1502.05087 [hep-ph]}
  \BibitemShut {NoStop}%
%%CITATION = ARXIV:1502.05087;%%
\bibitem [{\citenamefont {Jiang}\ \emph {et~al.}(2010)\citenamefont {Jiang},
  \citenamefont {Zhang}, \citenamefont {Li},\ and\ \citenamefont
  {Wang}}]{Jiang:2009uf}%
  \BibitemOpen
  \bibfield  {author} {\bibinfo {author} {\bibfnamefont {S.-Z.}\ \bibnamefont
  {Jiang}}, \bibinfo {author} {\bibfnamefont {Y.}~\bibnamefont {Zhang}},
  \bibinfo {author} {\bibfnamefont {C.}~\bibnamefont {Li}},\ and\ \bibinfo
  {author} {\bibfnamefont {Q.}~\bibnamefont {Wang}},\ }\bibfield  {title}
  {\bibinfo {title} {{Computation of the $p^6$ order chiral Lagrangian
  coefficients}},\ }\href {https://doi.org/10.1103/PhysRevD.81.014001}
  {\bibfield  {journal} {\bibinfo  {journal} {Phys. Rev. D}\ }\textbf {\bibinfo
  {volume} {81}},\ \bibinfo {pages} {014001} (\bibinfo {year} {2010})},\
  \Eprint {https://arxiv.org/abs/0907.5229} {arXiv:0907.5229 [hep-ph]}
  \BibitemShut {NoStop}%
%%CITATION = ARXIV:0907.5229;%%
\bibitem [{\citenamefont {Kampf}\ and\ \citenamefont
  {Moussallam}(2006)}]{Kampf:2006bn}%
  \BibitemOpen
  \bibfield  {author} {\bibinfo {author} {\bibfnamefont {K.}~\bibnamefont
  {Kampf}}\ and\ \bibinfo {author} {\bibfnamefont {B.}~\bibnamefont
  {Moussallam}},\ }\bibfield  {title} {\bibinfo {title} {{Tests of the
  naturalness of the coupling constants in ChPT at order $p^6$}},\ }\href
  {https://doi.org/10.1140/epjc/s2006-02606-7} {\bibfield  {journal} {\bibinfo
  {journal} {Eur. Phys. J. C}\ }\textbf {\bibinfo {volume} {47}},\ \bibinfo
  {pages} {723} (\bibinfo {year} {2006})},\ \Eprint
  {https://arxiv.org/abs/hep-ph/0604125} {arXiv:hep-ph/0604125 [hep-ph]}
  \BibitemShut {NoStop}%
%%CITATION = HEP-PH/0604125;%%
\bibitem [{\citenamefont {Jamin}\ \emph {et~al.}(2004)\citenamefont {Jamin},
  \citenamefont {Oller},\ and\ \citenamefont {Pich}}]{Jamin:2004re}%
  \BibitemOpen
  \bibfield  {author} {\bibinfo {author} {\bibfnamefont {M.}~\bibnamefont
  {Jamin}}, \bibinfo {author} {\bibfnamefont {J.~A.}\ \bibnamefont {Oller}},\
  and\ \bibinfo {author} {\bibfnamefont {A.}~\bibnamefont {Pich}},\ }\bibfield
  {title} {\bibinfo {title} {{Order $p^{6}$ chiral couplings from the scalar $K
  \pi$ form-factor}},\ }\href {https://doi.org/10.1088/1126-6708/2004/02/047}
  {\bibfield  {journal} {\bibinfo  {journal} {J.High Energy Phys.}\ }\textbf
  {\bibinfo {volume} {02}},\ \bibinfo {pages} {047} (\bibinfo {year} {2004})},\
  \Eprint {https://arxiv.org/abs/hep-ph/0401080} {arXiv:hep-ph/0401080
  [hep-ph]} \BibitemShut {NoStop}%
%%CITATION = HEP-PH/0401080;%%
\bibitem [{\citenamefont {Bijnens}\ and\ \citenamefont
  {Talavera}(2003)}]{Bijnens:2003uy}%
  \BibitemOpen
  \bibfield  {author} {\bibinfo {author} {\bibfnamefont {J.}~\bibnamefont
  {Bijnens}}\ and\ \bibinfo {author} {\bibfnamefont {P.}~\bibnamefont
  {Talavera}},\ }\bibfield  {title} {\bibinfo {title} {{$K_{\ell3}$ decays in
  chiral perturbation theory}},\ }\href
  {https://doi.org/10.1016/S0550-3213(03)00581-9} {\bibfield  {journal}
  {\bibinfo  {journal} {Nucl. Phys.}\ }\textbf {\bibinfo {volume} {B669}},\
  \bibinfo {pages} {341} (\bibinfo {year} {2003})},\ \Eprint
  {https://arxiv.org/abs/hep-ph/0303103} {arXiv:hep-ph/0303103 [hep-ph]}
  \BibitemShut {NoStop}%
%%CITATION = HEP-PH/0303103;%%
\bibitem [{\citenamefont {Cirigliano}\ \emph {et~al.}(2005)\citenamefont
  {Cirigliano}, \citenamefont {Ecker}, \citenamefont {Eidem\"{u}ller},
  \citenamefont {Kaiser}, \citenamefont {Pich},\ and\ \citenamefont
  {Portol\'{e}s}}]{Cirigliano:2005xn}%
  \BibitemOpen
  \bibfield  {author} {\bibinfo {author} {\bibfnamefont {V.}~\bibnamefont
  {Cirigliano}}, \bibinfo {author} {\bibfnamefont {G.}~\bibnamefont {Ecker}},
  \bibinfo {author} {\bibfnamefont {M.}~\bibnamefont {Eidem\"{u}ller}},
  \bibinfo {author} {\bibfnamefont {R.}~\bibnamefont {Kaiser}}, \bibinfo
  {author} {\bibfnamefont {A.}~\bibnamefont {Pich}},\ and\ \bibinfo {author}
  {\bibfnamefont {J.}~\bibnamefont {Portol\'{e}s}},\ }\bibfield  {title}
  {\bibinfo {title} {{The $\langle SPP \rangle$ Green function and $SU(3)$
  breaking in $K_{\ell3}$ decays}},\ }\href
  {https://doi.org/10.1088/1126-6708/2005/04/006} {\bibfield  {journal}
  {\bibinfo  {journal} {J.High Energy Phys.}\ }\textbf {\bibinfo {volume}
  {04}},\ \bibinfo {pages} {006} (\bibinfo {year} {2005})},\ \Eprint
  {https://arxiv.org/abs/hep-ph/0503108} {arXiv:hep-ph/0503108 [hep-ph]}
  \BibitemShut {NoStop}%
%%CITATION = HEP-PH/0503108;%%
\bibitem [{\citenamefont {Unterdorfer}\ and\ \citenamefont
  {Pichl}(2008)}]{Unterdorfer:2008zz}%
  \BibitemOpen
  \bibfield  {author} {\bibinfo {author} {\bibfnamefont {R.}~\bibnamefont
  {Unterdorfer}}\ and\ \bibinfo {author} {\bibfnamefont {H.}~\bibnamefont
  {Pichl}},\ }\bibfield  {title} {\bibinfo {title} {{On the radiative pion
  decay}},\ }\href {https://doi.org/10.1140/epjc/s10052-008-0584-8} {\bibfield
  {journal} {\bibinfo  {journal} {Eur. Phys. J. C}\ }\textbf {\bibinfo {volume}
  {55}},\ \bibinfo {pages} {273} (\bibinfo {year} {2008})},\ \Eprint
  {https://arxiv.org/abs/0801.2482} {arXiv:0801.2482 [hep-ph]} \BibitemShut
  {NoStop}%
%%CITATION = ARXIV:0801.2482;%%
\bibitem [{\citenamefont {Cirigliano}\ \emph {et~al.}(2006)\citenamefont
  {Cirigliano}, \citenamefont {Ecker}, \citenamefont {Eidem\"{u}ller},
  \citenamefont {Kaiser}, \citenamefont {Pich},\ and\ \citenamefont
  {Portol\'{e}s}}]{Cirigliano:2006hb}%
  \BibitemOpen
  \bibfield  {author} {\bibinfo {author} {\bibfnamefont {V.}~\bibnamefont
  {Cirigliano}}, \bibinfo {author} {\bibfnamefont {G.}~\bibnamefont {Ecker}},
  \bibinfo {author} {\bibfnamefont {M.}~\bibnamefont {Eidem\"{u}ller}},
  \bibinfo {author} {\bibfnamefont {R.}~\bibnamefont {Kaiser}}, \bibinfo
  {author} {\bibfnamefont {A.}~\bibnamefont {Pich}},\ and\ \bibinfo {author}
  {\bibfnamefont {J.}~\bibnamefont {Portol\'{e}s}},\ }\bibfield  {title}
  {\bibinfo {title} {{Towards a consistent estimate of the chiral low-energy
  constants}},\ }\href {https://doi.org/10.1016/j.nuclphysb.2006.07.010}
  {\bibfield  {journal} {\bibinfo  {journal} {Nucl. Phys.}\ }\textbf {\bibinfo
  {volume} {B753}},\ \bibinfo {pages} {139} (\bibinfo {year} {2006})},\ \Eprint
  {https://arxiv.org/abs/hep-ph/0603205} {arXiv:hep-ph/0603205 [hep-ph]}
  \BibitemShut {NoStop}%
%%CITATION = HEP-PH/0603205;%%
\bibitem [{\citenamefont {Bernard}\ and\ \citenamefont
  {Passemar}(2008)}]{Bernard:2007tk}%
  \BibitemOpen
  \bibfield  {author} {\bibinfo {author} {\bibfnamefont {V.}~\bibnamefont
  {Bernard}}\ and\ \bibinfo {author} {\bibfnamefont {E.}~\bibnamefont
  {Passemar}},\ }\bibfield  {title} {\bibinfo {title} {{Matching chiral
  perturbation theory and the dispersive representation of the scalar $K \pi$
  form-factor}},\ }\href {https://doi.org/10.1016/j.physletb.2008.02.004}
  {\bibfield  {journal} {\bibinfo  {journal} {Phys. Lett.}\ }\textbf {\bibinfo
  {volume} {B661}},\ \bibinfo {pages} {95} (\bibinfo {year} {2008})},\ \Eprint
  {https://arxiv.org/abs/0711.3450} {arXiv:0711.3450 [hep-ph]} \BibitemShut
  {NoStop}%
%%CITATION = ARXIV:0711.3450;%%
\bibitem [{\citenamefont {Moussallam}(2000)}]{Moussallam:2000zf}%
  \BibitemOpen
  \bibfield  {author} {\bibinfo {author} {\bibfnamefont {B.}~\bibnamefont
  {Moussallam}},\ }\bibfield  {title} {\bibinfo {title} {{Flavor stability of
  the chiral vacuum and scalar meson dynamics}},\ }\href
  {https://doi.org/10.1088/1126-6708/2000/08/005} {\bibfield  {journal}
  {\bibinfo  {journal} {J.High Energy Phys.}\ }\textbf {\bibinfo {volume}
  {08}},\ \bibinfo {pages} {005} (\bibinfo {year} {2000})},\ \Eprint
  {https://arxiv.org/abs/hep-ph/0005245} {arXiv:hep-ph/0005245 [hep-ph]}
  \BibitemShut {NoStop}%
%%CITATION = HEP-PH/0005245;%%
\end{thebibliography}%

\clearpage
\begin{center}
\large\bf Bayesian method for fitting the low-energy constants in chiral perturbation theory: Supplementary Material
\end{center}

\section{Some definitions}
\begin{align}
	\mathrm{Pct}_{\mathrm{model}}=&\frac{a_{i,\mathrm{model}}-a_{\mathrm{i,tr}}}{a_{i,\mathrm{tr}}}\times 100\%, \label{equ:2}\\
	\mathrm{Pct\sigma}_{\mathrm{model}}=&\frac{a_{i,\mathrm{model}}-a_{i,\mathrm{tr}}}{\sigma_{i,{\mathrm{model}}}}. \label{equ:3}
\end{align}
\begin{align}
	\mathrm{Pct}_{\mathrm{order}}=\frac{{\bar\mu}^{\mathrm{order}}_{\mathrm{model}}}{{\bar\mu}_{\mathrm{model}}}\times 100\%, \label{equ:1}
\end{align}
\begin{align}
	\mu^\mathrm{(NLO)}_\mathrm B&=\mu^\mathrm{LO}_\mathrm B(\bm{a}^\mathrm{LO})+\mu^\mathrm{NLO}_\mathrm B(\bm{a}^\mathrm{LO},\bm{a}^\mathrm{NLO})+(2s-1) e \mu^\mathrm{LO}_\mathrm B,\label{eq3}\\
	\mu^\mathrm{(NNLO)}_\mathrm B&=\mu^\mathrm{LO}_\mathrm B(\bm{a}^\mathrm{LO})+\mu^\mathrm{NLO}_\mathrm B(\bm{a}^\mathrm{LO},\bm{a}^\mathrm{NLO})+\mu^\mathrm{NNLO}_\mathrm B(\bm{a}^\mathrm{LO},\bm{a}^\mathrm{NLO},\bm{a}^\mathrm{NNLO})+(2s-1) e \mu^\mathrm{LO}_\mathrm B,\label{eq4}
\end{align}
\begin{align}
	\pr\{s=k\}=p^k(1-p)^{1-k},\quad k=0,1.\label{dp}
\end{align}
\begin{align}
	\mathrm{PM}_{\mathrm{model}}=\sqrt{{\sum_{i=1}^n\left(\frac{(a^\mathrm{NNLO}_{i,\mathrm{tr}}-\bar{a}^\mathrm{NNLO}_{i,\mathrm{model}})}{a^\mathrm{NNLO}_{i,\mathrm{tr}}}{\frac{\bar{\mu}_{a^\mathrm{NNLO}_{i,\mathrm{model}}}}{\mu_{i,\mathrm{tr}}}} \right)^2}\bigg/n}. \label{equ:4}
\end{align}
\clearpage

\section{Table of the example in the article}

\begin{table}[H]
\begin{longtable}[htbp]{lrrrrrrr}
\caption{The theoretical contributions of $O_i$ at each order from the NLO fit in examples from the article. The numbers marked by tr, B$_1$ and B$_2$ represent the truth values and the values fitted by Model B$_1$ and B$_2$, respectively. The fitting expected values at LO, NLO and HO are given in Columns 2, 4 and 6, respectively. The corresponding $\mathrm{Pct_{LO,NLO,HO}}$ defined in Eq. \eqref{equ:1} are also given in Columns 3, 5 and 7, respectively. Column 8 are the full theoretical values for comparison. All values about $O_i$ have been multiplied 100.}\label{table2}\\
	\hline\hline  $i$&\multicolumn{1}{c}{LO}&$\mathrm{Pct}_\mathrm{LO}$&\multicolumn{1}{c}{NLO}&$\mathrm{Pct}_\mathrm{NLO}$&\multicolumn{1}{c}{HO}&$\mathrm{Pct}_\mathrm{HO}$ &Theory\\
	\hline%\addlinespace[2pt]
	\endfirsthead
	\hline\hline Obs&\multicolumn{1}{c}{LO}&$\mathrm{Pct}_\mathrm{LO}$&\multicolumn{1}{c}{NLO}&$\mathrm{Pct}_\mathrm{NLO}$&\multicolumn{1}{c}{HO}&$\mathrm{Pct}_\mathrm{HO}$ &Theory\\
	\hline%\addlinespace[2pt]
	\endhead
	\hline\hline
	\endfoot
	\hline\hline
	\endlastfoot
	$1_\mathrm{tr}$     & $-$25.00 & 69.8\%  & $-$7.25  & 20.3\%  & $-$3.55 & 9.9\%  & $-$35.80 \\
	$1_{\mathrm{B}_1}$  & $-$25.00 & 74.0\%  & $-$7.18  & 21.2\%  & $-$1.60 & 4.7\%  & $-$33.78 \\
	$1_{\mathrm{B}_2}$  & $-$25.00 & 71.4\%  & $-$7.38  & 21.1\%  & $-$2.65 & 7.6\%  & $-$35.02 \\
	$2_\mathrm{tr}$     & 0.110  & 63.5\%  & 0.052  & 30.1\%  & 0.011 & 6.4\%  & 0.173  \\
	$2_{\mathrm{B}_1}$  & 0.110  & 65.4\%  & 0.056  & 33.3\%  & 0.002 & 1.3\%  & 0.168  \\
	$2_{\mathrm{B}_2}$  & 0.110  & 63.8\%  & 0.053  & 30.6\%  & 0.010 & 5.6\%  & 0.172  \\
	$3_\mathrm{tr}$     & $-$0.359 & 130.2\% & 0.078  & $-$28.2\% & 0.006 & $-$2.0\% & $-$0.276 \\
	$3_{\mathrm{B}_1}$  & $-$0.359 & 129.0\% & 0.081  & $-$29.1\% & 0.000 & 0.1\%  & $-$0.279 \\
	$3_{\mathrm{B}_2}$  & $-$0.359 & 128.9\% & 0.073  & $-$26.3\% & 0.007 & $-$2.6\% & $-$0.279 \\
	$4_\mathrm{tr}$     & 0.395  & 65.5\%  & 0.192  & 31.9\%  & 0.015 & 2.6\%  & 0.603  \\
	$4_{\mathrm{B}_1}$  & 0.395  & 67.6\%  & 0.187  & 32.0\%  & 0.003 & 0.4\%  & 0.584  \\
	$4_{\mathrm{B}_2}$  & 0.395  & 66.3\%  & 0.185  & 31.1\%  & 0.015 & 2.5\%  & 0.595  \\
	$5_\mathrm{tr}$     & 17.18  & 63.0\%  & 7.66   & 28.1\%  & 2.44  & 9.0\%  & 27.28  \\
	$5_{\mathrm{B}_1}$  & 17.18  & 63.3\%  & 7.53   & 27.8\%  & 2.42  & 8.9\%  & 27.12  \\
	$5_{\mathrm{B}_2}$  & 17.18  & 62.2\%  & 7.76   & 28.1\%  & 2.66  & 9.6\%  & 27.60  \\
	$6_\mathrm{tr}$     & $-$0.362 & 69.1\%  & $-$0.201 & 38.3\%  & 0.039 & $-$7.4\% & $-$0.524 \\
	$6_{\mathrm{B}_1}$  & $-$0.362 & 65.3\%  & $-$0.196 & 35.4\%  & 0.004 & $-$0.7\% & $-$0.554 \\
	$6_{\mathrm{B}_2}$  & $-$0.362 & 66.9\%  & $-$0.210 & 38.8\%  & 0.031 & $-$5.7\% & $-$0.541 \\
	$7_\mathrm{tr}$     & $-$1.05  & 70.9\%  & $-$0.32  & 21.6\%  & $-$0.11 & 7.5\%  & $-$1.49  \\
	$7_{\mathrm{B}_1}$  & $-$1.05  & 75.1\%  & $-$0.32  & 22.9\%  & $-$0.03 & 2.0\%  & $-$1.40  \\
	$7_{\mathrm{B}_2}$  & $-$1.05  & 72.2\%  & $-$0.32  & 22.1\%  & $-$0.08 & 5.7\%  & $-$1.46  \\
	$8_\mathrm{tr}$     & $-$0.62  & 64.7\%  & $-$0.25  & 26.2\%  & $-$0.09 & 9.1\%  & $-$0.96  \\
	$8_{\mathrm{B}_1}$  & $-$0.62  & 64.8\%  & $-$0.33  & 34.1\%  & $-$0.01 & 1.1\%  & $-$0.95  \\
	$8_{\mathrm{B}_2}$  & $-$0.62  & 64.1\%  & $-$0.26  & 26.5\%  & $-$0.09 & 9.4\%  & $-$0.96  \\
	$9_\mathrm{tr}$     & $-$0.201 & 88.4\%  & $-$0.040 & 17.6\%  & 0.013 & $-$5.9\% & $-$0.227 \\
	$9_{\mathrm{B}_1}$  & $-$0.201 & 87.0\%  & $-$0.037 & 16.0\%  & 0.007 & $-$3.0\% & $-$0.231 \\
	$9_{\mathrm{B}_2}$  & $-$0.201 & 88.5\%  & $-$0.041 & 18.2\%  & 0.015 & $-$6.6\% & $-$0.227 \\
	$10_\mathrm{tr}$    & $-$44.60 & 84.9\%  & $-$13.07 & 24.9\%  & 5.16  & $-$9.8\% & $-$52.51 \\
	$10_{\mathrm{B}_1}$ & $-$44.60 & 82.4\%  & $-$11.23 & 20.7\%  & 1.72  & $-$3.2\% & $-$54.11 \\
	$10_{\mathrm{B}_2}$ & $-$44.60 & 85.0\%  & $-$12.57 & 23.9\%  & 4.67  & $-$8.9\% & $-$52.49 \\
	$11_\mathrm{tr}$    & 38.38  & 85.4\%  & 10.64  & 23.7\%  & $-$4.08 & $-$9.1\% & 44.94  \\
	$11_{\mathrm{B}_1}$ & 38.38  & 81.1\%  & 10.64  & 22.5\%  & $-$1.72 & $-$3.6\% & 47.30  \\
	$11_{\mathrm{B}_2}$ & 38.38  & 83.9\%  & 10.54  & 23.0\%  & $-$3.19 & $-$7.0\% & 45.73  \\
	$12_\mathrm{tr}$    & $-$5.00  & 118.4\% & 0.84   & $-$19.9\% & $-$0.06 & 1.5\%  & $-$4.22  \\
	$12_{\mathrm{B}_1}$ & $-$5.00  & 112.9\% & 0.55   & $-$12.5\% & 0.02  & $-$0.5\% & $-$4.43  \\
	$12_{\mathrm{B}_2}$ & $-$5.00  & 113.8\% & 0.66   & $-$15.0\% & $-$0.05 & 1.1\%  & $-$4.39  \\
	$13_\mathrm{tr}$    & $-$9.51  & 64.8\%  & $-$3.84  & 26.2\%  & $-$1.32 & 9.0\%  & $-$14.67 \\
	$13_{\mathrm{B}_1}$ & $-$9.51  & 65.3\%  & $-$4.92  & 33.8\%  & $-$0.14 & 0.9\%  & $-$14.57 \\
	$13_{\mathrm{B}_2}$ & $-$9.51  & 64.4\%  & $-$3.98  & 26.9\%  & $-$1.29 & 8.7\%  & $-$14.78 \\
	$14_\mathrm{tr}$    & $-$16.22 & 66.0\%  & $-$6.10  & 24.8\%  & $-$2.26 & 9.2\%  & $-$24.58 \\
	$14_{\mathrm{B}_1}$ & $-$16.22 & 66.6\%  & $-$6.19  & 25.4\%  & $-$1.94 & 8.0\%  & $-$24.35 \\
	$14_{\mathrm{B}_2}$ & $-$16.22 & 65.5\%  & $-$6.24  & 25.2\%  & $-$2.30 & 9.3\%  & $-$24.75 \\
	$15_\mathrm{tr}$    & $-$19.59 & 123.5\% & 3.26   & $-$20.5\% & 0.47  & $-$3.0\% & $-$15.86 \\
	$15_{\mathrm{B}_1}$ & $-$19.59 & 122.6\% & 3.07   & $-$19.2\% & 0.54  & $-$3.4\% & $-$15.97 \\
	$15_{\mathrm{B}_2}$ & $-$19.59 & 126.9\% & 3.77   & $-$24.4\% & 0.38  & $-$2.5\% & $-$15.43 \\
	$16_\mathrm{tr}$    & 3.26   & 85.0\%  & 0.77   & 20.1\%  & $-$0.20 & $-$5.1\% & 3.83   \\
	$16_{\mathrm{B}_1}$ & 3.26   & 84.7\%  & 0.73   & 18.9\%  & $-$0.14 & $-$3.6\% & 3.85   \\
	$16_{\mathrm{B}_2}$ & 3.26   & 86.3\%  & 0.73   & 19.3\%  & $-$0.21 & $-$5.6\% & 3.77   \\
	$17_\mathrm{tr}$    & $-$2.77  & 66.0\%  & $-$1.03  & 24.6\%  & $-$0.40 & 9.4\%  & $-$4.19  \\
	$17_{\mathrm{B}_1}$ & $-$2.77  & 67.3\%  & $-$1.09  & 26.6\%  & $-$0.25 & 6.1\%  & $-$4.11  \\
	$17_{\mathrm{B}_2}$ & $-$2.77  & 65.8\%  & $-$1.04  & 24.8\%  & $-$0.40 & 9.4\%  & $-$4.20
\end{longtable}
\end{table}

\begin{table}[H]
\begin{longtable}{lrrrrrrrrr}
\caption{Same as Table \ref{table2}, except for the NNLO fit.}\label{table8}\\
	\hline\hline  $i$&\multicolumn{1}{c}{LO}&$\mathrm{Pct}_\mathrm{LO}$&\multicolumn{1}{c}{NLO}&$\mathrm{Pct}_\mathrm{NLO}$&\multicolumn{1}{c}{NNLO}&$\mathrm{Pct}_\mathrm{NNLO}$&\multicolumn{1}{c}{HO}&$\mathrm{Pct}_\mathrm{HO}$ &Theory\\
	\hline%\addlinespace[2pt]
	\endfirsthead
	\hline\hline Obs&\multicolumn{1}{c}{LO}&$\mathrm{Pct}_\mathrm{LO}$&\multicolumn{1}{c}{NLO}&$\mathrm{Pct}_\mathrm{NLO}$&\multicolumn{1}{c}{NNLO}&$\mathrm{Pct}_\mathrm{NNLO}$&\multicolumn{1}{c}{HO}&$\mathrm{Pct}_\mathrm{HO}$ &Theory\\
	\hline%\addlinespace[2pt]
	\endhead
	\hline\hline
	\endfoot
	\hline\hline
	\endlastfoot
	$1_\mathrm{tr}$     & $-$25.00 & 69.8\%  & $-$7.25  & 20.3\%  & $-$2.54  & 7.1\%   & $-$1.01  & 2.8\%   & $-$35.80 \\
	$1_{\mathrm{B}_1}$  & $-$25.00 & 73.1\%  & $-$7.00  & 20.5\%  & $-$2.07  & 6.0\%   & $-$0.15  & 0.4\%   & $-$34.22 \\
	$1_{\mathrm{B}_2}$  & $-$25.00 & 71.0\%  & $-$7.12  & 20.2\%  & $-$2.33  & 6.6\%   & $-$0.74  & 2.1\%   & $-$35.19 \\
	$2_\mathrm{tr}$     & 0.110  & 63.5\%  & 0.052  & 30.1\%  & 0.015  & 8.9\%   & $-$0.004 & $-$2.5\%  & 0.173  \\
	$2_{\mathrm{B}_1}$  & 0.110  & 64.6\%  & 0.054  & 32.0\%  & 0.005  & 3.2\%   & 0.000  & 0.2\%   & 0.170  \\
	$2_{\mathrm{B}_2}$  & 0.110  & 64.0\%  & 0.052  & 30.6\%  & 0.014  & 8.1\%   & $-$0.005 & $-$2.7\%  & 0.172  \\
	$3_\mathrm{tr}$     & $-$0.359 & 130.2\% & 0.078  & $-$28.2\% & 0.041  & $-$15.0\% & $-$0.036 & 13.0\%  & $-$0.276 \\
	$3_{\mathrm{B}_1}$  & $-$0.359 & 128.9\% & 0.083  & $-$29.8\% & $-$0.003 & 0.9\%   & 0.000  & 0.0\%   & $-$0.279 \\
	$3_{\mathrm{B}_2}$  & $-$0.359 & 129.0\% & 0.079  & $-$28.5\% & 0.038  & $-$13.8\% & $-$0.037 & 13.3\%  & $-$0.279 \\
	$4_\mathrm{tr}$     & 0.395  & 65.5\%  & 0.192  & 31.9\%  & $-$0.03  & $-$5.1\%  & 0.046  & 7.6\%   & 0.603  \\
	$4_{\mathrm{B}_1}$  & 0.395  & 67.6\%  & 0.193  & 33.0\%  & $-$0.01  & $-$0.9\%  & 0.002  & 0.3\%   & 0.584  \\
	$4_{\mathrm{B}_2}$  & 0.395  & 66.9\%  & 0.188  & 31.8\%  & $-$0.04  & $-$6.6\%  & 0.047  & 7.9\%   & 0.590  \\
	$5_\mathrm{tr}$     & 17.175 & 63.0\%  & 7.66   & 28.1\%  & 1.408  & 5.2\%   & 1.034  & 3.8\%   & 27.28  \\
	$5_{\mathrm{B}_1}$  & 17.175 & 63.4\%  & 7.54   & 27.8\%  & 2.137  & 7.9\%   & 0.244  & 0.9\%   & 27.09  \\
	$5_{\mathrm{B}_2}$  & 17.175 & 62.2\%  & 7.66   & 27.8\%  & 1.669  & 6.0\%   & 1.097  & 4.0\%   & 27.60  \\
	$6_\mathrm{tr}$     & $-$0.362 & 69.1\%  & $-$0.201 & 38.3\%  & 0.03   & $-$5.8\%  & 0.009  & $-$1.6\%  & $-$0.524 \\
	$6_{\mathrm{B}_1}$  & $-$0.362 & 65.9\%  & $-$0.196 & 35.7\%  & 0.01   & $-$1.6\%  & 0.000  & $-$0.1\%  & $-$0.550 \\
	$6_{\mathrm{B}_2}$  & $-$0.362 & 66.2\%  & $-$0.200 & 36.6\%  & 0.01   & $-$1.6\%  & 0.007  & $-$1.2\%  & $-$0.547 \\
	$7_\mathrm{tr}$     & $-$1.054 & 70.9\%  & $-$0.321 & 21.6\%  & $-$0.048 & 3.3\%   & $-$0.063 & 4.3\%   & $-$1.49  \\
	$7_{\mathrm{B}_1}$  & $-$1.054 & 73.5\%  & $-$0.320 & 22.3\%  & $-$0.060 & 4.2\%   & 0.000  & 0.0\%   & $-$1.43  \\
	$7_{\mathrm{B}_2}$  & $-$1.054 & 71.7\%  & $-$0.320 & 21.8\%  & $-$0.049 & 3.3\%   & $-$0.046 & 3.1\%   & $-$1.47  \\
	$8_\mathrm{tr}$     & $-$0.618 & 64.7\%  & $-$0.251 & 26.2\%  & $-$0.092 & 9.6\%   & 0.005  & $-$0.5\%  & $-$0.955 \\
	$8_{\mathrm{B}_1}$  & $-$0.618 & 63.8\%  & $-$0.327 & 33.8\%  & $-$0.022 & 2.3\%   & $-$0.001 & 0.1\%   & $-$0.968 \\
	$8_{\mathrm{B}_2}$  & $-$0.618 & 63.6\%  & $-$0.262 & 27.0\%  & $-$0.097 & 9.9\%   & 0.006  & $-$0.6\%  & $-$0.971 \\
	$9_\mathrm{tr}$     & $-$0.201 & 88.4\%  & $-$0.040 & 17.6\%  & $-$0.012 & 5.3\%   & 0.025  & $-$11.2\% & $-$0.227 \\
	$9_{\mathrm{B}_1}$  & $-$0.201 & 88.7\%  & $-$0.037 & 16.2\%  & 0.011  & $-$4.8\%  & 0.000  & $-$0.1\%  & $-$0.226 \\
	$9_{\mathrm{B}_2}$  & $-$0.201 & 89.0\%  & $-$0.039 & 17.2\%  & $-$0.012 & 5.2\%   & 0.026  & $-$11.3\% & $-$0.225 \\
	$10_\mathrm{tr}$    & $-$44.60 & 84.9\%  & $-$13.07 & 24.9\%  & 2.79   & $-$5.3\%  & 2.37   & $-$4.5\%  & $-$52.51 \\
	$10_{\mathrm{B}_1}$ & $-$44.60 & 83.9\%  & $-$11.60 & 21.8\%  & 2.89   & $-$5.4\%  & 0.17   & $-$0.3\%  & $-$53.14 \\
	$10_{\mathrm{B}_2}$ & $-$44.60 & 85.4\%  & $-$11.83 & 22.6\%  & 2.36   & $-$4.5\%  & 1.85   & $-$3.5\%  & $-$52.21 \\
	$11_\mathrm{tr}$    & 38.38  & 85.4\%  & 10.64  & 23.7\%  & $-$1.96  & $-$4.4\%  & $-$2.12  & $-$4.7\%  & 44.94  \\
	$11_{\mathrm{B}_1}$ & 38.38  & 82.6\%  & 10.20  & 21.9\%  & $-$2.02  & $-$4.3\%  & $-$0.09  & $-$0.2\%  & 46.47  \\
	$11_{\mathrm{B}_2}$ & 38.38  & 83.6\%  & 11.22  & 24.4\%  & $-$1.65  & $-$3.6\%  & $-$2.04  & $-$4.4\%  & 45.90  \\
	$12_\mathrm{tr}$    & $-$5.00  & 118.4\% & 0.84   & $-$19.9\% & 0.12   & $-$2.7\%  & $-$0.18  & 4.2\%   & $-$4.22  \\
	$12_{\mathrm{B}_1}$ & $-$5.00  & 113.5\% & 0.57   & $-$12.9\% & 0.03   & $-$0.6\%  & 0.00   & 0.0\%   & $-$4.41  \\
	$12_{\mathrm{B}_2}$ & $-$5.00  & 113.9\% & 0.87   & $-$19.9\% & $-$0.06  & 1.3\%   & $-$0.21  & 4.7\%   & $-$4.39  \\
	$13_\mathrm{tr}$    & $-$9.51  & 64.8\%  & $-$3.84  & 26.2\%  & $-$0.71  & 4.9\%   & $-$0.61  & 4.1\%   & $-$14.67 \\
	$13_{\mathrm{B}_1}$ & $-$9.51  & 64.7\%  & $-$4.84  & 32.9\%  & $-$0.34  & 2.3\%   & $-$0.03  & 0.2\%   & $-$14.71 \\
	$13_{\mathrm{B}_2}$ & $-$9.51  & 64.4\%  & $-$4.52  & 30.6\%  & $-$0.15  & 1.0\%   & $-$0.58  & 3.9\%   & $-$14.76 \\
	$14_\mathrm{tr}$    & $-$16.22 & 66.0\%  & $-$6.10  & 24.8\%  & $-$1.11  & 4.5\%   & $-$1.15  & 4.7\%   & $-$24.58 \\
	$14_{\mathrm{B}_1}$ & $-$16.22 & 66.7\%  & $-$6.16  & 25.3\%  & $-$1.76  & 7.2\%   & $-$0.16  & 0.7\%   & $-$24.30 \\
	$14_{\mathrm{B}_2}$ & $-$16.22 & 65.7\%  & $-$6.10  & 24.7\%  & $-$1.21  & 4.9\%   & $-$1.17  & 4.7\%   & $-$24.69 \\
	$15_\mathrm{tr}$    & $-$19.59 & 123.5\% & 3.26   & $-$20.5\% & 1.12   & $-$7.1\%  & $-$0.65  & 4.1\%   & $-$15.86 \\
	$15_{\mathrm{B}_1}$ & $-$19.59 & 125.5\% & 2.88   & $-$18.5\% & 1.03   & $-$6.6\%  & 0.07   & $-$0.4\%  & $-$15.61 \\
	$15_{\mathrm{B}_2}$ & $-$19.59 & 126.0\% & 3.63   & $-$23.3\% & 0.95   & $-$6.1\%  & $-$0.53  & 3.4\%   & $-$15.54 \\
	$16_\mathrm{tr}$    & 3.26   & 85.0\%  & 0.77   & 20.1\%  & $-$0.35  & $-$9.3\%  & 0.16   & 4.1\%   & 3.83   \\
	$16_{\mathrm{B}_1}$ & 3.26   & 86.7\%  & 0.76   & 20.3\%  & $-$0.26  & $-$6.9\%  & 0.00   & $-$0.1\%  & 3.76   \\
	$16_{\mathrm{B}_2}$ & 3.26   & 86.6\%  & 0.73   & 19.5\%  & $-$0.38  & $-$10.1\% & 0.15   & 4.1\%   & 3.76   \\
	$17_\mathrm{tr}$    & $-$2.77  & 66.0\%  & $-$1.03  & 24.6\%  & $-$0.27  & 6.5\%   & $-$0.12  & 3.0\%   & $-$4.19  \\
	$17_{\mathrm{B}_1}$ & $-$2.77  & 66.9\%  & $-$1.07  & 25.9\%  & $-$0.27  & 6.6\%   & $-$0.03  & 0.6\%   & $-$4.14  \\
	$17_{\mathrm{B}_2}$ & $-$2.77  & 65.7\%  & $-$1.04  & 24.7\%  & $-$0.27  & 6.4\%   & $-$0.14  & 3.3\%   & $-$4.21
\end{longtable}
\end{table}

\section{An extra example}

Eq. \eqref{A2} gives the functions of an extra example. For convenience, the parameters with the same name in the different functions are different. The values of $b_i$ and $a_i^\mathrm{LO}$ can be found in Table \ref{pA2}. The values of $a_i^\mathrm{NLO}$ and $a_i^\mathrm{NNLO}$ are given in the second column in Tables \ref{table4} and \ref{table10}, which are marked by a subscript ``tr''.
\begin{align}
	\mathrm O_1={}&-b_1\exp(b_2(a^\mathrm{NNLO}_{1}t^3b_7 - a^\mathrm{NLO}_{3}t^2b_4 - a^\mathrm{NLO}_{6}t^2b_3 - b_5\ln(-a^\mathrm{NLO}_{1}t^2b_6 + 1) - a^\mathrm{LO}_1t)) + b_1, \label{A2}\\
	\mathrm O_2={}&-b_1\exp(b_2(-a^\mathrm{NNLO}_{2}t^3b_7 + a^\mathrm{NLO}_{1}t^2b_5 - a^\mathrm{NLO}_{2}t^2b_3 - a^\mathrm{NLO}_{3}t^2b_4 - a^\mathrm{NLO}_{4}t^2b_6 + a^\mathrm{LO}_2t)) + b_1,\notag\\
	\mathrm O_3={}&-b_1\exp(b_2) + b_1\exp(a^\mathrm{NLO}_{7}t^2b_6 + b_2\exp(a^\mathrm{NNLO}_{3}t^3b_3) + b_4\sin(a^\mathrm{NLO}_{8}b_5) + a^\mathrm{LO}_3t),\notag\\
	\mathrm O_4={}&b_1\ln(-b_2b_3(a^\mathrm{NNLO}_{4}t^3b_5 + a^\mathrm{NLO}_{3}t^2b_4 - a^\mathrm{NLO}_{5}t^2b_7 - a^\mathrm{LO}_4t) + 1),\notag\\
	\mathrm O_5={}&b_1\exp(a^\mathrm{NLO}_{6}t^2b_2 - a^\mathrm{LO}_5t + \exp(-a^\mathrm{NNLO}_{5}t^3b_3)) - \exp(1)b_1,\notag\\
	\mathrm O_6={}&b_1\sin(b_2\exp(b_3)) - b_1\sin(b_2\exp(a^\mathrm{NLO}_{2}t^2b_6b_7 + a^\mathrm{NLO}_{2}t^2b_8 + b_3(-b_4a^\mathrm{LO}_6t + \exp(a^\mathrm{NNLO}_{6}t^3b_5)))),\notag\\
	\mathrm O_7={}&b_1\ln(-\exp(1)b_2) - b_1\ln(-b_2\exp(-a^\mathrm{NNLO}_{7}t^3b_6 + b_3\ln(a^\mathrm{NLO}_{1}t^2b_4 + 1) + \exp(-a^\mathrm{NLO}_{4}t^2b_5)) + a^\mathrm{LO}_7t),\notag\\
	\mathrm O_8={}&b_1\exp(\sin(a^\mathrm{NNLO}_{8}t^3b_3) + \sin(a^\mathrm{NNLO}_{8}t^3b_4 + a^\mathrm{NLO}_{2}t^2b_5 + a^\mathrm{LO}_8t)) - b_1,\notag\\
	\mathrm O_9={}&-b_1\exp(-a^\mathrm{NNLO}_{9}t^3b_9 - a^\mathrm{NLO}_{4}t^2b_8 + a^\mathrm{NLO}_{8}b_{10} + b_2(a^\mathrm{NNLO}_{9}t^3b_6 - a^\mathrm{NNLO}_{9}t^3b_7 + a^\mathrm{NLO}_{2}t^2b_5 + a^\mathrm{NLO}_{7}t^2b_3\notag\\
	& - a^\mathrm{NLO}_{8}t^2b_4) - a^\mathrm{LO}_9t) + b_1,\notag\\
	\mathrm O_{10}={}&-b_1\exp(a^\mathrm{NNLO}_{10}t^3b_4 + a^\mathrm{NNLO}_{10}t^3 + a^\mathrm{NLO}_{3}t^2 + a^\mathrm{NLO}_{5}t^2b_5 - b_2\sin(a^\mathrm{NNLO}_{10}t^3b_3 - a^\mathrm{NLO}_{7}) - b_6\ln(a^\mathrm{NLO}_{8}b_7 + 1) \notag\\
	&+ a^\mathrm{LO}_{10}t) + b_1,\notag\\
	\mathrm O_{11}={}&b_1\ln(a^\mathrm{NNLO}_{11}t^3b_2b_3 + a^\mathrm{NNLO}_{11}t^3b_7 + a^\mathrm{NLO}_{7}t^2 - a^\mathrm{NLO}_{8}t^2b_6 - b_4\ln(a^\mathrm{NLO}_{4}b_5 + 1) + a^\mathrm{LO}_{11}t + 1),\notag\\
	\mathrm O_{12}={}&b_1\ln(-a^\mathrm{NNLO}_{12}t^3b_3 - a^\mathrm{NLO}_{3}t^2b_2 - a^\mathrm{NLO}_{6}t^2b_4 - a^\mathrm{LO}_{12}t + 1),\notag\\
	\mathrm O_{13}={}&b_1\exp(b_2(-a^\mathrm{NLO}_{4}t^2b_3 - a^\mathrm{NLO}_{4}t^2b_7 - a^\mathrm{NLO}_{5}t^2b_6 - b_4\ln(-a^\mathrm{NNLO}_{13}t^3b_5 + 1) - a^\mathrm{LO}_{13}t) + \sin(a^\mathrm{NNLO}_{13}t^3b_8)) - b_1,\notag\\
	\mathrm O_{14}={}&b_1\sin(a^\mathrm{NNLO}_{14}t^3b_6 + a^\mathrm{NLO}_{3}t^2b_2b_3 - a^\mathrm{NLO}_{3}t^2b_4 + a^\mathrm{NLO}_{6} - a^\mathrm{LO}_{14}t + \exp(a^\mathrm{NLO}_{7}b_5)) - b_1\sin(1),\notag\\
	\mathrm O_{15}={}&-b_1\exp(a^\mathrm{NLO}_{4}t^2b_2 - a^\mathrm{NLO}_{6}t^2b_3 - b_4\exp(-a^\mathrm{NNLO}_{15}t^3b_5) - a^\mathrm{LO}_{15}t) + b_1\exp(-b_4),\notag\\
	\mathrm O_{16}={}&-b_1\exp(a^\mathrm{NLO}_{1}t^2b_5 - a^\mathrm{NLO}_{2}t^2b_6 + b_2\ln(b_3\sin(a^\mathrm{NNLO}_{16}t^3b_4) + 1) + b_7a^\mathrm{LO}_{16}t) + b_1,\notag\\
	\mathrm O_{17}={}&-b_1\sin(a^\mathrm{NLO}_{6}t^2b_2 - a^\mathrm{NLO}_{6}t^2b_4b_5 + a^\mathrm{NLO}_{8}t^2b_3 + a^\mathrm{NLO}_{8}t^2b_6 + a^\mathrm{NLO}_{8}t^2 + b_7\sin(a^\mathrm{NLO}_{6}b_8) - a^\mathrm{LO}_{17}t \notag\\
	&+ \exp(b_9\exp(a^\mathrm{NNLO}_{17}t^3b_{10})))+ b_1\sin(\exp(b_9)).\notag
\end{align}

\begin{table}[!h]
	\caption{The values of parameters $b_i$ and $a_i^\mathrm{LO}$ in Eq. \eqref{A2}. Because the values are exact, more significant digits are given.}\label{pA2}
	\begin{ruledtabular}
		\begin{tabular}{lrrrrrrrrrrr}
			& $10^2b_{1}$ & $10^2b_{2}$ & $10^2b_{3}$ & $10^2b_{4}$ & $10^2b_{5}$ & $10^2b_{6}$ & $10^2b_{7}$ & $10^2b_{8}$ & $10^2b_{9}$ & $10^2b_{10}$ & $10^2a_i^\mathrm{LO}$    \\\hline
			O$_{1}$  & 63.05265      & 32.00250      & 81.10103      & -4.21326      & 9.76488       & 9.75977       & -10.50120     &               &               &                & 87.53844   \\
			O$_{2}$  & 64.53738      & 37.56237      & 76.09650      & 23.19692      & 12.04466      & 8.60590       & -17.97096     &               &               &                & 93.73702   \\
			O$_{3}$  & -57.42876     & -59.35961     & -8.96750      & 0.87049       & -0.32251      & -7.91228      &               &               &               &                & 28.99127   \\
			O$_{4}$  & 97.98420      & 77.16734      & 45.41523      & 9.96670       & -16.83649     &               & 6.74978       &               &               &                & 99.84650   \\
			O$_{5}$  & 0.03801       & -5.95901      & 8.09908       &               &               &               &               &               &               &                & -53.93489  \\
			O$_{6}$  & 98.71049      & 47.71969      & 46.26914      & 79.66420      & 1.22979       & 9.55114       & 9.55114       & 5.48347       &               &                & -79.66420  \\
			O$_{7}$  & -1.32834      & -10.47658     & -11.67140     & -11.66142     & -10.10252     & -2.75218      &               &               &               &                & -13.43084  \\
			O$_{8}$  & 28.98778      & 30.00000      & -14.09222     & 10.12459      & -9.62538      &               &               &               &               &                & 65.10724   \\
			O$_{9}$  & 12.64510      & 48.89734      & 54.66916      & 58.93231      & 32.94741      & 58.84268      & 41.15732      & 62.40324      & 31.81133      & 31.80857       & 28.76825   \\
			O$_{10}$ & -32.96860     & -56.84774     & -21.40418     & -62.38096     & 283.66063     & 1.77682       & 1.76278       &               &               &                & -29.48017  \\
			O$_{11}$ & -26.27893     & -56.60695     & -56.60695     & -45.23316     & -44.77793     & -62.16478     & -30.87118     &               &               &                & -15.35752  \\
			O$_{12}$ & -1.11391      & 7.94288       & 7.25529       & 21.70861      &               &               &               &               &               &                & -46.92748  \\
			O$_{13}$ & 77.33092      & 19.27071      & 13.39227      & 19.97001      & 9.94301       & 20.40295      & 13.39227      & 7.72936       &               &                & 94.86854   \\
			O$_{14}$ & 51.83380      & 54.40546      & 14.59436      & 2.86290       & 11.28245      & -7.39685      &               &               &               &                & 42.14291   \\
			O$_{15}$ & 40.74084      & 10.05237      & 77.29493      & 45.99355      & -7.61843      &               &               &               &               &                & 35.41254   \\
			O$_{16}$ & -79.89735     & 49.99007      & 19.97474      & 9.94983       & 10.85088      & 8.37558       & 10.56758      &               &               &                & -100.05312 \\
			O$_{17}$ & -29.37166     & -41.98590     & -33.97179     & -46.05588     & -46.05588     & -33.97179     & -54.06307     & -54.04960     & -51.26040     & -51.23417      & -23.43300
		\end{tabular}
	\end{ruledtabular}
\end{table}

\begin{table}[htbp]
	\caption{The fitting parameters in Model B$_2$. The subscripts NLO and NNLO represent the NLO and NNLO fit, respectively. The definitions of these parameters are in Eqs. \eqref{eq3} -- \eqref{dp} and the text below them.}\label{app table1}
	\begin{ruledtabular}
		\begin{tabular}{lcccccc}
			$i$
			&$\mu_{e,\mathrm{NLO}}$&$\sigma_{e,\mathrm{NLO}}$&$p_\mathrm{NLO}$
			&$\mu_{e,\mathrm{NNLO}}$&$\sigma_{e,\mathrm{NNLO}}$&$p_\mathrm{NNLO}$\\\hline
			1  & 0.010 & 0.050 & 1 & 0.040 & 0.020 & 0 \\
			2  & 0.070 & 0.050 & 0 & 0.030 & 0.020 & 0 \\
			3  & 0.050 & 0.050 & 0 & 0.020 & 0.020 & 0 \\
			4  & 0.020 & 0.050 & 1 & 0.030 & 0.020 & 0 \\
			5  & 0.110 & 0.050 & 0 & 0.050 & 0.020 & 0 \\
			6  & 0.070 & 0.050 & 0 & 0.030 & 0.020 & 0 \\
			7  & 0.080 & 0.050 & 0 & 0.050 & 0.020 & 0 \\
			8  & 0.030 & 0.050 & 0 & 0.020 & 0.020 & 1 \\
			9  & 0.030 & 0.050 & 1 & 0.060 & 0.020 & 0 \\
			10 & 0.020 & 0.050 & 1 & 0.050 & 0.020 & 0 \\
			11 & 0.030 & 0.050 & 0 & 0.020 & 0.020 & 1 \\
			12 & 0.050 & 0.050 & 1 & 0.070 & 0.021 & 0 \\
			13 & 0.030 & 0.050 & 1 & 0.020 & 0.020 & 0 \\
			14 & 0.030 & 0.050 & 0 & 0.030 & 0.020 & 1 \\
			15 & 0.070 & 0.050 & 0 & 0.030 & 0.020 & 0 \\
			16 & 0.040 & 0.050 & 1 & 0.010 & 0.020 & 0 \\
			17 & 0.010 & 0.050 & 0 & 0.040 & 0.020 & 0
		\end{tabular}
	\end{ruledtabular}
\end{table}

\begin{table}[htbp]
	\caption{The prior of LECs in Model B$_2$. The subscripts NLO and NNLO represent the NLO and NNLO fit, respectively.}\label{app table2}
	\begin{ruledtabular}
		\begin{tabular}{lrrrr}
			$i$
			&$\mu_{a_i^\mathrm{NLO}}$&$\sigma_{a_i^\mathrm{NLO}}$					&$\mu_{a_i^\mathrm{NNLO}}$&$\sigma_{a_i^\mathrm{NNLO}}$\\\hline
			1  & 0.511  & 0.256 & 1.314  & 0.657 \\
			2  & 0.789  & 0.394 & $-$  0.446 & 0.223 \\
			3  & $-$  2.990 & 1.495 & $-$  0.379 & 0.190 \\
			4  & 0.268  & 0.134 & 0.670  & 0.335 \\
			5  & 1.000  & 0.500 & 0.652  & 0.326 \\
			6  & 0.169  & 0.085 & $-$  1.981 & 0.990 \\
			7  & $-$  0.190 & 0.095 & 1.433  & 0.716 \\
			8  & 0.407  & 0.203 & $-$  0.570 & 0.285 \\
			9  &        &       & 0.227  & 0.114 \\
			10 &        &       & $-$  0.174 & 0.087 \\
			11 &        &       & 1.169  & 0.585 \\
			12 &        &       & $-$  1.737 & 0.868 \\
			13 &        &       & $-$  0.227 & 0.114 \\
			14 &        &       & 0.250  & 0.125 \\
			15 &        &       & 0.880  & 0.440 \\
			16 &        &       & $-$  0.844 & 0.422 \\
			17 &        &       & 0.118  & 0.059
		\end{tabular}
	\end{ruledtabular}
\end{table}

\begin{table}[h]
	\renewcommand\arraystretch{1.3}
	\caption{The NLO and NNLO fitting results of $a^\mathrm{NLO}_i$ in extra examples. Row 2 is the truth value of $a^\mathrm{NLO}_{i}$. Rows 3, 6 and 9 are the NLO fitting results of Model A, B$_1$ and B$_2$, respectively. Rows 12, 15 and 18 are the NNLO fitting results of Model A, B$_1$ and B$_2$, respectively. The percentage $\mathrm{Pct}_{\mathrm{A, B_1, B_2}}$ is defined in Eq. \eqref{equ:2}, and the ratio $\mathrm{Pct\sigma}_{\mathrm{A, B_1, B_2}}$ is defined in Eq. \eqref{equ:3}.}\label{table4}
	\begin{ruledtabular}
		\begin{tabular}{lcccccccccc}
			$i$                                 & 1          & 2          & 3             & 4          & 5          & 6          & 7             & 8          & WAIC     & LOO      \\\hline
			$a^\mathrm{NLO}_{i,\mathrm{tr}}$    & 0.44       & 0.84       & $-$2.84       & 0.3        & 0.92       & 0.22       & $-$0.23       & 0.44       &          &          \\\hline

			&            &            &               &            & NLO        &            &               &            &          &          \\
			$a^\mathrm{NLO}_{i,\mathrm{tr}}$    & 0.44       & 0.84       & $-$2.84       & 0.3        & 0.92       & 0.22      & $-$0.23       & 0.44       &           &           \\
			$a^\mathrm{NLO}_{i,\mathrm{A}}$     & 0.494(30)  & 0.984(30)  & $-$3.068(70)  & 0.374(12)  & 1.004(25)  & 0.222(6)  & $-$0.202(10)  & 0.421(14)  & $-$42.087 & $-$49.194 \\
			$\mathrm{Pct}_{\mathrm{A}}$         & 12.3\%     & 17.1\%     & 8.0\%         & 24.7\%     & 9.1\%      & 0.9\%     & $-$12.2\%     & $-$4.3\%   &           &           \\
			$\mathrm{Pct\sigma}_{\mathrm{A}}$   & 1.8        & 4.8        & $-$3.3        & 6.2        & 3.4        & 0.3       & 2.8           & $-$1.4     &           &           \\
			$a^\mathrm{NLO}_{i,\mathrm{B_1}}$   & 0.421(116) & 0.888(116) & $-$2.801(256) & 0.341(46)  & 0.909(90)  & 0.213(23) & $-$0.206(39)  & 0.416(57)  & 20.389    & 14.007    \\
			$\mathrm{Pct}_{\mathrm{B_1}}$       & $-$4.3\%   & 5.7\%      & $-$1.4\%      & 13.7\%     & $-$1.2\%   & $-$3.2\%  & $-$10.4\%     & $-$5.5\%   &           &           \\
			$\mathrm{Pct\sigma}_{\mathrm{B_1}}$ & $-$0.2     & 0.4        & 0.2           & 0.9        & $-$0.1     & $-$0.3    & 0.6           & $-$0.4     &           &           \\
			$a^\mathrm{NLO}_{i,\mathrm{B_2}}$   & 0.449(38)  & 0.877(40)  & $-$2.962(81)  & 0.310(14)  & 0.962(28)  & 0.227(7)  & $-$0.235(11)  & 0.445(15)  & 21.448    & 18.136    \\
			$\mathrm{Pct}_{\mathrm{B_2}}$       & 2.0\%      & 4.4\%      & 4.3\%         & 3.3\%      & 4.6\%      & 3.2\%     & 2.2\%         & 1.1\%      &           &           \\
			$\mathrm{Pct\sigma}_{\mathrm{B_2}}$ & 0.2        & 0.9        & $-$1.5        & 0.7        & 1.5        & 1.0       & $-$0.5        & 0.3        &           &           \\ \hline
			&            &            &               &            & NNLO       &            &               &            &          &          \\
			$a^\mathrm{NLO}_{i,\mathrm{A}}$     & 0.271(127) & 0.733(95)  & $-$2.391(467) & 0.638(172) & 0.737(212) & 0.223(45) & $-$0.241(247) & 0.574(399) & 8.610     & 2.528     \\
			$\mathrm{Pct}_{\mathrm{A}}$         & $-$38.41\% & $-$12.74\% & $-$15.81\%    & 112.67\%   & $-$19.89\% & 1.36\%    & 4.78\%        & 30.45\%    &           &           \\
			$\mathrm{Pct\sigma}_{\mathrm{A}}$   & $-$1.33    & $-$1.13    & 0.96          & 1.97       & $-$0.86    & 0.07      & $-$0.04       & 0.34       &           &           \\
			$a^\mathrm{NLO}_{i,\mathrm{B_1}}$   & 0.410(71)  & 0.878(63)  & $-$2.866(147) & 0.345(31)  & 0.932(52)  & 0.215(16) & $-$0.205(24)  & 0.419(37)  & 50.309    & 44.176    \\
			$\mathrm{Pct}_{\mathrm{B_1}}$       & $-$6.82\%  & 4.52\%     & 0.92\%        & 15.00\%    & 1.30\%     & $-$2.27\% & $-$10.87\%    & $-$4.77\%  &           &           \\
			$\mathrm{Pct\sigma}_{\mathrm{B_1}}$ & $-$0.42    & 0.60       & $-$0.18       & 1.45       & 0.23       & $-$0.31   & 1.04          & $-$0.57    &           &           \\
			$a^\mathrm{NLO}_{i,\mathrm{B_2}}$   & 0.447(53)  & 0.877(46)  & $-$2.933(121) & 0.328(30)  & 0.950(43)  & 0.220(9)  & $-$0.223(19)  & 0.431(28)  & 54.235    & 49.085    \\
			$\mathrm{Pct}_{\mathrm{B_2}}$       & 1.6\%      & 4.4\%      & 3.3\%         & 9.3\%      & 3.3\%      & 0.0\%     & $-$3.0\%      & $-$2.0\%   &           &           \\
			$\mathrm{Pct\sigma}_{\mathrm{B_2}}$ & 0.13       & 0.80       & $-$0.77       & 0.93       & 0.70       & 0.00      & 0.37          & $-$0.32    &           &

		\end{tabular}
	\end{ruledtabular}
\end{table}

\begin{table}[H]
	\caption{The comparison of the NLO fitting values for extra examples. The subscripts tr, exp, B$_1$ and B$_2$ in the first column represent the truth values, the experimental values, the theoretical values from Model B$_1$ and Model B$_2$, respectively. The experimental values in the third column are sampled from the truth values.  $O_i$ is defined in Eq. \eqref{A2}.}\label{table6}
	\begin{ruledtabular}
		\begin{tabular}{lrrrr}
			$i$ & $10^2O_{i,\mathrm {tr}}$ & $10^2O_{i,\mathrm {exp}}$~~~~~ & $10^2O_{i,\mathrm {B_1}}$~~~~~ & $10^2O_{i,\mathrm {B_2}}$~~~~~ \\
			\hline
			1  & 21.297  & 21.816  $\pm$  0.319 & 21.398  $\pm$  0.617  & 21.522  $\pm$  0.144 \\
			2  & $-$26.253 & $-$26.494 $\pm$  0.394 & $-$26.568 $\pm$  2.603  & $-$26.408 $\pm$  0.984 \\
			3  & $-$10.613 & $-$10.697 $\pm$  0.159 & $-$10.897 $\pm$  0.204  & $-$10.674 $\pm$  0.108 \\
			4  & 40.713  & 40.058  $\pm$  0.611 & 40.004  $\pm$  0.901  & 41.210  $\pm$  0.347 \\
			5  & 0.063   & 0.064   $\pm$  0.001 & 0.065   $\pm$  0.001  & 0.064   $\pm$  0.001 \\
			6  & $-$18.401 & $-$19.036 $\pm$  0.276 & $-$19.417 $\pm$  0.528  & $-$18.837 $\pm$  0.267 \\
			7  & $-$0.675  & $-$0.658  $\pm$  0.010 & $-$0.661  $\pm$  0.013  & $-$0.663  $\pm$  0.009 \\
			8  & 22.194  & 21.941  $\pm$  0.333 & 22.357  $\pm$  0.423  & 21.991  $\pm$  0.191 \\
			9  & 4.492   & 4.513   $\pm$  0.067 & 4.527   $\pm$  0.453  & 4.510   $\pm$  0.146 \\
			10 & $-$11.726 & $-$11.770 $\pm$  0.176 & $-$11.782 $\pm$  11.942 & $-$11.768 $\pm$  3.746 \\
			11 & 4.687   & 4.789   $\pm$  0.070 & 4.781   $\pm$  1.406  & 4.780   $\pm$  0.388 \\
			12 & $-$0.624  & $-$0.644  $\pm$  0.009 & $-$0.633  $\pm$  0.032  & $-$0.638  $\pm$  0.010 \\
			13 & $-$17.245 & $-$17.328 $\pm$  0.259 & $-$17.088 $\pm$  0.372  & $-$17.412 $\pm$  0.154 \\
			14 & $-$13.943 & $-$14.023 $\pm$  0.209 & $-$14.211 $\pm$  0.766  & $-$13.939 $\pm$  0.251 \\
			15 & 10.468  & 10.646  $\pm$  0.157 & 10.770  $\pm$  0.482  & 10.617  $\pm$  0.206 \\
			16 & $-$10.156 & $-$10.323 $\pm$  0.152 & $-$10.305 $\pm$  1.270  & $-$10.318 $\pm$  0.437 \\
			17 & 6.774   & 6.756   $\pm$  0.102 & 6.731   $\pm$  0.483  & 6.763   $\pm$  0.131
		\end{tabular}
	\end{ruledtabular}
\end{table}

\begin{table}[h]
	\caption{The NNLO fitting results of extra examples. Column 2 is the truth value of $a^\mathrm{NNLO}_{i}$. Column 3, 6 and 9 are the results of Model A, B$_1$ and B$_2$, respectively. The percentage $\mathrm{Pct}_{\mathrm{A, B_1, B_2}}$ is defined in Eq. \eqref{equ:2}, and the ratio $\mathrm{Pct\sigma}_{\mathrm{A, B_1, B_2}}$ is defined in Eq. \eqref{equ:3}. PM is defined in Eq. \eqref{equ:4}.}\label{table10}
	\begin{ruledtabular}
		\begin{tabular}{lrrrrrrrrrr}
			$i$         &    $a^\mathrm{NNLO}_{i,\mathrm{tr}}$    &    $a^\mathrm{NNLO}_{i,\mathrm{A}}$ &    $\mathrm{Pct}_{\mathrm{A}}$    &    $\mathrm{Pct\sigma}_{\mathrm{A}}$    & $a^\mathrm{NNLO}_{i,\mathrm{B_1}}$ &    $\mathrm{Pct}_{\mathrm{B_1}}$    & $\mathrm{Pct\sigma}_{\mathrm{B_1}}$ & $a^\mathrm{NNLO}_{i,\mathrm{B_2}}$  &    $\mathrm{Pct}_{\mathrm{B_2}}$    &    $\mathrm{Pct\sigma}_{\mathrm{B_2}}$ \\ \hline
			1    & 1.13  & 1.070  (348) & $-$5.3\%   & $-$0.2 & 0.947  (273) & $-$16.2\% & $-$0.7 & 1.266  (200) & 12.0\%  & 0.7  \\
			2    & $-$0.42 & 0.039  (786) & $-$109.3\% & 0.6  & $-$0.266 (359) & $-$36.7\% & 0.4  & $-$0.336 (187) & $-$20.0\% & 0.4  \\
			3    & $-$0.36 & $-$0.416 (459) & 15.6\%   & $-$0.1 & $-$0.357 (166) & $-$0.8\%  & 0.0  & $-$0.325 (110) & $-$9.7\%  & 0.3  \\
			4    & 0.75  & 0.662  (247) & $-$11.7\%  & $-$0.4 & 0.429  (210) & $-$42.8\% & $-$1.5 & 0.618  (154) & $-$17.6\% & $-$0.9 \\
			5    & 0.6   & 0.834  (123) & 39.0\%   & 1.9  & 0.735  (180) & 22.5\%  & 0.8  & 0.546  (146) & $-$9.0\%  & $-$0.4 \\
			6    & $-$1.61 & 0.084  (913) & $-$105.2\% & 1.9  & $-$0.482 (891) & $-$70.1\% & 1.3  & $-$1.329 (799) & $-$17.5\% & 0.4  \\
			7    & 1.22  & 0.716  (902) & $-$41.3\%  & $-$0.6 & 1.683  (428) & 38.0\%  & 1.1  & 1.424  (365) & 16.7\%  & 0.6  \\
			8    & $-$0.53 & $-$0.250 (448) & $-$52.8\%  & 0.6  & $-$0.635 (470) & 19.8\%  & $-$0.2 & $-$0.552 (235) & 4.2\%   & $-$0.1 \\
			9    & 0.22  & $-$0.546 (361) & $-$348.2\% & $-$2.1 & 0.104  (60) & $-$52.7\% & $-$1.9 & 0.199  (57) & $-$9.5\%  & $-$0.4 \\
			10   & $-$0.17 & 0.052  (997) & $-$130.6\% & 0.2  & $-$0.104 (61) & $-$38.8\% & 1.1  & $-$0.146 (54) & $-$14.1\% & 0.4  \\
			11   & 1.02  & 0.020  (992) & $-$98.0\%  & $-$1.0 & 0.124  (802) & $-$87.8\% & $-$1.1 & 1.056  (493) & 3.5\%   & 0.1  \\
			12   & $-$1.44 & $-$1.457 (281) & 1.2\%    & $-$0.1 & $-$1.106 (190) & $-$23.2\% & 1.8  & $-$1.609 (159) & 11.7\%  & $-$1.1 \\
			13   & $-$0.22 & $-$0.089 (135) & $-$59.5\%  & 1.0  & $-$0.151 (81) & $-$31.4\% & 0.9  & $-$0.207 (53) & $-$5.9\%  & 0.2  \\
			14   & 0.26  & 1.028  (872) & 295.4\%  & 0.9  & 0.425  (379) & 63.5\%  & 0.4  & 0.242  (115) & $-$6.9\%  & $-$0.2 \\
			15   & 0.79  & 1.263  (668) & 59.9\%   & 0.7  & 0.794  (356) & 0.5\%   & 0.0  & 1.005  (250) & 27.2\%  & 0.9  \\
			16   & $-$0.76 & $-$0.023 (988) & $-$97.0\%  & 0.7  & $-$0.259 (564) & $-$65.9\% & 0.9  & $-$0.775 (337) & 2.0\%   & 0.0  \\
			17   & 0.12  & $-$0.164 (652) & $-$236.7\% & $-$0.4 & 0.084  (74) & $-$30.0\% & $-$0.5 & 0.126  (41) & 5.0\%   & 0.1  \\
			WAIC & $-$     & 8.610   & $-$        & $-$    & 50.309  & $-$       &      & 54.235  & $-$       & $-$    \\
			LOO  & $-$     & 2.528   & $-$        & $-$    & 44.176  & $-$       &      & 49.085  & $-$       & $-$    \\
			PM   & $-$     & 0.140   & $-$        & $-$    & 0.035   & $-$       &      & 0.012   & $-$       & $-$
		\end{tabular}
	\end{ruledtabular}
\end{table}

\begin{table}[h]
	\caption{The comparison of the NNLO fitting values for extra examples. The subscripts tr, exp, B$_1$ and B$_2$ in the first row represent the truth values, the experimental values, the theoretical values from Model B$_1$ and Model B$_2$, respectively. The experimental values in the third column are sampled from the truth values. $O_i$ is defined in Eq. \eqref{A2}.}\label{table12}
	\begin{ruledtabular}
		\begin{tabular}{lrrrr}
			$i$ &$10^2O_{i,\mathrm {tr}}$&$10^2O_{i,\mathrm {exp}}$~~~~~&$10^2O_{i,\mathrm {B_1}}$~~~~~&$10^2O_{i,\mathrm {B_2}}$~~~~~\\\hline
			1  & 21.297  & 21.816  $\pm$  0.319 & 21.706  $\pm$  0.654 & 21.773  $\pm$  0.570 \\
			2  & $-$26.253 & $-$26.494 $\pm$  0.394 & $-$26.540 $\pm$  2.188 & $-$26.411 $\pm$  1.490 \\
			3  & $-$10.613 & $-$10.697 $\pm$  0.159 & $-$10.703 $\pm$  0.302 & $-$10.669 $\pm$  0.260 \\
			4  & 40.713  & 40.058  $\pm$  0.611 & 39.987  $\pm$  1.333 & 40.117  $\pm$  1.194 \\
			5  & 0.063   & 0.064   $\pm$  0.001 & 0.065   $\pm$  0.002 & 0.064   $\pm$  0.002 \\
			6  & $-$18.401 & $-$19.036 $\pm$  0.276 & $-$19.248 $\pm$  0.378 & $-$18.865 $\pm$  0.393 \\
			7  & $-$0.675  & $-$0.658  $\pm$  0.010 & $-$0.669  $\pm$  0.019 & $-$0.659  $\pm$  0.018 \\
			8  & 22.194  & 21.941  $\pm$  0.333 & 21.793  $\pm$  0.587 & 21.961  $\pm$  0.432 \\
			9  & 4.492   & 4.513   $\pm$  0.067 & 4.514   $\pm$  0.348 & 4.517   $\pm$  0.328 \\
			10 & $-$11.726 & $-$11.770 $\pm$  0.176 & $-$11.761 $\pm$  7.191 & $-$11.757 $\pm$  5.937 \\
			11 & 4.687   & 4.789   $\pm$  0.070 & 4.776   $\pm$  0.933 & 4.787   $\pm$  0.724 \\
			12 & $-$0.624  & $-$0.644  $\pm$  0.009 & $-$0.638  $\pm$  0.022 & $-$0.643  $\pm$  0.020 \\
			13 & $-$17.245 & $-$17.328 $\pm$  0.259 & $-$17.317 $\pm$  0.548 & $-$17.347 $\pm$  0.457 \\
			14 & $-$13.943 & $-$14.023 $\pm$  0.209 & $-$14.046 $\pm$  0.987 & $-$14.040 $\pm$  0.494 \\
			15 & 10.468  & 10.646  $\pm$  0.157 & 10.604  $\pm$  0.474 & 10.624  $\pm$  0.350 \\
			16 & $-$10.156 & $-$10.323 $\pm$  0.152 & $-$10.310 $\pm$  0.874 & $-$10.326 $\pm$  0.639 \\
			17 & 6.774   & 6.756   $\pm$  0.102 & 6.752   $\pm$  0.427 & 6.743   $\pm$  0.300
		\end{tabular}
	\end{ruledtabular}
\end{table}

\begin{table}[H]
\begin{longtable}{lrrrrrrr}
	\caption{Same as Table \ref{table2}, except for the extra example.}\label{table5}\\
	\hline\hline  $i$&\multicolumn{1}{c}{LO}&$\mathrm{Pct}_\mathrm{LO}$&\multicolumn{1}{c}{NLO}&$\mathrm{Pct}_\mathrm{NLO}$&\multicolumn{1}{c}{HO}&$\mathrm{Pct}_\mathrm{HO}$ &Theory\\
	\hline%\addlinespace[2pt]
	\endfirsthead
	\hline\hline Obs&\multicolumn{1}{c}{LO}&$\mathrm{Pct}_\mathrm{LO}$&\multicolumn{1}{c}{NLO}&$\mathrm{Pct}_\mathrm{NLO}$&\multicolumn{1}{c}{HO}&$\mathrm{Pct}_\mathrm{HO}$ &Theory\\
	\hline%\addlinespace[2pt]
	\endhead
	\hline\hline
	\endfoot
	\hline\hline
	\endlastfoot
	$1_\mathrm{tr}$     & 17.66  & 82.9\% & 3.46   & 16.2\% & 0.18   & 0.8\%  & 21.30  \\
	$1_{\mathrm{B}_1}$  & 17.66  & 82.5\% & 3.31   & 15.5\% & 0.42   & 2.0\%  & 21.40  \\
	$1_{\mathrm{B}_2}$  & 17.66  & 82.1\% & 3.67   & 17.1\% & 0.19   & 0.9\%  & 21.52  \\
	$2_\mathrm{tr}$     & $-$22.72 & 86.6\% & $-$5.134 & 19.6\% & 1.604  & $-$6.1\% & $-$26.25 \\
	$2_{\mathrm{B}_1}$  & $-$22.72 & 85.5\% & $-$3.888 & 14.6\% & 0.044  & $-$0.2\% & $-$26.57 \\
	$2_{\mathrm{B}_2}$  & $-$22.72 & 86.0\% & $-$5.143 & 19.5\% & 1.459  & $-$5.5\% & $-$26.41 \\
	$3_\mathrm{tr}$     & $-$9.20  & 86.6\% & $-$1.91  & 18.0\% & 0.49   & $-$4.6\% & $-$10.61 \\
	$3_{\mathrm{B}_1}$  & $-$9.20  & 84.4\% & $-$1.85  & 17.0\% & 0.15   & $-$1.4\% & $-$10.90 \\
	$3_{\mathrm{B}_2}$  & $-$9.20  & 86.2\% & $-$1.92  & 18.0\% & 0.44   & $-$4.2\% & $-$10.67 \\
	$4_\mathrm{tr}$     & 34.29  & 84.2\% & 5.85   & 14.4\% & 0.57   & 1.4\%  & 40.71  \\
	$4_{\mathrm{B}_1}$  & 34.29  & 85.7\% & 5.69   & 14.2\% & 0.02   & 0.1\%  & 40.00  \\
	$4_{\mathrm{B}_2}$  & 34.29  & 83.2\% & 6.37   & 15.5\% & 0.56   & 1.3\%  & 41.21  \\
	$5_\mathrm{tr}$     & 0.056  & 87.8\% & 0.01   & 21.5\% & $-$0.006 & $-$9.4\% & 0.063  \\
	$5_{\mathrm{B}_1}$  & 0.056  & 86.3\% & 0.01   & 21.3\% & $-$0.005 & $-$7.6\% & 0.065  \\
	$5_{\mathrm{B}_2}$  & 0.056  & 87.0\% & 0.01   & 21.3\% & $-$0.005 & $-$8.3\% & 0.064  \\
	$6_\mathrm{tr}$     & $-$15.96 & 86.7\% & $-$3.58  & 19.5\% & 1.14   & $-$6.2\% & $-$18.40 \\
	$6_{\mathrm{B}_1}$  & $-$15.96 & 82.2\% & $-$3.75  & 19.3\% & 0.29   & $-$1.5\% & $-$19.42 \\
	$6_{\mathrm{B}_2}$  & $-$15.96 & 84.7\% & $-$3.71  & 19.7\% & 0.83   & $-$4.4\% & $-$18.84 \\
	$7_\mathrm{tr}$     & $-$0.626 & 92.8\% & $-$0.100 & 14.7\% & 0.051  & $-$7.5\% & $-$0.675 \\
	$7_{\mathrm{B}_1}$  & $-$0.626 & 94.8\% & $-$0.094 & 14.3\% & 0.060  & $-$9.1\% & $-$0.661 \\
	$7_{\mathrm{B}_2}$  & $-$0.626 & 94.5\% & $-$0.098 & 14.8\% & 0.062  & $-$9.3\% & $-$0.663 \\
	$8_\mathrm{tr}$     & 18.87  & 85.0\% & 3.80   & 17.1\% & $-$0.48  & $-$2.2\% & 22.19  \\
	$8_{\mathrm{B}_1}$  & 18.87  & 84.4\% & 3.67   & 16.4\% & $-$0.18  & $-$0.8\% & 22.36  \\
	$8_{\mathrm{B}_2}$  & 18.87  & 85.8\% & 3.70   & 16.8\% & $-$0.58  & $-$2.6\% & 21.99  \\
	$9_\mathrm{tr}$     & 3.64   & 81.0\% & 0.74   & 16.6\% & 0.11   & 2.5\%  & 4.49   \\
	$9_{\mathrm{B}_1}$  & 3.64   & 80.4\% & 0.90   & 19.8\% & $-$0.01  & $-$0.2\% & 4.53   \\
	$9_{\mathrm{B}_2}$  & 3.64   & 80.7\% & 0.76   & 16.9\% & 0.11   & 2.4\%  & 4.51   \\
	$10_\mathrm{tr}$    & $-$9.72  & 82.9\% & $-$1.85  & 15.8\% & $-$0.15  & 1.3\%  & $-$11.73 \\
	$10_{\mathrm{B}_1}$ & $-$9.72  & 82.5\% & $-$2.05  & 17.4\% & $-$0.02  & 0.1\%  & $-$11.78 \\
	$10_{\mathrm{B}_2}$ & $-$9.72  & 82.6\% & $-$1.86  & 15.8\% & $-$0.19  & 1.6\%  & $-$11.77 \\
	$11_\mathrm{tr}$    & 4.04   & 86.1\% & 0.76   & 16.3\% & $-$0.11  & $-$2.4\% & 4.69   \\
	$11_{\mathrm{B}_1}$ & 4.04   & 84.4\% & 0.74   & 15.5\% & 0.00   & 0.1\%  & 4.78   \\
	$11_{\mathrm{B}_2}$ & 4.04   & 84.4\% & 0.87   & 18.1\% & $-$0.12  & $-$2.5\% & 4.78   \\
	$12_\mathrm{tr}$    & $-$0.52  & 83.7\% & $-$0.08  & 12.1\% & $-$0.03  & 4.2\%  & $-$0.62  \\
	$12_{\mathrm{B}_1}$ & $-$0.52  & 82.6\% & $-$0.07  & 11.6\% & $-$0.04  & 5.7\%  & $-$0.63  \\
	$12_{\mathrm{B}_2}$ & $-$0.52  & 82.0\% & $-$0.08  & 13.3\% & $-$0.03  & 4.8\%  & $-$0.64  \\
	$13_\mathrm{tr}$    & $-$14.14 & 82.0\% & $-$2.70  & 15.7\% & $-$0.40  & 2.3\%  & $-$17.24 \\
	$13_{\mathrm{B}_1}$ & $-$14.14 & 82.7\% & $-$2.83  & 16.6\% & $-$0.12  & 0.7\%  & $-$17.09 \\
	$13_{\mathrm{B}_2}$ & $-$14.14 & 81.2\% & $-$2.87  & 16.5\% & $-$0.40  & 2.3\%  & $-$17.41 \\
	$14_\mathrm{tr}$    & $-$11.80 & 84.7\% & $-$2.48  & 17.8\% & 0.34   & $-$2.4\% & $-$13.94 \\
	$14_{\mathrm{B}_1}$ & $-$11.80 & 83.1\% & $-$2.54  & 17.9\% & 0.13   & $-$0.9\% & $-$14.21 \\
	$14_{\mathrm{B}_2}$ & $-$11.80 & 84.7\% & $-$2.47  & 17.7\% & 0.33   & $-$2.4\% & $-$13.94 \\
	$15_\mathrm{tr}$    & 9.11   & 87.0\% & 1.99   & 19.0\% & $-$0.63  & $-$6.0\% & 10.47  \\
	$15_{\mathrm{B}_1}$ & 9.11   & 84.6\% & 1.74   & 16.2\% & $-$0.08  & $-$0.7\% & 10.77  \\
	$15_{\mathrm{B}_2}$ & 9.11   & 85.8\% & 2.10   & 19.8\% & $-$0.59  & $-$5.6\% & 10.62  \\
	$16_\mathrm{tr}$    & $-$8.45  & 83.2\% & $-$1.36  & 13.4\% & $-$0.35  & 3.4\%  & $-$10.16 \\
	$16_{\mathrm{B}_1}$ & $-$8.45  & 82.0\% & $-$1.85  & 17.9\% & $-$0.01  & 0.1\%  & $-$10.30 \\
	$16_{\mathrm{B}_2}$ & $-$8.45  & 81.9\% & $-$1.53  & 14.8\% & $-$0.34  & 3.3\%  & $-$10.32 \\
	$17_\mathrm{tr}$    & 5.68   & 83.9\% & 1.15   & 17.0\% & $-$0.06  & $-$1.0\% & 6.77   \\
	$17_{\mathrm{B}_1}$ & 5.68   & 84.5\% & 1.02   & 15.2\% & 0.02   & 0.3\%  & 6.73   \\
	$17_{\mathrm{B}_2}$ & 5.68   & 84.1\% & 1.13   & 16.8\% & $-$0.06  & $-$0.8\% & 6.76
\end{longtable}
\end{table}

\begin{table}[H]
\begin{longtable}{lrrrrrrrrr}
	\caption{Same as Table \ref{table2}, except for the extra example for the NNLO fit.}\label{table11}\\
	\hline\hline  $i$&\multicolumn{1}{c}{LO}&$\mathrm{Pct}_\mathrm{LO}$&\multicolumn{1}{c}{NLO}&$\mathrm{Pct}_\mathrm{NLO}$&\multicolumn{1}{c}{NNLO}&$\mathrm{Pct}_\mathrm{NNLO}$&\multicolumn{1}{c}{HO}&$\mathrm{Pct}_\mathrm{HO}$ &Theory\\
	\hline%\addlinespace[2pt]
	\endfirsthead
	\hline\hline Obs&\multicolumn{1}{c}{LO}&$\mathrm{Pct}_\mathrm{LO}$&\multicolumn{1}{c}{NLO}&$\mathrm{Pct}_\mathrm{NLO}$&\multicolumn{1}{c}{NNLO}&$\mathrm{Pct}_\mathrm{NNLO}$&\multicolumn{1}{c}{HO}&$\mathrm{Pct}_\mathrm{HO}$ &Theory\\
	\hline%\addlinespace[2pt]
	\endhead
	\hline\hline
	\endfoot
	\hline\hline
	\endlastfoot
	$1_\mathrm{tr}$     & 17.66  & 82.9\% & 3.46   & 16.2\% & 0.96   & 4.5\%  & $-$0.79   & $-$3.7\% & 21.30  \\
	$1_{\mathrm{B}_1}$  & 17.66  & 81.4\% & 3.40   & 15.7\% & 0.59   & 2.7\%  & 0.05    & 0.2\%  & 21.71  \\
	$1_{\mathrm{B}_2}$  & 17.66  & 81.1\% & 3.53   & 16.2\% & 1.23   & 5.7\%  & $-$0.66   & $-$3.0\% & 21.77  \\
	$2_\mathrm{tr}$     & $-$22.72 & 86.6\% & $-$5.134 & 19.6\% & 0.961  & $-$3.7\% & 0.64    & $-$2.4\% & $-$26.25 \\
	$2_{\mathrm{B}_1}$  & $-$22.72 & 85.6\% & $-$4.398 & 16.6\% & 0.549  & $-$2.1\% & 0.03    & $-$0.1\% & $-$26.54 \\
	$2_{\mathrm{B}_2}$  & $-$22.72 & 86.0\% & $-$4.936 & 18.7\% & 0.665  & $-$2.5\% & 0.58    & $-$2.2\% & $-$26.41 \\
	$3_\mathrm{tr}$     & $-$9.20  & 86.6\% & $-$1.91  & 18.0\% & 0.33   & $-$3.1\% & 0.16    & $-$1.5\% & $-$10.61 \\
	$3_{\mathrm{B}_1}$  & $-$9.20  & 85.9\% & $-$1.85  & 17.3\% & 0.34   & $-$3.2\% & $-$0.002  & 0.0\%  & $-$10.70 \\
	$3_{\mathrm{B}_2}$  & $-$9.20  & 86.2\% & $-$1.89  & 17.7\% & 0.27   & $-$2.5\% & 0.15    & $-$1.4\% & $-$10.67 \\
	$4_\mathrm{tr}$     & 34.29  & 84.2\% & 5.85   & 14.4\% & 1.59   & 3.9\%  & $-$1.02   & $-$2.5\% & 40.71  \\
	$4_{\mathrm{B}_1}$  & 34.29  & 85.7\% & 5.97   & 14.9\% & $-$0.31  & $-$0.8\% & 0.04    & 0.1\%  & 39.99  \\
	$4_{\mathrm{B}_2}$  & 34.29  & 85.5\% & 6.24   & 15.6\% & 0.69   & 1.7\%  & $-$1.10   & $-$2.7\% & 40.12  \\
	$5_\mathrm{tr}$     & 0.056  & 87.8\% & 0.01   & 21.5\% & $-$0.003 & $-$4.8\% & $-$0.003  & $-$4.6\% & 0.063  \\
	$5_{\mathrm{B}_1}$  & 0.056  & 85.9\% & 0.01   & 21.1\% & $-$0.004 & $-$6.4\% & $-$0.0004 & $-$0.6\% & 0.065  \\
	$5_{\mathrm{B}_2}$  & 0.056  & 86.8\% & 0.01   & 21.3\% & $-$0.003 & $-$4.0\% & $-$0.003  & $-$4.1\% & 0.064  \\
	$6_\mathrm{tr}$     & $-$15.96 & 86.7\% & $-$3.58  & 19.5\% & 0.65   & $-$3.5\% & 0.48    & $-$2.6\% & $-$18.40 \\
	$6_{\mathrm{B}_1}$  & $-$15.96 & 82.9\% & $-$3.71  & 19.3\% & 0.29   & $-$1.5\% & 0.13    & $-$0.7\% & $-$19.25 \\
	$6_{\mathrm{B}_2}$  & $-$15.96 & 84.6\% & $-$3.71  & 19.7\% & 0.55   & $-$2.9\% & 0.25    & $-$1.3\% & $-$18.87 \\
	$7_\mathrm{tr}$     & $-$0.626 & 92.8\% & $-$0.100 & 14.7\% & 0.021  & $-$3.1\% & 0.030   & $-$4.4\% & $-$0.675 \\
	$7_{\mathrm{B}_1}$  & $-$0.626 & 93.6\% & $-$0.094 & 14.0\% & 0.040  & $-$6.0\% & 0.011   & $-$1.6\% & $-$0.669 \\
	$7_{\mathrm{B}_2}$  & $-$0.626 & 95.1\% & $-$0.096 & 14.5\% & 0.030  & $-$4.6\% & 0.033   & $-$5.0\% & $-$0.659 \\
	$8_\mathrm{tr}$     & 18.87  & 85.0\% & 3.80   & 17.1\% & $-$0.92  & $-$4.1\% & 0.44    & 2.0\%  & 22.19  \\
	$8_{\mathrm{B}_1}$  & 18.87  & 86.6\% & 3.69   & 17.0\% & $-$0.86  & $-$4.0\% & 0.09    & 0.4\%  & 21.79  \\
	$8_{\mathrm{B}_2}$  & 18.87  & 85.9\% & 3.70   & 16.8\% & $-$0.96  & $-$4.4\% & 0.35    & 1.6\%  & 21.96  \\
	$9_\mathrm{tr}$     & 3.64   & 81.0\% & 0.74   & 16.6\% & 0.33   & 7.3\%  & $-$0.22   & $-$4.9\% & 4.49   \\
	$9_{\mathrm{B}_1}$  & 3.64   & 80.6\% & 0.94   & 20.9\% & $-$0.07  & $-$1.5\% & $-$0.001  & 0.0\%  & 4.51   \\
	$9_{\mathrm{B}_2}$  & 3.64   & 80.5\% & 0.87   & 19.2\% & 0.23   & 5.1\%  & $-$0.22   & $-$4.9\% & 4.52   \\
	$10_\mathrm{tr}$    & $-$9.72  & 82.9\% & $-$1.85  & 15.8\% & $-$0.60  & 5.1\%  & 0.45    & $-$3.8\% & $-$11.73 \\
	$10_{\mathrm{B}_1}$ & $-$9.72  & 82.6\% & $-$2.06  & 17.5\% & 0.02   & $-$0.1\% & 0.00    & 0.0\%  & $-$11.76 \\
	$10_{\mathrm{B}_2}$ & $-$9.72  & 82.7\% & $-$2.25  & 19.1\% & $-$0.28  & 2.4\%  & 0.49    & $-$4.2\% & $-$11.76 \\
	$11_\mathrm{tr}$    & 4.04   & 86.1\% & 0.76   & 16.3\% & $-$0.21  & $-$4.5\% & 0.10    & 2.2\%  & 4.69   \\
	$11_{\mathrm{B}_1}$ & 4.04   & 84.5\% & 0.69   & 14.4\% & 0.05   & 1.1\%  & 0.00    & 0.0\%  & 4.78   \\
	$11_{\mathrm{B}_2}$ & 4.04   & 84.3\% & 0.88   & 18.3\% & $-$0.21  & $-$4.3\% & 0.08    & 1.7\%  & 4.79   \\
	$12_\mathrm{tr}$    & $-$0.52  & 83.7\% & $-$0.08  & 12.1\% & $-$0.06  & 9.9\%  & 0.04    & $-$5.7\% & $-$0.624 \\
	$12_{\mathrm{B}_1}$ & $-$0.52  & 81.9\% & $-$0.08  & 12.4\% & $-$0.03  & 5.2\%  & $-$0.003  & 0.5\%  & $-$0.638 \\
	$12_{\mathrm{B}_2}$ & $-$0.52  & 81.3\% & $-$0.08  & 13.0\% & $-$0.07  & 11.1\% & 0.04    & $-$5.5\% & $-$0.643 \\
	$13_\mathrm{tr}$    & $-$14.14 & 82.0\% & $-$2.70  & 15.7\% & $-$0.73  & 4.2\%  & 0.32    & $-$1.9\% & $-$17.24 \\
	$13_{\mathrm{B}_1}$ & $-$14.14 & 81.6\% & $-$2.92  & 16.9\% & $-$0.26  & 1.5\%  & $-$0.005  & 0.0\%  & $-$17.32 \\
	$13_{\mathrm{B}_2}$ & $-$14.14 & 81.5\% & $-$2.91  & 16.7\% & $-$0.61  & 3.5\%  & 0.31    & $-$1.8\% & $-$17.35 \\
	$14_\mathrm{tr}$    & $-$11.80 & 84.7\% & $-$2.48  & 17.8\% & 0.73   & $-$5.2\% & $-$0.39   & 2.8\%  & $-$13.94 \\
	$14_{\mathrm{B}_1}$ & $-$11.80 & 84.0\% & $-$2.57  & 18.3\% & 0.32   & $-$2.3\% & 0.01    & $-$0.1\% & $-$14.05 \\
	$14_{\mathrm{B}_2}$ & $-$11.80 & 84.1\% & $-$2.59  & 18.4\% & 0.69   & $-$4.9\% & $-$0.34   & 2.4\%  & $-$14.04 \\
	$15_\mathrm{tr}$    & 9.11   & 87.0\% & 1.99   & 19.0\% & $-$0.37  & $-$3.6\% & $-$0.25   & $-$2.4\% & 10.47  \\
	$15_{\mathrm{B}_1}$ & 9.11   & 85.9\% & 1.77   & 16.7\% & $-$0.29  & $-$2.8\% & 0.02    & 0.2\%  & 10.60  \\
	$15_{\mathrm{B}_2}$ & 9.11   & 85.7\% & 1.91   & 18.0\% & $-$0.15  & $-$1.4\% & $-$0.25   & $-$2.3\% & 10.62  \\
	$16_\mathrm{tr}$    & $-$8.45  & 83.2\% & $-$1.36  & 13.4\% & $-$0.43  & 4.2\%  & 0.08    & $-$0.8\% & $-$10.16 \\
	$16_{\mathrm{B}_1}$ & $-$8.45  & 81.9\% & $-$1.87  & 18.2\% & 0.02   & $-$0.2\% & $-$0.01   & 0.1\%  & $-$10.31 \\
	$16_{\mathrm{B}_2}$ & $-$8.45  & 81.8\% & $-$1.55  & 15.0\% & $-$0.42  & 4.1\%  & 0.09    & $-$0.9\% & $-$10.33 \\
	$17_\mathrm{tr}$    & 5.68   & 83.9\% & 1.15   & 17.0\% & 0.15   & 2.2\%  & $-$0.21   & $-$3.1\% & 6.77   \\
	$17_{\mathrm{B}_1}$ & 5.68   & 84.2\% & 1.03   & 15.3\% & 0.03   & 0.5\%  & 0.005   & 0.1\%  & 6.75   \\
	$17_{\mathrm{B}_2}$ & 5.68   & 84.3\% & 1.08   & 16.1\% & 0.18   & 2.7\%  & $-$0.21   & $-$3.1\% & 6.74
\end{longtable}
\end{table}

\section{Table of the NLO fitting $L_i^r$ with 12 inputs}
\begin{table}[H]
	\caption{The NLO fitting results of $L_i^r$ with 12 inputs. Columns 2 and 3 are fitted by Model A and B, respectively. Columns 4 to 7 are the NLO and NNLO fitting results in Refs. \cite{Yang:2020eif,Bijnens:2014lea}, respectively.}\label{table24}
	\begin{ruledtabular}
		\begin{tabular}{lcccccc}
			LECs          &   NLO Model A    &   NLO Model B    & NLO fit \cite{Bijnens:2014lea} & NNLO fit \cite{Bijnens:2014lea} &NLO Fit 2 \cite{Yang:2020eif}&NNLO Fit 2 \cite{Yang:2020eif}\\ \hline
			$10^3L_1^r$      &  $ 0.92(09)$   &  $ 0.51(15)$  & $1.00(09)$  &  $ 0.53        (06)$ &  $ 0.44(05)$   &  $ 0.43(05)$   \\
			$10^3L_2^r$      & $ 1.41  (08)$  & $ 1.08  (22)$ & $1.48(09)$  &  $ 0.81        (04)$ & $ 0.84  (10)$  & $ 0.74  (04)$  \\
			$10^3L_3^r$      & $ -3.52 (28)$  & $ -3.36 (61)$ & $-3.82(30)$ &  $ -3.07       (20)$ & $ -2.84 (16)$  & $ -2.74 (17)$  \\
			$10^3L_4^r$      & $ \equiv0.3  $ & $ 0.19  (18)$ & $\equiv0.3$ &     $ \equiv0.3 $    & $ 0.30  (33)$  & $ 0.33  (08)$  \\
			$10^3L_5^r$      & $ 1.24  (03)$  & $ 1.10  (37)$ & $1.23(06)$  &  $ 1.01        (06)$ & $ 0.92  (02)$  & $ 0.95  (04)$  \\
			$10^3L_6^r$      & $ 0.13  (06)$  & $ 0.05  (22)$ & $0.14(06)$  &  $ 0.14        (05)$ & $ 0.22  (08)$  & $ 0.20  (03)$  \\
			$10^3L_7^r$      & $ -0.23 (14)$  & $ -0.26 (17)$ & $-0.27(14)$ &  $ -0.34       (09)$ & $ -0.23 (12)$  & $ -0.23 (08)$  \\
			$10^3L_8^r$      & $ 0.53  (12)$  & $ 0.51  (22)$ & $0.55(12)$  &  $ 0.47        (10)$ & $ 0.44  (10)$  & $ 0.42  (09)$  \\
			WAIC             &  $ -3.68   $   &  $ 25.77   $  &      --     &          --          &       --       &      --        \\
			LOOCV            &   $ -5.74 $    &   $ 24.00 $   &      --     &          --          &       --       &      --        \\
			$\chi^2$(d.o.f.) &       --       &      --       &     --(5)   & $             1.0(9)$& $   4.2(4)   $ &   $4.3(9) $
		\end{tabular}
	\end{ruledtabular}
\end{table}

\begin{table}[H]
	\caption{The convergences of 12 observables. The LECs are adopted the NLO fitting results obtained by Model B in Table \ref{table24}. The second to the fourth columns are the contributions at the LO, NLO and HO, respectively. The percentage $\mathrm{Pct}_{\mathrm{LO, NLO, HO}}$ is defined in Eq. \eqref{equ:1}. The last two columns are the theoretical estimates and the experimental inputs, respectively. }\label{table25}
	\begin{ruledtabular}
		\begin{tabular}{lccccc}
			\multicolumn{1}{c}{Observables}&\multicolumn{1}{c}{LO$|\mathrm{Pct}_\mathrm{LO}$}&\multicolumn{1}{c}{NLO$|\mathrm{Pct}_\mathrm{NLO}$}&\multicolumn{1}{c}{HO$|\mathrm{Pct}_\mathrm{HO}$}&Theory&Experiment\\\hline
			$m_s/\hat m|_1$            &$ 25.84  ( 93.8\%  )$&$ 1.84  ( 6.7\%   )$&$ -0.12  ( -0.4\%  )$&$ 27.6   \pm 3.26  $& $27.3_{-1.3}^{+0.7}$ \\
			$m_s/\hat m|_2$            &$ 24.21  ( 87.8\%  )$&$ 3.46  ( 12.6\%  )$&$ -0.11  ( -0.4\%  )$&$ 27.6   \pm 6.55  $& $27.3_{-1.3}^{+0.7}$ \\
			$F_K/F_\pi$                &$ 1.000  ( 84.1\%  )$&$ 0.185 ( 15.5\%  )$&$ 0.004  ( 0.4\%   )$&$ 1.189  \pm 0.040 $& 1.199$\pm$ 0.003     \\
			$f_s$                      &$ 3.782  ( 66.2\%  )$&$ 1.348 ( 23.6\%  )$&$ 0.582  ( 10.2\%  )$&$ 5.712  \pm 0.407 $& $5.712\pm 0.032$     \\
			$g_p$                      &$ 3.782  ( 77.5\%  )$&$ 1.035 ( 21.2\%  )$&$ 0.063  ( 1.3\%   )$&$ 4.880  \pm 0.188 $& $4.958\pm 0.085$     \\
			$a_0^0$                    &$ 0.159  ( 72.4\%  )$&$ 0.044 ( 20.1\%  )$&$ 0.017  ( 7.5\%   )$&$ 0.2197  \pm 0.006 $& $0.2196\pm 0.0034$   \\
			$10a_0^2$                  &$ -0.455 ( 104.0\% )$&$ 0.020 ( -4.5\%  )$&$ -0.002 ( 0.5\%   )$&$ -0.437 \pm 0.018 $& $-0.444\pm 0.012$    \\
			$a_0^{1/2}m_\pi$           &$ 0.142  ( 63.2\%  )$&$ 0.033 ( 14.7\%  )$&$ 0.049  ( 22.0\%  )$&$ 0.224  \pm 0.015 $& $0.224\pm 0.022$     \\
			$10a_0^{3/2}m_\pi$         &$ -0.709 ( 159.0\% )$&$ 0.087 ( -19.4\% )$&$ 0.177  ( -39.6\% )$&$ -0.446 \pm 0.105 $& $-0.448\pm 0.077$    \\
			$f_s^{\prime}$             &&&&$ 0.495  \pm 0.354 $& $ 0.868  \pm 0.049 $ \\
			$g^{\prime}$               &&&&$ 0.397  \pm 0.043 $& $ 0.508  \pm 0.122 $ \\
			$\langle r^2\rangle_S^\pi$ &&&&$ 0.60   \pm 0.18  $& $ 0.61   \pm 0.04  $
			\\
		\end{tabular}
	\end{ruledtabular}
\end{table}

\section{Correlation coefficient matrices}
Here gives the matrices of correlation coefficients for $L_i^r$.

\begin{table}[H]
	\caption{The 17-input NLO fit correlation coefficient matrix.}
	\begin{ruledtabular}
	\begin{tabular}{lrrrrrrrr}
		& $10^3L_1^r$ & $10^3L_2^r$ & $10^3L_3^r$ & $10^3L_4^r$ & $10^3L_5^r$ & $10^3L_6^r$ & $10^3L_7^r$ & $10^3L_8^r$ \\\hline
		$10^3L_1^r$ & 1.000       & 0.497       & $-$  0.635      & $-$  0.023      & $-$  0.024      & $-$  0.030      & 0.015       & $-$  0.026      \\
		$10^3L_2^r$ & 0.497       & 1.000       & $-$  0.746      & $-$  0.098      & $-$  0.042      & $-$  0.080      & 0.033       & $-$  0.045      \\
		$10^3L_3^r$ & $-$  0.635      & $-$  0.746      & 1.000       & 0.019       & 0.062       & 0.015       & $-$  0.011      & 0.042       \\
		$10^3L_4^r$ & $-$  0.023      & $-$  0.098      & 0.019       & 1.000       & $-$  0.375      & 0.585       & $-$  0.004      & $-$  0.213      \\
		$10^3L_5^r$ & $-$  0.024      & $-$  0.042      & 0.062       & $-$  0.375      & 1.000       & $-$  0.101      & $-$  0.134      & 0.637       \\
		$10^3L_6^r$ & $-$  0.030      & $-$  0.080      & 0.015       & 0.585       & $-$  0.101      & 1.000       & 0.410       & $-$  0.459      \\
		$10^3L_7^r$ & 0.015       & 0.033       & $-$  0.011      & $-$  0.004      & $-$  0.134      & 0.410       & 1.000       & $-$  0.722      \\
		$10^3L_8^r$ & $-$  0.026      & $-$  0.045      & 0.042       & $-$  0.213      & 0.637       & $-$  0.459      & $-$  0.722      & 1.000
	\end{tabular}
	\end{ruledtabular}
\end{table}

\begin{table}[h]
	\caption{The NNLO Fit correlation coefficient matrix.}
	\begin{ruledtabular}
		\begin{tabular}{lrrrrrrrr}
			            & $10^3L_1^r$ & $10^3L_2^r$ & $10^3L_3^r$ & $10^3L_4^r$ & $10^3L_5^r$ & $10^3L_6^r$ & $10^3L_7^r$ & $10^3L_8^r$ \\\hline
			$10^3L_1^r$ & 1.000       & 0.477       & $-$  0.678      & 0.060       & $-$  0.012      & $-$  0.013      & 0.018       & $-$  0.026      \\
			$10^3L_2^r$ & 0.477       & 1.000       & $-$  0.718      & $-$  0.041      & $-$  0.026      & $-$  0.026      & 0.006       & $-$  0.030      \\
			$10^3L_3^r$ & $-$  0.678      & $-$  0.718      & 1.000       & 0.044       & 0.143       & 0.002       & 0.030       & 0.044       \\
			$10^3L_4^r$ & 0.060       & $-$  0.041      & 0.044       & 1.000       & $-$  0.584      & 0.292       & 0.017       & $-$  0.351      \\
			$10^3L_5^r$ & $-$  0.012      & $-$  0.026      & 0.143       & $-$  0.584      & 1.000       & 0.163       & 0.008       & 0.543       \\
			$10^3L_6^r$ & $-$  0.013      & $-$  0.026      & 0.002       & 0.292       & 0.163       & 1.000       & 0.123       & $-$  0.093      \\
			$10^3L_7^r$ & 0.018       & 0.006       & 0.030       & 0.017       & 0.008       & 0.123       & 1.000       & $-$  0.610      \\
			$10^3L_8^r$ & $-$  0.026      & $-$  0.030      & 0.044       & $-$  0.351      & 0.543       & $-$  0.093      & $-$  0.610      & 1.000
		\end{tabular}
	\end{ruledtabular}
\end{table}s
\end{document}